\documentclass[aps,prl,twocolumn,a4paper,10pt,notitlepage,footinbib,superscriptaddress,longbibliography]{revtex4-1}
\usepackage[normalem]{ulem}
\usepackage[english]{babel}
\usepackage{lmodern}
\usepackage[utf8]{inputenc}
\usepackage{endnotes}
\usepackage{amssymb,amsmath,amsfonts}
\usepackage{textcomp}
\usepackage{braket}
\usepackage{ulem}
\usepackage{soul}
\usepackage{comment}
\usepackage{graphicx}
\usepackage{dcolumn}
\usepackage{bm}
\usepackage{xcolor}
\usepackage{soul}
\usepackage{tikz}
\usetikzlibrary{positioning, shapes.geometric, arrows.meta, shadows, decorations.pathmorphing, calc, backgrounds,decorations.markings}

\newcommand{\de}{\mathrm{d}}      

\newcommand{\comments}[1]{}   

\definecolor{bostonuniversityred}{rgb}{0.8, 0.0, 0.0}
\definecolor{massimiliano}{RGB}{0,0,255}
\definecolor{marco}{rgb}{0.8, 0.1, 0.7}

\definecolor{blueprl}{RGB}{13.0, 18.0, 180.0 }
\usepackage[colorlinks=true, allcolors=blueprl]{hyperref}

\begin{document}

\title{Testing the Localization Landscape Theory on the Bethe Lattice
}

\date{\today}
\author{Lorenzo Tonetti}
\affiliation{Sorbonne Universit\'e, Laboratoire de Physique Th\'eorique et Hautes Energies, CNRS-UMR 7589, 4 Place Jussieu, 75252 Paris Cedex 05, France}

\author{Leticia F. Cugliandolo}
\affiliation{Sorbonne Universit\'e, Laboratoire de Physique Th\'eorique et Hautes Energies, CNRS-UMR 7589, 4 Place Jussieu, 75252 Paris Cedex 05, France}

\author{Marco Tarzia}
\affiliation{Sorbonne Universit\'e, Laboratoire de Physique Th\'eorique de la Mati\`ere Condens\'ee, 
CNRS-UMR 7600, 4 Place Jussieu, 75252 Paris Cedex 05, France}

\begin{abstract}
The Localization Landscape Theory (LLT) provides a classical picture of Anderson localization by introducing an effective confining potential whose percolation is proposed to coincide with the mobility edge. Although this proposal shows remarkable numerical agreement in three dimensions,
its fundamental validity remains unsettled. Here we test the LLT analytically on the Bethe lattice, where both the Anderson localization transition and the LLT percolation problem are exactly solvable. We find that the two transitions do not coincide, and their critical behaviors differ markedly. In particular, LLT percolation displays standard mean-field percolation criticality that is fundamentally distinct from the peculiar critical behavior of the Anderson transition on the Bethe lattice.
Our results provide an exact benchmark showing that, while geometrically intuitive, the LLT does not capture the true quantum critical properties of localization.
	\end{abstract}
\maketitle

Anderson localization refers to the suppression of wave propagation caused by disorder-induced interference effects. 
Originally proposed by Anderson in the context of electronic transport~\cite{anderson1958absence}, 
the phenomenon arises when multiple scattering from random impurities leads to destructive interference that inhibits diffusion~\cite{lee1985disordered,evers2008anderson,lagendijk2009fifty}.
Anderson localization has been recently observed in a broad range of physical systems, including light~\cite{lagendijk2009fifty, segev2013anderson}, cold atomic gases~\cite{aspect2009anderson,roati2008anderson,billy2008direct,kondov2011three,jendrzejewski2012three,semeghini2015measurement}, kicked rotors~\cite{chabe2008experimental}, and classical sound elastic waves~\cite{hu2008localization}.
The effect is particularly pronounced in low-dimensional systems, where even infinitesimal disorder can exponentially 
localize all eigenstates, and inhibit a true metallic phase~\cite{mott1961theory,gor1996particle,Abrahams79}.
In higher dimensions, localization occurs only beyond a critical disorder strength, 
the mobility edge, giving rise to a transition between localized (insulating phase) and extended (metallic phase) states~\cite{Kramer93,wegner1979mobility,evers2008anderson}. In the last decades the field has thrived due to two major developments: the interplay between 
localization and topology (topological insulators~\cite{Ludwig_2016,PhysRevLett.81.862,PhysRevLett.81.4704,PhysRevLett.82.4524,PhysRevB.60.4245,PhysRevB.61.10267,PhysRevB.63.235318}) and interactions (many-body localization~\cite{Basko2006,gornyi2005interacting,Nandkishore2015, abanin-colloquium-2019, alet-many-body-2018, sierant-many-body-2025}).

The tight-binding Anderson model (AM) is the minimal framework for disorder-induced localization~\cite{anderson1958absence}. It describes noninteracting electrons hopping on a lattice with random on-site energies drawn from a bounded distribution. In one dimension, all eigenstates are exponentially localized, as shown by transfer-matrix and recursive Green’s function methods~\cite{anderson1958absence,mott1961theory}. In two dimensions, scaling theory similarly predicts localization of all states for any finite disorder~\cite{Abrahams79}. In contrast, in three (and higher) dimensions, a true localization–delocalization transition occurs, but the mobility edge and critical exponents--such as the divergence of the localization 
length--are known only from numerical studies~\cite{PhysRevLett.105.046403,Kramer93,evers2008anderson,lagendijk2009fifty,tarquini2017critical}.

One obstacle to progress is that localization emerges from intricate interference patterns rather than from obvious features of the disorder potential. Since localization centers do not correlate simply with potential extrema, no natural classical limit exists from which to develop perturbation theory. In this context, 
a fascinating and influential proposal by Filoche and Mayboroda~\cite{filoche2012universal} proposes a new framework for understanding 
Anderson localization. Their approach, known as the Localization Landscape Theory (LLT), 
introduces a scalar field--the Localization Landscape--whose inverse defines an effective classical confining potential which delineates regions in which quantum states are spatially trapped~\cite{filoche2012universal,PhysRevLett.116.056602,arnold2019localization,arnold2019computing,david2021landscape}. The delocalization transition is then interpreted as a geometric percolation process: the mobility edge corresponds to the energy at which the classically allowed regions of the effective potential first percolate through the system. The numerical evaluation of the mobility edge 
in the three-dimensional AM and the percolation transition of the LLT compare favourably for low-energy states close
to the bottom of the spectrum~\cite{filoche2024anderson}.

This intriguing proposal naturally calls for a test in a setting where both the AM and the percolation of the LLT can be solved exactly. Such a test can indeed be realized on the 
Bethe lattice--an infinite, loopless, and homogeneous graph in which each site is connected to 
$K+1$ nearest neighbors. The AM on the Bethe lattice exhibits a localization transition at a finite disorder strength~\cite{abou1973selfconsistent}, 
characterized by mean-field critical exponents (e.g., a localization-length exponent $\nu=1$) and nontrivial fractal properties of the wave functions at criticality~\cite{abou1973selfconsistent,zirnbauer1986localization,verbaarschot1988graded,mirlin1991localization,mirlin1991universality,fyodorov1991localization,fyodorov1992novel,mirlin1994statistical,tikhonov2019statistics,tikhonov2019critical,biroli2010anderson,biroli2022critical}. In this work, we investigate the percolation transition of the LLT and its critical properties, and we demonstrate that they do not coincide with those of the AM localization transition on the same geometry. Although the LLT provides an excellent approximation to the mobility edge of the AM in three dimensions, it fails to 
capture  the precise transition curve and the critical behavior in more general geometries.

\begin{figure}[b!]
    \begin{center}
\vspace{-0.2cm}
\scalebox{0.62}{

\begin{tikzpicture}[
  >=latex,
  thick,
  font=\large,
  roundnode/.style={
    circle,
    draw=black,
    fill=black,
    inner sep=1pt,
    minimum size=1mm
  },
  arrowMid/.style={
    postaction={
      decorate,
      decoration={
        markings,
        mark=at position 0.5 with {\arrow{latex}}
      }
    }
  }
]

\node[roundnode, minimum size=5mm, fill=white] (iLeft) at (0,0) {$i$};
\node at (0.6,0.6){$\mathcal{G}_{ii},\eta_i,u_i$};

\node[] (c1l) at (135:3.5) {};
\node[] (c1r) at (105:3.5) {};
\draw[fill=lightgray] (c1l) .. controls (150:0.6) and (90:0.6) .. (c1r);
\node[] (c2l) at (195:3.5) {};
\node[] (c2r) at (165:3.5) {};
\draw[fill=red, fill opacity = 0.4] (c2l) .. controls (210:0.6) and (150:0.6) .. (c2r);
\node[] (c3l) at (255:3.5) {};
\node[] (c3r) at (225:3.5) {};
\draw[fill=lightgray] (c3l) .. controls (270:0.6) and (210:0.6) .. (c3r);
\node[] (c4l) at (15:3.5) {};
\node[] (c4r) at (-15:3.5) {};
\draw[fill=lightgray] (c4l) .. controls (30:0.6) and (-30:0.6) .. (c4r);
\node[roundnode] (n1) at (120:1.5) {};
\node[] at (120:2.5) {\small$j=1$};
\node[roundnode] (n2) at (180:1.5) {};
\node[] at (179:2.5) {\small$j=k$};
\node[roundnode] (n3) at (240:1.5) {};
\node[] at (240:2.5) {\small$j=K$};

\node[roundnode] (nExtra) at (0:1.5) {};
\node at (1:2.5) {\small$l=K+1$};
\draw[arrowMid] (n1) -- (iLeft);
\draw[arrowMid] (n2) -- (iLeft);
\draw[arrowMid] (n3) -- (iLeft);
\draw[arrowMid] (nExtra) -- (iLeft); 

\foreach \a in {142,150,158,202,210,218}{%
  \draw[fill=black] (iLeft) ++(\a:2.6) circle (1.0pt);
}

\node at (-3.0,3.0) {(a)};

\begin{scope}[shift={(7.5,0)}]
    \node[roundnode, minimum size=5mm, fill=white] (iRight) at (0,0) {$k$};
    \node[] (cR1l) at (135:3.5) {};
    \node[] (cR1r) at (105:3.5) {};
    \draw[fill=red, fill opacity=0.4] (cR1l) .. controls (150:0.6) and (90:0.6) .. (cR1r);
    \node[] (cR3l) at (255:3.5) {};
    \node[] (cR3r) at (225:3.5) {};
    \draw[fill=red, fill opacity=0.4] (cR3l) .. controls (270:0.6) and (210:0.6) .. (cR3r);
    \node[roundnode] (nR1) at (120:1.5) {};
    \node[] at (120:2.5) {$1$};
    \node[roundnode] (nR3) at (240:1.5) {};
    \node[] at (240:2.5) {$K$};
    \node at (0.75,-0.8) {$\mathcal{G}_{k \to i}$};
    \node[roundnode,dotted,minimum size=5mm, fill=white] (nRExtra) at (0:1.5) {$i$};
    \node[above=1mm of nRExtra]{$\mathcal{G}_{ii},\eta_i,u_i$};

    \node at (-3.0,3.0) {(b)};

    \draw[arrowMid] (nR1) -- (iRight);
    \draw[arrowMid] (nR3) -- (iRight);

    \draw[dotted,arrowMid] (iRight) -- node[below=0.1] {$\eta_{k \to i}$} (nRExtra);

  \foreach \a in {155,180, 205}{
    \draw[fill=black] (iRight) ++(\a:2.6) circle (1.0pt);
  }
\end{scope}

\end{tikzpicture}
}
\end{center}
\vspace{-0.2cm}
    \caption{
    (a) Bethe lattice centered on site $i$; colored regions mark the subtrees rooted at its 
$K+1$ nearest neighbors.
(b) Same structure with site $i$ removed (dotted link and site), showing only 
the $k$-th branch. Each disconnected branch 
$j=1, \dots, K+1$ forms an infinite tree with coordination $K+1$, except at the root. 
}
\label{fig:tree}
\end{figure}
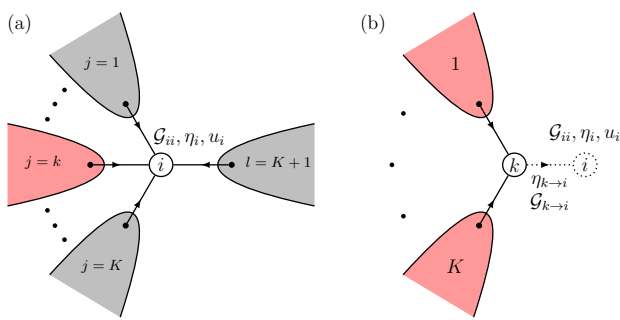

We focus on the {\it Anderson  model} of a non-interacting (spin-less) quantum particle on a lattice
under a disordered potential. The  Hamiltonian is
\begin{equation}
    \hat{\mathcal{H}} = \sum_{i} \epsilon_i \hat{c}_i^\dagger \hat{c}_i - t \sum_{\langle i,j \rangle} \left( \hat{c}_i^\dagger \hat{c}_j + \hat{c}_j^\dagger \hat{c}_i \right) \; , 
    \label{eq:Anderson}
\end{equation}
where $\hat{c}_i^\dagger$ and $\hat{c}_i$ are the creation and annihilation operators at site $i$, 
and $i$ runs from $1$ to the number of lattice sites $N$. 
The on-site disorder potentials $\epsilon_i$ are independent identically distributed 
random variables drawn from the uniform distribution
\begin{equation}
\label{eq:gammas-main}
    \gamma(\epsilon)=\begin{cases}
        1/W &\text{if} \,\,\epsilon\in [-W/2,W/2]\,,\\
        0 &\text{else.}
    \end{cases}
\end{equation}
In the second term in Eq.~(\ref{eq:Anderson}), $t$ is
the hopping amplitude. We take the sum to run over nearest-neighbors on an infinite Bethe lattice, 
i.e., an infinite tree in which each site is connected to $K+1$ nearest neighbors, 
$j=1, \dots, K+1$, as sketched in Fig.~\ref{fig:tree}(a). 
The Bethe lattice can be formally defined as the $N\to\infty$
limit of a random regular graph with $N$ nodes and fixed degree 
$K+1$ (see~\cite{wormald1999models} and the SM for details).
In the absence of disorder ($W=0$), $\hat{\mathcal{H}}$ reduces to the adjacency matrix of the Bethe lattice, 
whose spectrum spans $E \in [-2t\sqrt{K},\,2t\sqrt{K}]$
with an isolated eigenvalue at $E_{\rm iso}(W=0)=-t(K+1)$ corresponding to the uniform mode~\cite{biroli2010anderson}. 
Turning on disorder broadens the spectrum to $E\in [-2t\sqrt{K}-W/2,\,2t\sqrt{K}+W/2]$~\cite{klein1998extended,acosta1992analyticity,aizenman2006stability,bapst2011lifshitz}. 

The {\it Localization Landscape Theory}~\cite{filoche2012universal,PhysRevLett.116.056602,arnold2019localization,arnold2019computing,david2021landscape}
introduces an $N$-dimensional  vector ${\mathbf u}$ as the solution of the 
equation $[ \hat{\mathcal{H}} - E_{\rm min} \hat{\mathcal{I}}] {\mathbf u} = {\mathbf 1}$ with ${\mathbf 1}$ the 
uniform vector with components identically equal to one, and  
defines a real-space potential, $1/u_i$. 
The diagonal shift of $\hat{\cal H}$ by $E_{\rm min}$ is introduced to guarantee that $\hat{\mathcal{H}} - E_{\rm min} \hat{\mathcal{I}}$ is positive definite, thereby ensuring a positive Localization Landscape (the shift is arbitrary, but the choice of $-E_{\rm min}$ corresponds to the minimal and optimal one~\cite{filoche2017localization}). The central idea of the LLT is that $1/u_i$ acts as an effective (correlated) classical potential which is significantly smoother than the disordered potential $\epsilon_i$ (as it satisfies a second-order PDE). The minima of this landscape identify the regions where the low-energy states would localize~\cite{PhysRevLett.116.056602,arnold2019localization}.

The set of nearest-neighboring sites on the lattice 
with lower potential than the electron's energy, $\Omega_E = \{ i \; {{\rm such \; that}} \;  1/u_i \leq E \}$,
are the spatial regions where a particle is classically confined. 
The LLT proposes that, when a 
macroscopic (``giant'') cluster of this kind exists, that is, when the set $\Omega_E$ forms 
a connected path spanning the system,  the quantum particle of energy $E$ delocalizes. Under this intriguing proposal, finding the mobility edge and critical properties 
of the Anderson model simplifies considerably: instead of solving a Schr\"odinger equation, one only needs to analyze a classical 
percolation problem once the potential $1/u_i$  is known. 

    The hierarchical structure of the {\it Bethe lattice} 
    allows for an exact 
determination of the Anderson localization transition and critical behavior
in the thermodynamic limit~\cite{abou1973selfconsistent,zirnbauer1986localization,verbaarschot1988graded,mirlin1991localization,mirlin1991universality,fyodorov1991localization,fyodorov1992novel,mirlin1994statistical,tikhonov2019statistics,tikhonov2019critical,biroli2010anderson}. 
Similarly, the potential $1/u_i$ can be computed analytically, and the associated classical correlated percolation problem can be solved exactly, thereby providing an ideal framework to test the LLT proposal. 
The solution is obtained through a recursive approach~\cite{mezard2009information}:
removing the central site 
$i$ disconnects the lattice into 
$K+1$ independent, semi-infinite subtrees, each rooted at one of $i$'s neighboring sites, see 
Fig.~\ref{fig:tree}(b). These ``cavity'' subtrees are statistically identical to the original Bethe lattice, except that the root site  
has coordination $K$ instead of 
$K+1$. 
Under the assumption (which becomes asymptotically exact in the thermodynamic limit~\cite{bordenave2010resolvent,mezard2009information,biroli2018delocalization}) 
that, in absence of the central node, the probability measures on its neighbors factorize, one obtains a self-consistent equation for the probability measure, 
and then expresses local observables 
in terms of conditionally independent quantities defined at the roots of the cavity subtrees.
We apply this method to the Anderson localization and Localization Landscape percolation.

{\it Anderson localization.} The spectral properties of the eigenvectors and eigenvalues of $\hat{\cal H}$ at energy $E$ are encoded in the statistics of the  elements of the resolvent matrix, ${\cal G} (E) = ({\cal H} - E - {\rm i} 0^+)^{-1} \! ,$ where the last term is an infinitesimal imaginary regulator that softens the pole singularities in the denominator of ${\cal G}$. 
On the Bethe lattice the diagonal elements of ${\cal G}$ verify a set of self-consistent recursion relation~\cite{abou1973selfconsistent,tikhonov2019critical,biroli2010anderson,biroli2018delocalization}:
\begin{eqnarray}
    \label{eq:g-main}
    \mathcal{G}_{ii}^{-1} (E) & = & \epsilon_i  -E - {\rm i} 0^+ - t^2 \sum_{k \in \partial i} \mathcal{G}_{k \to i} \, , \\
    \label{eq:gcav-main}
    \mathcal{G}_{k\to i}^{-1} (E) & = &  \epsilon_k  - E - {\rm i} 0^+ - t^2 \sum_{l \in \partial k \setminus i} \mathcal{G}_{l \to k} \, . 
\end{eqnarray}
The notation $k \to i$ is  standard in the literature, and 
represents the ``message'' that the branch $k$ would  pass
on $i$~\cite{mezard2009information} (see Fig.~\ref{fig:tree}), with $\partial i$ the neighbours of $i$ 
and $\partial k \mkern-2mu \setminus \ \mkern-5mu i $ the neighbours of $k$ excluding $i$. 
Equation~\eqref{eq:gcav-main} should be  interpreted as a self-consistent integral equation for the probability distribution of the cavity Green's functions, whose fixed point 
yields the probability distribution of the diagonal elements through Eq.~\eqref{eq:g-main}~\cite{abou1973selfconsistent,biroli2010anderson,tikhonov2019critical,rizzo2024localized}.
The imaginary part of the Green's functions yield the order parameter distribution function for Anderson localization, from which all the relevant observables can be computed.
In the metallic phase $P({\rm Im} {\cal G})$  converges to a stable distribution, while in the localized phase $P({\rm Im} {\cal G})$ tends to zero in a singular way. 
The critical threshold and  the critical properties can thus be obtained by studying the linear stability of Eqs.~(\ref{eq:g-main})-(\ref{eq:gcav-main}) with respect to a small imaginary part (see the SM)~\cite{abou1974self,tikhonov2019critical,rizzo2024localized}.

{\it Localization Landscape percolation.} From the inversion of the Green's function definition, 
one easily obtains  
$(\hat{\mathcal H} - E_{\rm  min} \hat{\mathcal{I}}){\mathbf u} = {\rm Re} \, {\mathcal G}^{-1} (E_{\rm min}) {\mathbf u}  =  {\mathbf 1} \implies  {\mathbf u} = 
{\rm Re} \, {\mathcal G} (E_{\rm min} ){\mathbf 1}$, i.e. $u_i = \sum_j {\rm Re} \, {\cal G}_{ij} (E_{\rm min})$.
Two methods allow us to obtain the components of the Localization Landscape. The simplest one (devised in the End Matter) uses
a Gaussian representation which includes an auxiliary field $\eta$, that satisfies the recursive equations
\begin{eqnarray}
    \eta_{i}  &=&  1 + t \sum_{k \in \partial i} {\rm Re} \, \mathcal{G}_{k \to i} (E_{\rm min} ) \, \eta_{k \to i}   
   \; , 
 \label{eq:eta-main}
  \\
    \eta_{k \to i}  & = & 1 + t \!\! \sum_{l \in \partial k \setminus i} 
    {\rm Re} \, \mathcal{G}_{l \to k} (E_{\rm min}) \, \eta_{l \to k}   \, ,  
    \label{eq:etaitoj}
  \end{eqnarray}
where the Green's fuctions are solutions of Eqs.~\eqref{eq:gcav-main}. 
The components of the vector ${\mathbf u}$ are given by
\begin{equation}
\label{eq:u-main}
u_i = \sum_j {\rm Re} \, {\mathcal G}_{ij} \, (E_{\rm min}) = {\rm Re} \, {\mathcal G}_{ii} (E_{\rm min}) \, \eta_i
\; .
\end{equation}
An alternative method for the derivation of Eq.~\eqref{eq:u-main} can be found in the SM. 
On the Bethe lattice,  the probability $p_i$ that  site $i$ belongs to the giant cluster is 
\begin{eqnarray}
p_i = \theta(u_i-1/E) \Big[ 1 - \prod_{k \in \partial i} (1-p_{k \to i}) \Big]
\; , 
\end{eqnarray}
with  $p_{k \to i}$ the probability that the 
root of the semi-infinite branch originating at $k \in \partial i$ is connected to the infinite cluster in absence of $i$, which verifies
\begin{eqnarray}
   p_{k\to i} &=& \theta(u_k - 1/E) \Big[ 1 - \prod_{\ell \in \partial k \setminus i } (1 -  p_{\ell \to k}) \Big] \, .
   \label{eq:linear}
\end{eqnarray}
Since the percolation transition is expected to be continuous, the percolation 
threshold can be found studying the linear stability of Eq.~(\ref{eq:linear})  varying $E$ (see SM).

\begin{figure}[!ht]
    \centering
    \includegraphics[width=0.45\textwidth]{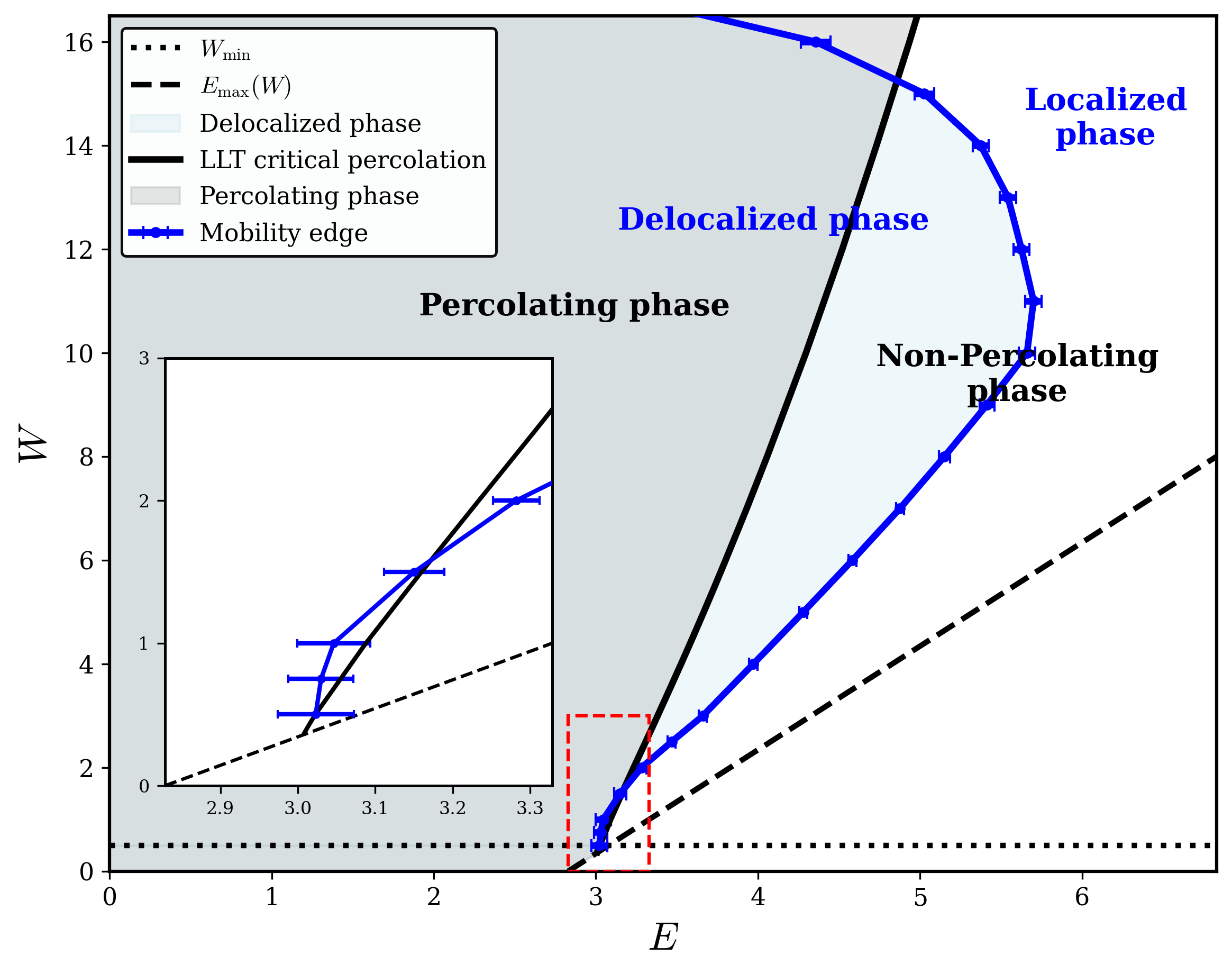}
    \vspace{-0.1cm}
    \caption{\small Phase diagram of the AM on the Bethe lattice with  $t=1$ and $K+1=3$ in the $(E>0,W)$ half-plane 
    (the spectrum is statistically symmetric with respect to $E=0$). 
   Upper bound of the spectrum, $E_{\text{max}} = 2t\sqrt{K} + W/2$ (black dashed line), 
    mobility edge (blue solid line),  
    LLT critical percolation curve (black solid line),  
    and lower bound, $W_{\rm min} \sim 0.3$, below which 
    all $1/u_i$ vanish (black dotted horizontal line). Within the numerical accuracy we have so far, the dashed line for 
    $W_{\rm min} \sim 0.3$ vanishes at a point indistinguishable from the spectral boundary.
    }
    \label{fig:PD}
\end{figure}

From Eqs.~\eqref{eq:g-main}-\eqref{eq:linear}, we determine the critical percolation parameter, $E_c^{\rm perc}(W)$, 
as well as the mobility edge of the AM, $E_c^{\rm loc}(W)$~\cite{abou1973selfconsistent,biroli2010anderson}. 
In order to compare these two quantities, one must subtract the spectral shift $-E_{\rm min}$ introduced when defining the effective potential $1/u_i$. This subtraction yields a negative percolation threshold. However, exploiting the statistical symmetry of the spectrum, one can equivalently express the percolation threshold on the positive-energy side as 
${E_c^{\rm perc}}(W) \mapsto -E_{\rm min} - E_c^{\rm perc}(W)$. 

In Fig.~\ref{fig:PD}
we display the $(E>0, W)$ phase diagram. The black dashed line locates the upper bound of the spectrum,  
$E_{\text{max}} = 2t\sqrt{K} + W/2$~\cite{klein1998extended,acosta1992analyticity,aizenman2006stability,bapst2011lifshitz}. 
The  black solid line is the  LLT critical percolation curve, while the blue 
solid line is the mobility edge.
The two thresholds are close to each other at weak disorder, $W \lesssim 2$ (see the inset of Fig.~\ref{fig:PD}),  
but they depart at 
larger $W$ and cross twice. This is in striking contrast with the $3d$ behavior, where the percolation line is found to provide an upper bound 
for the mobility edge~\cite{filoche2024anderson}.
The horizontal dotted line indicates the disorder strength $W_{\rm min}$  
at which the percolation line meets the spectral edge.
At this point, the effective potential $1/u_i$ vanishes, 
implying that for any value of $E$, the system is in the percolating phase.
 $W_{\rm min}$ coincides with the point where the isolated eigenvalue, which originates from $-t(K+1)$ 
 at zero disorder, merges with the continuous part of the spectrum.
Hence, for $W < W_{\rm min}$, even in the presence of disorder, the smallest eigenvalue of $\hat{\mathcal{H}}$ remains separated from the continuum by a gap.
As a consequence, to ensure that $(\hat{\mathcal{H}} - E_{\rm min} \hat{\mathcal{I}})$ is strictly positive definite, the spectral shift in this region should not be set to $-2t\sqrt{K} - W/2$, but rather to the value of the isolated eigenvalue (which can be computed numerically following the method of Ref.~\cite{biroli2010anderson}).
We argue that implementing this correction would make the percolation threshold follow the spectral edge smoothly down to zero disorder. Details are reported in the SM.
 
Similarly, the Anderson localization threshold reaches the spectral edge at a finite value of $W$ close to $W_{\rm min}$, 
implying that for weaker disorder all eigenstates are delocalized, in agreement with the exact result of Ref.~\cite{aizenman2011extended}.

\begin{figure}[b!]
    \centering
    \includegraphics[width=0.45\textwidth]{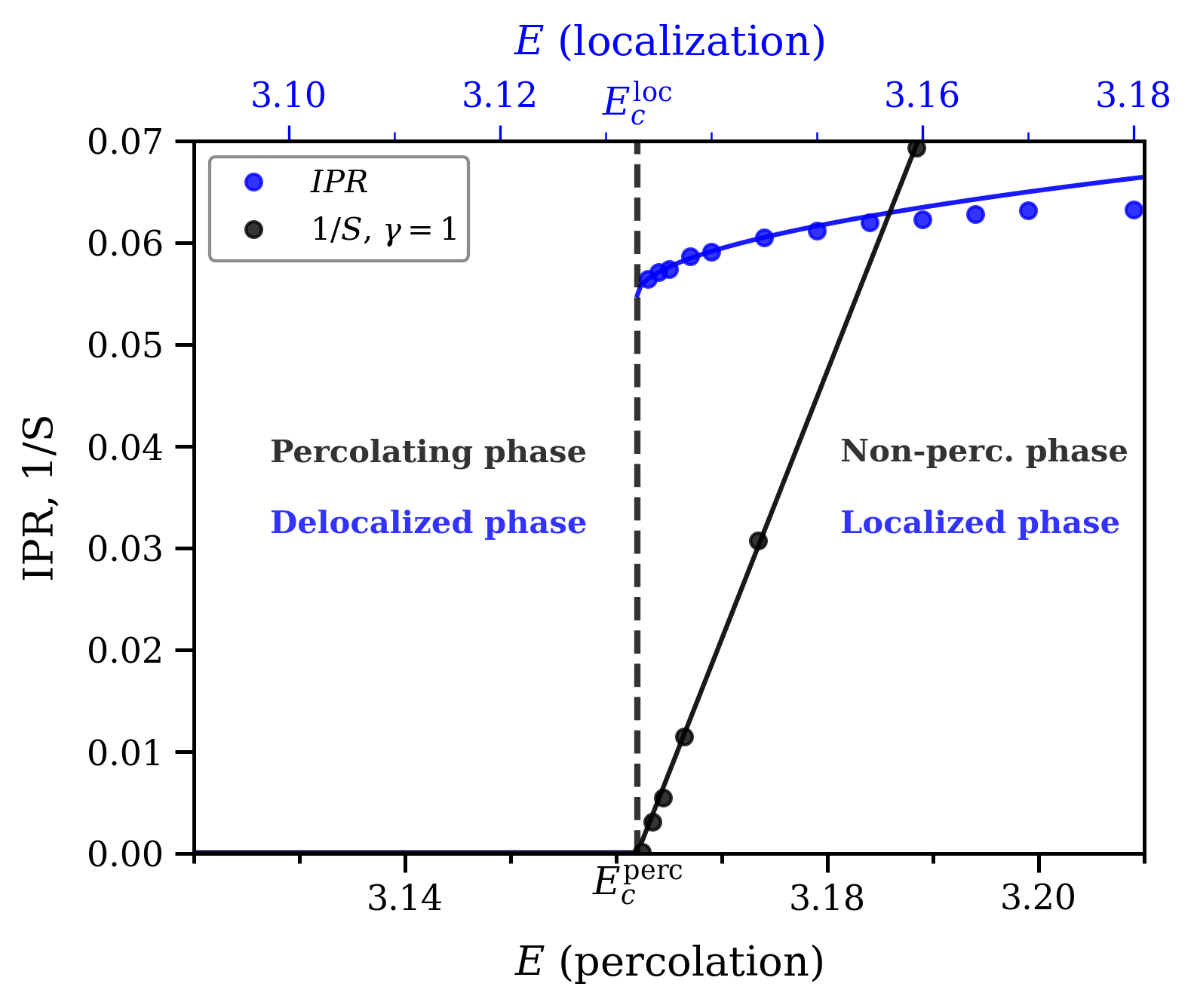}
    \caption{
    Critical behavior of the Anderson model IPR (blue datapoints and square root fit~\cite{rizzo2024localized}, top axis scale) 
    and LLT percolation inverse cluster size $1/S$ (black datapoints and linear fit, 
    bottom axis scale).
 $W = 1.5, K = 2$ and $t = 1$. 
     }
    \label{fig:IPR-S}
\end{figure}

Although the critical percolation curve and the mobility edge do not coincide, 
one can wonder whether the critical properties of Anderson localization are well captured by the percolation problem in the effective potential. 
The natural counterpart of the eigenstates' inverse participation 
rate (IPR) of the AM (a measure of the inverse volume occupied by an eigenstate) 
is the inverse of the average cluster size, $1/S$, in the percolation 
problem. Both quantities are determined with the method detailed in~\cite{rizzo2024localized} and the SM. The IPR and $1/S$  turn out to have very 
different critical behavior. While the IPR jumps from zero to a finite value at the critical energy, $E_c^{\rm loc}$, and 
then grows as a square root of the  distance from $E_c^{\rm loc}$ as conjectured in~\cite{rizzo2024localized},  
$1/S$ grows linearly from zero at $E_c^{\rm perc}$, with the same critical exponent, $\gamma=1$, 
as the one for conventional random percolation in large dimensions.  These results are shown in Fig.~\ref{fig:IPR-S} for a 
choice of parameters such that  $E_c^{\rm perc}$ and $E_c^{\rm loc}$ are very close. 

\begin{figure}[t!]
    \centering
      \includegraphics[width=0.48\textwidth]{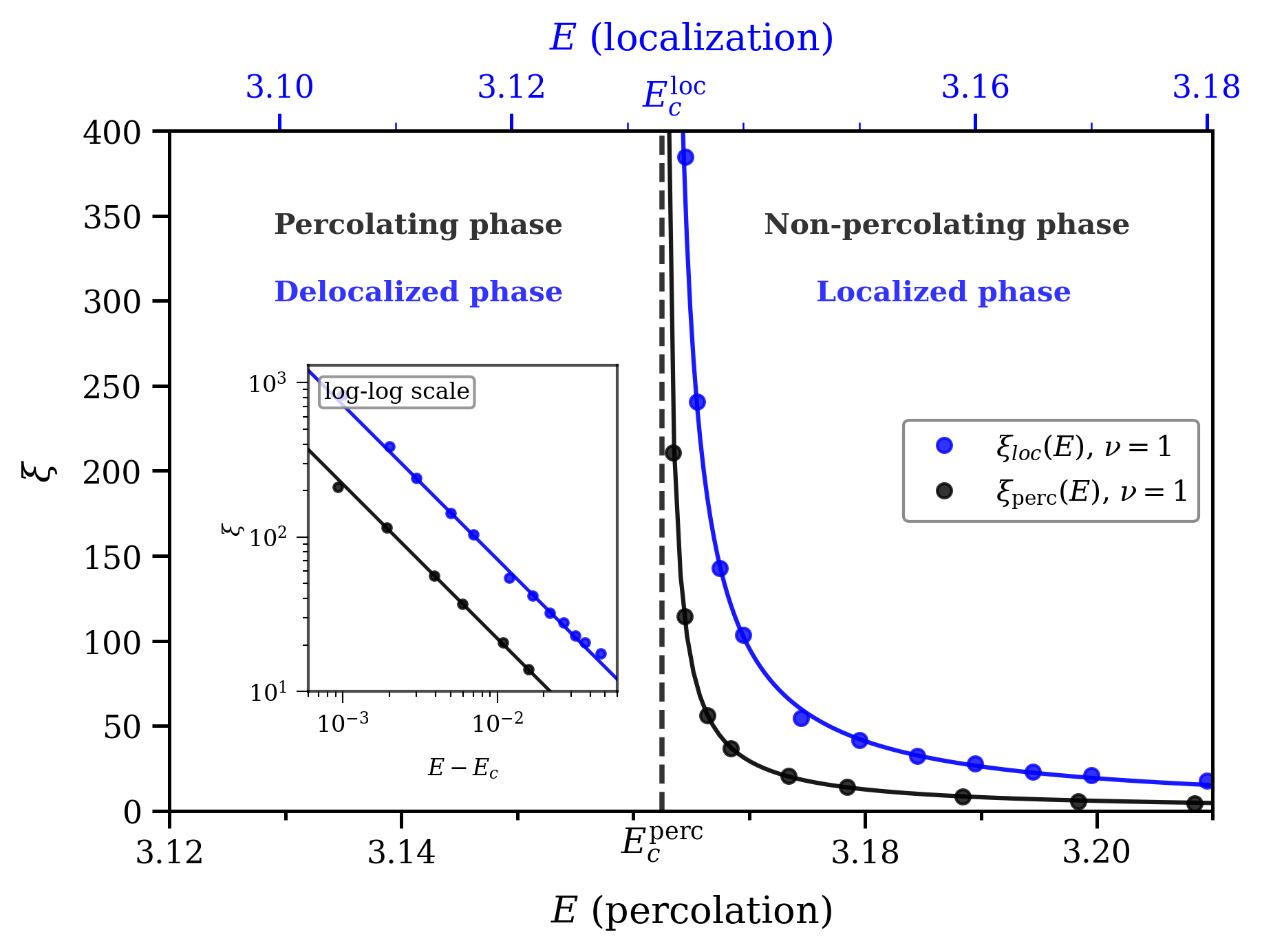}
   \caption{The correlation length of the Anderson model, $\xi_{\rm loc}$ (blue, top axis scale), and LLT critical percolation, $\xi_{\rm perc}$ 
   (black, bottom axis scale),  respectively. 
   Both diverge with critical exponent $\nu=1$ (inset) at the corresponding transitions. $W = 1.5, K = 2$ and  $t = 1$.
  }
    \label{fig:xi}
\end{figure}

Figure~\ref{fig:xi} compares the correlation length of the Anderson transition towards delocalization
and the one of the LLT critical percolation, for the same parameters used in Fig.~\ref{fig:IPR-S}. 
They both diverge, $\xi\simeq (E-E_c)^{-\nu}$, with exponent $\nu=1$, 
but with quite different prefactors. All in all,  the regions where $1/u_i \le E$ largely underestimate 
the actual spatial extent of the 
localized eigenstates of energy $E$.

{\it Conclusion.} 
Although the LLT offers an appealing classical percolation-based interpretation of delocalization, it fails to reproduce 
the critical behavior of the Anderson transition on the Bethe lattice. 
Moreover, although the critical percolation curve lies close to the mobility edge at small $W$, differently from the three-dimensional case, 
it does not provide an upper bound to the Anderson localization's critical energies $E_c^{\rm loc}(W)$, since the two critical curves cross twice. 
These discrepancies suggest that, despite its apparent accuracy in the prediction of the position of the mobility edge at the bottom of the spectrum on the cubic 
lattice~\cite{filoche2024anderson}, the theory does not provide a quantitatively reliable
description of Anderson localization on generic geometries. 
We believe that the discrepancy stems from the different nature of the two transitions. 
Anderson localization is driven by quantum interference effects, whereas the LLT percolation transition is purely classical. 
The correlations induced by the spatial structure of the Localization Landscape are not sufficient to alter the universality class of the underlying percolation problem. 
In particular, while the upper critical dimension for short-range percolation is $d_u^{\rm perc} = 6$, the upper critical dimension for Anderson localization is $d_u^{\rm AL} = \infty$~\cite{baroni2024corrections,tarquini2017critical}. 
This could be related to the deterioration of  the qualitative agreement between localization and the LLT 
as the dimension increases, and in particular on tree-like structures. 

Intriguingly, the percolation threshold of the Bethe lattice with local degree 
$K+1=6$ provides an excellent approximation to the mobility edge in three dimensions 
(SM). It would also be interesting to investigate whether adjusting the spectral shift could help avoid the crossings and bring the LLT percolation line closer to the mobility edge at larger $W$.

{\it Acknowledgments.} This work is supported by a grant from ``Fondation CFM pour la recherche''. 
Numerical calculations were performed at the LPTMC cluster.
We acknowledge funding from the ANR research
grant ManyBodyNet ANR-24-CE30-5851. We thank M. Filoche and M. Vrech for very useful discussions.

\vspace{1em}
\begin{center}
\textbf{End Matter: Cavity Equations for the Localization Landscape}
\end{center}
\vspace{1em}
As anticipated in the main text, following the original formulation of the LLT, we work with a positive definite Hamiltonian.
In the usual Anderson tight-binding Hamiltonian the spectrum is symmetric, and it has been shown in 
Refs.~\cite{klein1998extended,aizenman2006stability} that on the Bethe lattice the energies $E$ take values in the interval $[-2t\sqrt{K}-W/2,2t\sqrt{K}+W/2]$. 
Moreover, there exists an isolated eigenvalue that lies outside this interval~\cite{biroli2010anderson}. In the  $W=0$ case this eigenvalue is $E_{\rm iso}(W=0)=-t(K+1)$, and the corresponding eigenvector $\psi_{\rm iso} = (1/\sqrt{N}, \ldots, 1/\sqrt{N})$ is the one associated to the uniform occupation of all sites. As $W$ increases the eigenvalue $E_{\rm iso}(W)$ gets closer to $-2t\sqrt{K}-W/2$, reaching the left border of the interval $[-2t\sqrt{K}-W/2,2t\sqrt{K}+W/2]$ at $W=W_{\rm min}$.
In order to make $\hat{\mathcal{H}}$ positive definite we need to translate all the random on-site energies $\epsilon_i$ by a fixed amount $-E$. 
The choice of $E$ is arbitrary, but the optimal value has 
been found to be the minimal one~\cite{filoche2017localization}. 
For the Anderson model on the Bethe lattice this is given by
\begin{equation}
E_{\rm min}(W) =
    \begin{cases}
        -2t\sqrt{K}-W/2 &\text{if} \quad W>W_{\rm min}\\
        E_{\rm iso}(W) &\text{if} \quad W\leq W_{\rm min}
    \end{cases}
\end{equation}
The functional form of the Hamiltonian in Eq.~(\ref{eq:Anderson}) remains the same as long as we define the on-site energies as random variables $\varepsilon_i = \epsilon_i -E_{\rm min}(W)$, i.e. drawn uniformly from $[-W/2 -E_{\rm min}(W),\,W/2-E_{\rm min}(W)]$. 
From now on we will omit the explicit $W$-dependence in $E_{\rm min}(W)$.
Having translated the diagonal elements, the Hamiltonian now falls in the class of non-singular M-matrices \cite{plemmons1977m}. This ensures that its inverse exists, and that it has all real non-negative components.

From the inversion of the Green's function definition, $( \hat {\mathcal H}-z \hat {\mathcal I}) \hat {\mathcal G}(z) = \hat {\mathcal I}$
one easily relates the Localization Landscape ${\mathbf u}$ to the system's Green function: 
\begin{equation}
\hat{\mathcal H}_+ {\mathbf u} = ( \hat {\mathcal H}-E_{\rm min} \hat {\mathcal I} )\mathbf{u}= \hat {\mathcal G}^{-1}(E_{\rm min})  {\mathbf u}  =  {\mathbf 1} 
\; .
\end{equation}
$E_{\rm min}$ correspond to the lower edge of the spectrum, where the average and local density of states are strictly equal to zero. Hence, setting $z=E_{\rm min}$ in the definition of ${\cal G}$, the imaginary part of the Green's functions vanish identically, and the cavity equations reduce to the ones for the real part alone. The components of the vector $\mathbf{u}$ 
are given by
\begin{equation}
u_i = \sum_{j=1}^N {\rm Re} \, \mathcal{G}_{ij}(E_{\rm min})\,,
\label{eq:u-G}
\end{equation}
and are all non-negative. 
This is precisely the form adopted in the main text. Henceforth, we will omit the explicit $E_{\rm min}$-dependence and 
the $\rm Re$ symbol.

The element of the Green's function can be written as the expectation values of the product of two Gaussian variables, 
\begin{equation}
    \label{eq:gGaussian-main}
    \mathcal{G}_{ij} =  \frac{1}{Z_0}\int {\cal D}{\mathbf x} \;  x_i x_j \; e^{-S_0[{\mathbf x}]}  
    \equiv \langle x_i x_j\rangle
    \, , 
\end{equation}
where
\begin{eqnarray}
    S_0[{\mathbf x}]  &=& \frac{1}{2}  \sum_i \varepsilon_i x_i^2 - t \sum_{\langle i,j \rangle} x_i x_j =\frac{1}{2}\mathbf{x^t}{\mathcal H}_+\mathbf{x}
    \; , 
\\
    Z_0  &=& \int {\cal D} x \, e^{-S_0[{\mathbf x}]}
    \; .
\end{eqnarray}
The angular brackets indicate an average over such Gaussian measure.
This expression is justified because $\mathcal{H}_+$ is a positive definite matrix, thus $e^{-S_0[{\mathbf x}]}/Z_0$ is a properly normalized Gaussian measure.

The variables $u_i$ in the LLT are given by Eq.~(\ref{eq:u-G}), which in terms of the Gaussian variables becomes
\begin{equation}
    u_i = \frac{1}{Z_0} \int {\cal D}{\mathbf x} \; x_i \sum\limits_j x_j \, e^{-S_0[{\mathbf x}]}  = \langle x_i \sum\limits_j x_j \rangle\, .
\end{equation}
In order to compute this object, it is convenient to linearly couple a source to the sum of Gaussian variables:
\begin{equation}
\label{eq:action}
S_J[{\mathbf x}] = S_0[{\mathbf x}] - J \sum_j x_j 
\;\implies \;
    u_i = \left. \frac{\partial \langle x_i \rangle_J}{\partial J} \right \vert_{J=0} \, .
\end{equation}
In absence of the source term, all the variables $x_i$ have zero mean, and their marginal distributions (the probability
 distribution derived from the  joint one by integrating over all other variables)
 are just Gaussians centered in zero.
In order to compute $\langle x_i \rangle_J$ we need a marginal distribution with non-zero mean. The most general such form is
\begin{equation}
\label{eq:measure}
    \mu_i(x_i) \propto e^{-\frac{x_i^2}{2 \, \mathcal{G}_{ii}} + \xi_i x_i} \,.
\end{equation}
Analogously, for the cavity lattices rooted in $k\in \partial i$, we can write the marginal distribution on a site $k$ as
\begin{equation}
\label{eq:cavmeasure}
    \mu_{k \to i}(x_k) \propto e^{-\frac{x_k^2}{2 \, \mathcal{G}_{k \to i}} + \xi_{k \to i} x_k} \, .
\end{equation}
To find the recursion relations for the Green's function $\mathcal{G}_{ii}$ and the field $\xi_i$, we have to write the marginal distribution
on site $i$ as the integral of the full measure $e^{-S_0[{\mathbf x}]}/Z_0$ over all the variables $\{x_k\}_{k \neq i}$. Since the only variables  coupled to $x_i$ in $S_0[{\mathbf x}]$ are the nearest neighbors $\{x_k\}_{k\in \partial i}$, the marginal distribution on site $i$ can be rewritten as an integral over $\{x_k\}_{k \in \partial i}$ of the terms in $e^{-S_0[{\mathbf x}]}$ which depend on $x_i$, weighted with the joint probability distribution of $\{x_k\}_{k \in \partial i}$. Moreover, since $\{x_k\}_{k \in \partial i}$ are decoupled in $S_0[{\mathbf x}]$, the joint probability distribution is separable as the product of the cavity marginals on each nearest neighbor in absence of $i$. The resulting equation is
\begin{align}
\label{eq:measureeq}
    \nonumber \mu_i(x_i) &= e^{ -\frac{1}{2} \varepsilon_i x_i^2 + J x_i } \int \prod_{k \in \partial i}  \left[ {\rm d} x_k \, \mu_{k \to i} (x_k) \right] e^{t x_i \sum\limits_{k \in \partial i}x_k}\\  
    &\propto 
    e^{- \frac{1}{2} \varepsilon_i x_i^2 + J x_i + \frac{1}{2} \sum\limits_{k \in \partial i} (\xi_{k \to i} + t x_i)^2 \mathcal{G}_{k \to i}} \, .
\end{align}
Analogously, we can relate the marginal distribution on site $k$ in the cavity lattice to the ones on sites $l\in \partial k\setminus i$ on their respective cavity lattices:
\begin{align}
\label{eq:cavmeasureeq} 
    \hspace{-0.1cm}
    \mu_{k \to i}(x_k)  &\propto e^{- \frac{1}{2} \varepsilon_i x_k^2 + J x_k + \frac{1}{2} \sum\limits_{l \in \partial k \setminus i} (\xi_{l \to k} + t x_k)^2 \mathcal{G}_{l \to k}} \, .
\end{align}
Enforcing in Eqs.~(\ref{eq:measureeq}) and (\ref{eq:cavmeasureeq}) the functional forms of Eqs.~(\ref{eq:measure}) and (\ref{eq:cavmeasure}), we immediately obtain
\begin{align}
    \label{eq:gENDMATTER}
    \mathcal{G}_{ii}^{-1} & =  \varepsilon_i - t^2 \sum_{k \in \partial i} \mathcal{G}_{k \to i} \, ,\\
    \xi_{i} & = J + t \sum_{k \in \partial i} \mathcal{G}_{k \to i} \xi_{k \to i}   \, ,\\
    \label{eq:cavg-main}
    \mathcal{G}_{k \to i}^{-1} & = \varepsilon_k - t^2 \sum_{l \in \partial k \setminus i} \mathcal{G}_{l \to k} \,  ,\\
    \xi_{k \to i} & = J + t \sum_{l \in \partial k \setminus i}   \mathcal{G}_{l \to k} \xi_{l \to k} \, .
\end{align}
It is now convenient to introduce the rescaled fields $\eta_{k \to i}$ such that $J\eta_{k \to i} = \xi_{k \to i}$, in terms of which the second and fourth equation read
\begin{align}
    \eta_{i} &= 1 + t \sum_{k \in \partial i} \mathcal{G}_{k \to i} \eta_{k \to i}   
    \, ,  \\
    \eta_{k \to i}  &= 1 + t \sum_{l \in \partial k \setminus i} 
    \mathcal{G}_{l \to k} \eta_{l \to k}   \, , 
\end{align}
which are now completely independent of the value of the source $J$.

The last step is to express $\langle x_i \rangle_J$ in terms of the cavity fields. Since we know its marginal distribution~(\ref{eq:measure}) we just need to compute
\begin{equation}
    \langle x_i \rangle_J =  \int {\rm d} x_i\, x_i \, \mu_i(x_i)=\mathcal{G}_{ii} J \eta_i \,,
\end{equation}
Finally, taking the derivative with respect to the source $J$, we obtain 
\begin{equation}
    \label{eq:etau-main}
    u_i = \mathcal{G}_{ii}\eta_i\,.
\end{equation}
\bibliography{references.bib}


\clearpage

\onecolumngrid

\setcounter{equation}{0}
\renewcommand{\theequation}{S\arabic{equation}}

\setcounter{table}{0}
\renewcommand{\thetable}{S\arabic{table}}

\noindent
\begin{center}
\bfseries\large Supplemental Material\\[6pt]
\end{center}

\vspace{5pt}

\noindent
\begin{center}
\begin{minipage}{0.75\textwidth}

\small
This document constitutes the Supplemental Material for the paper Testing the Localization Landscape Theory on the Bethe Lattice.
We provide details on the definition of the Bethe lattice as the limit of random regular graphs, present the derivation of the cavity equations for Anderson localization and for the Localization Landscape Theory, describe the numerical methods employed, and discuss the solution of the cavity equations in an independent-site approximation
and in the high-connectivity limit.
\end{minipage}
\end{center}

\setcounter{secnumdepth}{3}
\setcounter{figure}{0}
\renewcommand{\thefigure}{S\arabic{figure}}
\setcounter{table}{0}
\renewcommand{\thetable}{S\arabic{table}}
\setcounter{section}{0}
\renewcommand{\thesection}{\Roman{section}}
\setcounter{subsection}{0}
\renewcommand{\thesubsection}{\Alph{subsection}}  
\setcounter{subsubsection}{0}
\renewcommand{\thesubsubsection}{\arabic{subsubsection}}  
\setcounter{equation}{0}
\renewcommand{\theequation}{S\arabic{equation}}

\tableofcontents
\clearpage

In this Supplemental Material, we introduce random regular graphs and the Bethe lattice definition. Then, we summarize some specific features of the well-known solution of Anderson localization on the Bethe lattice, and next we present the theoretical framework developed to study the Localization Landscape percolation problem on the same structure.
In Section~\ref{sec:RRG} we present the definition of the Bethe lattice and its properties.
In Section~\ref{sec:Anderson}, we recall the definition of the Anderson model and the main tools used to analyze its localization properties on the Bethe lattice.
We then end with a description of numerical and analytic methods.

The organization of the document is the following.
Section~\ref{sec:EQGU} introduces the Localization Landscape percolation problem and derives the equations needed to characterize its critical behavior also on the Bethe lattice.
Section~\ref{sec:numerics} provides details of the numerical methods employed.
Finally, Section~\ref{sec:analytic} describes the analytical approaches used to study the relevant equations within the cavity framework, including an assumption of statistical independence and the large-connectivity limit.

\section{Random Regular Graphs}
\label{sec:RRG}

Random regular graphs are an important class of random graphs in which each node has exactly the same degree $K+1$, ensuring uniform connectivity across the network. They are random in the sense that a random regular graph with $N$ vertices is a graph drawn uniformly from the set of all graphs with $N$ nodes and fixed degree~\cite{Bollobas,wormald1999models}. These graphs serve as useful models of space in statistical physics since, in the thermodynamic limit, they represent the infinite dimensional limit for Euclidean lattices, while preserving a finite local connectivity. In contrast, the usual mean-field approximation performed making the graph fully connected, has infinite connectivity in the thermodynamic limit, and loses the notion of distance, since all pairs of sites are connected by a path of length $1$. Moreover, for the Anderson localization problem one cannot observe the localization transition in the fully-connected geometry~\cite{EversMirlin}. 

An important feature of random regular graphs is that they are locally tree-like. Indeed, it can be shown that the typical loop length scales as 
$\ln N/\ln K$~\cite{wormald1999models}.
The infinite graph associated with the thermodynamic limit of a random regular graph is the so-called Bethe lattice. Since the loop length diverges as 
$N\to\infty$, from the perspective of a generic reference site the lattice appears as an infinite tree with fixed connectivity.
Here, its definition as the thermodynamic limit of a random regular graph is essential. Because random regular graphs exhibit no intrinsic hierarchy among nodes, taking the limit $N\to\infty$ implies that each site sees the remainder of the lattice as an infinite tree and the system becomes translationally invariant.

\begin{figure}[ht]
    \centering
    \includegraphics[width=0.49\textwidth]{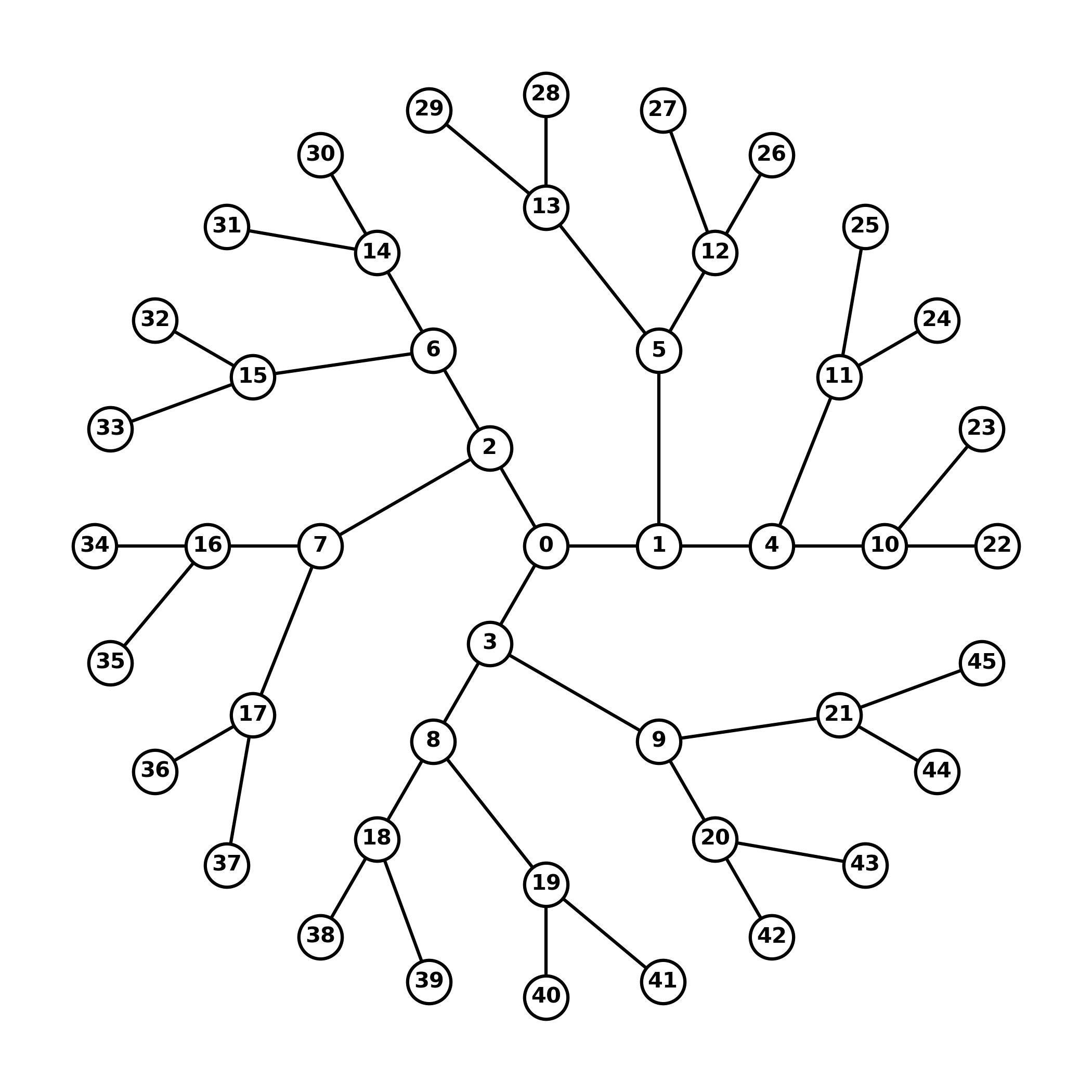}
    \hfill
    \includegraphics[width=0.49\textwidth]{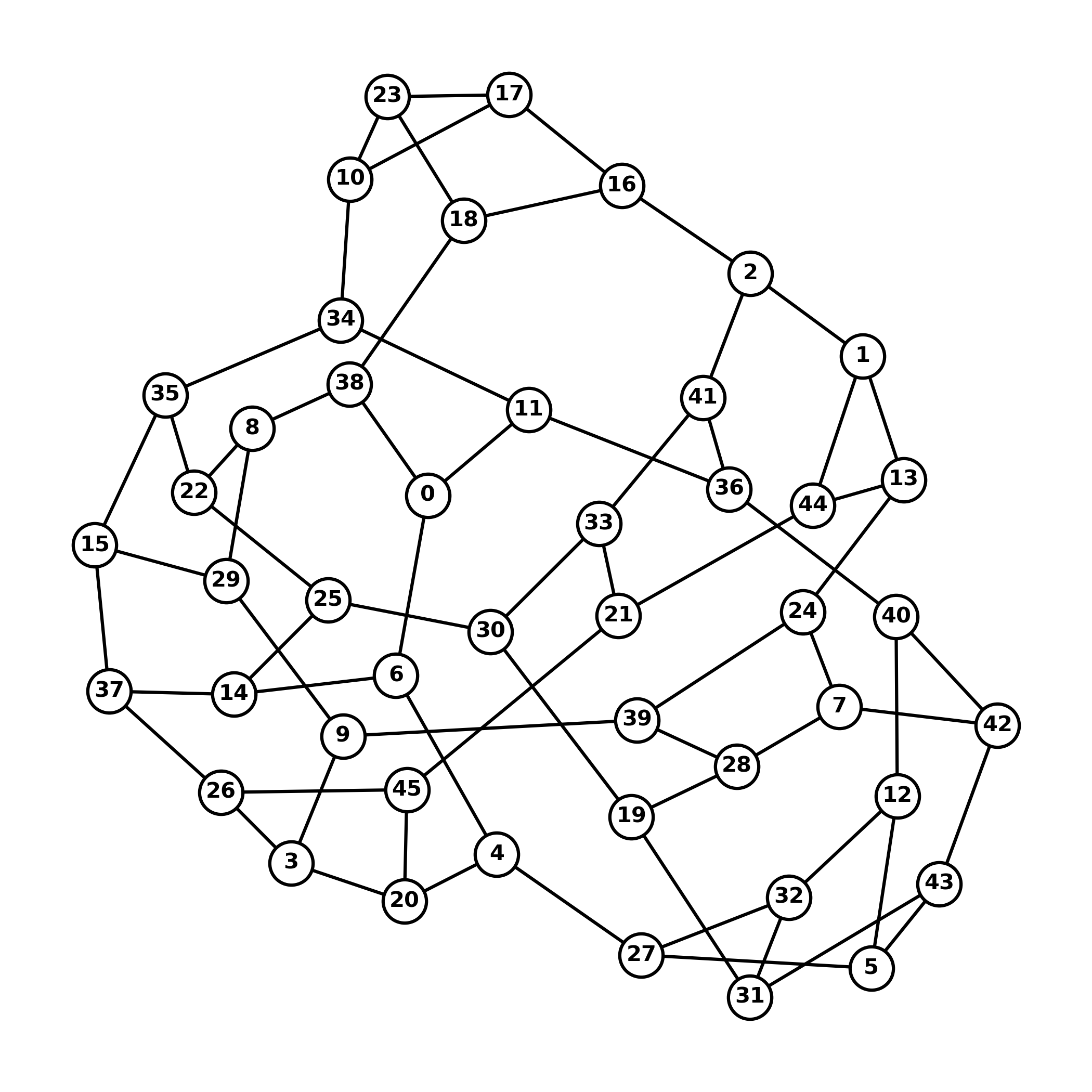}
    \caption{\small{Schematic representation of a Cayley tree and a random regular graph. Left: Cayley 
    tree with $4$ generations and coordination number $K+1 = 3$ (thus having $46$ nodes). Right: realization of a random regular graph also 
    with $K+1=3$ and the same number of nodes.}}
    \label{fig:RRGCT}
\end{figure}

A common source of confusion is identifying the Bethe lattice with the thermodynamic limit of a Cayley tree. A Cayley tree of size
\begin{equation}
    N=1+\frac{K+1}{K-1}(K^{n}-1)
    \end{equation}
    is a loopless tree with $n$ generations, where the root site has $K+1$
offspring sites in the first generation, and all sites from generation $1$ to $n-1$ have $K$ offspring sites in the next generation. 
Unlike a random regular graph, the Cayley tree is not translationally invariant in the thermodynamic limit, is bipartite, 
and possesses a boundary that contains a macroscopic number of sites. In Fig.~\ref{fig:RRGCT} we compare a Cayley tree to 
a random regular graph for $N=46$.

The most important properties of the Bethe lattice, which add to the ones of random regular graphs, can be summarized as:
\begin{itemize}
    \item It is statistically translational invariant in the thermodynamic limit.
    \item Conditioning on one of the vertices, the $K+1$ branches of the remaining graph become statistically independent~\cite{mezard2009information}.
    \item There is only one simple path of finite length between two vertices.
\end{itemize}
These properties simplify the analysis of the statistical properties of models defined on Bethe lattices.

\section{Anderson Localization on the Bethe Lattice}
\label{sec:Anderson}

In this Section we summarize the analysis of the  Anderson model on the Bethe lattice and we 
present some results relevant to the comparison to the predictions of the Localization Landscape Theory.
In Sec.~\ref{sec:Anderson-def} we recall the definition of the Anderson model and the local density of states (LDoS), 
in Sec.~\ref{sec:Anderson-Bethe}  we present the Green's function formalism used to analyze the LDoS and phase transition,
in Sec.~\ref{sec:SCanderson} we describe the distribution analysis of the Green's function,  
and in Sec.~\ref{sec:ALeigenval} a linear analysis of the  distribution equation which allows one to 
locate the phase transition.

\subsection{Model and main observables}
\label{sec:Anderson-def}

The tight-binding Anderson model~\cite{anderson1958absence} reads
\begin{equation}
\label{eq:ham}
    \hat{\mathcal{H}} = \sum_{i} \epsilon_i \hat{c}_i^\dagger \hat{c}_i - t \sum_{\langle i,j \rangle} \left( \hat{c}_i^\dagger \hat{c}_j + \hat{c}_j^\dagger \hat{c}_i \right)
\end{equation}
where $\hat{c}_i^\dagger$ and $\hat{c}_i$ are 
creation and annihilation (spin-less) fermion operators at site $i$ in the second-quantization formalism. The parameters 
$\epsilon_i$ are the on-site disorder potentials, and they are independent and identically distributed random
variables drawn from the uniform distribution
\begin{equation}
\label{eq:gammas}
    \gamma(\epsilon)=\begin{cases}
        1/W &\text{if} \,\,\epsilon\in [-W/2,W/2]
        \;,
        \\
        0 &\text{else} \; .
    \end{cases}
\end{equation}
 The hopping amplitude between neighboring sites is denoted $t$, and $\sum_{\langle i,j \rangle}$ runs over nearest-neighbor pairs.
 
 The mobility edge is the disorder strength dependent critical energy, $E^{\rm loc}_c(W)$,  which 
 separates localized from extended states. Below the mobility edge, the wave functions are spatially 
 localized due to interference from disorder, inhibiting transport; above it, states remain delocalized 
 and can carry current. As disorder increases, the mobility edge shifts until all states become localized, 
 marking the Anderson transition to an insulating phase. 

A key mathematical tool for analyzing the localization properties of this (and similar) models 
is the resolvent or Green's function, defined as
\begin{equation}
\label{eq:defG}
    \hat{\mathcal{G}}(z) \equiv (\hat{\mathcal{H}}-z\mathcal{I} )^{-1},
\end{equation}
where $\hat{\mathcal{H}}$ is the Hamiltonian of the system, and $z=E-{\rm i}\alpha$ is a complex number 
with $E$ the energy and imaginary part $\alpha$ useful for regularization reasons. 
The imaginary part of the diagonal elements of $\hat{\mathcal{G}}(z)$, 
\begin{equation}
\label{eq:diagGF}
    \mathcal{G}_{ii}(z)=\langle i | \hat{\mathcal{G}}(z)|i\rangle
    \,,
\end{equation}
where the kets $\ket{i}$ identify the quantum states in which an electron is localized on site $i$, 
is directly related to the local density of states to be defined below, while its decay properties characterize the spatial extent of the 
wavefunctions.

The local density of states (LDoS) is a position-resolved version of the density of states, and 
describes how quantum states are distributed in both energy $E$ and space $i$, 
\begin{equation}
\label{eq:rhoi-def}
    \rho_i(E) =\sum_n \; |\psi_i^{(n)}|^2 \; \delta(E-E_n) 
    \; , 
\end{equation}
where $E_n$ are the eigenvalues and $\psi^{(n)}$ the 
eigenstates of the Hamiltonian $\hat{\mathcal H}$.
 In terms of the imaginary part of the Green's function, the LDoS is given by
\begin{equation}
\label{eq:rhoi}
    \rho_i(E) =\sum_n \; |\psi_i^{(n)}|^2 \; \delta(E-E_n)= \frac{1}{\pi} \; {\rm Im} \, \mathcal{G}_{ii}(E+ {\rm i} 0^+) \,\,
    \; ,
\end{equation}
where $ \mathcal{G}_{ii}(E+{\rm i}\alpha) $ is the diagonal element of the advanced Green's function (the distinction between ``advanced" and ``retarded" Green's functions is determined by the choice of $z = E \pm {\rm i} \alpha$ with $\alpha >0$ in Eq.~(\ref{eq:defG})). The expression of $\rho_i$ in terms of $\mathcal{G}_{ii}$ is a well-known result, and 
the derivation can be found, for example,  in Ref.~\cite{economou1972existence}.

The LDoS serves as the order parameter for the delocalization-localization transition. Physically, it represents the inverse lifetime of a particle of energy $E$ created in $i$. In the localized phase, $\rho_i(E)$ is exponentially small on $O(N)$ sites and of order unity on only $O(1)$ sites. In contrast, in the delocalized phase, $\rho_i(E)$ remains nonzero on $O(N)$ sites, enabling transport across the system.

Another order parameter for the delocalizarion-localization transition is the Inverse Participation Ratio or IPR, which is defined as
\begin{equation}
\label{eq:IPR}
    I_2(E)=\mathbb{E}\Bigg[\frac{\sum_n \sum_i \big|\psi^{(n)}_i\big|^4 \delta(E_n-E)}{\sum_n\delta(E_n-E)}\Bigg] = \mathbb{E}\Bigg[\lim_{\alpha \to 0^+} \frac{\alpha \sum_i |\mathcal{G}_{ii}(E+{\rm i}\alpha)|^2}{ \sum_i {\rm Im} \, \mathcal{G}_{ii}(E+{\rm i}\alpha)  }\Bigg]\,, 
\end{equation}
and is essentially a measure of the average inverse volume occupied by an eigenstate. 
Henceforth, the expectation value ${\mathbb E}[ \dots]$ is taken with respect to the disorder distribution. 
In the delocalized phase $I_2(E) \sim O(1/N)$, while in the localized one $I_2(E) \sim O(1)$.

Finally, the last quantity we are interested in is the eigenstate correlation function, which is defined as 
\begin{equation}
\label{eq:corrfuncAL}
    C_{\rm loc}(|i-j|\,;E)=
    \mathbb{E}\Bigg[\frac{\sum_n  \big|\psi^{(n)}_i\big|^2 \big| \psi^{(n)}_j\big|^2 \delta(E-E_n)}{\sum_n\delta(E-E_n)}\Bigg] 
    =
    \mathbb{E} \Bigg[ \lim_{\alpha \to 0^+} \frac{\alpha |\mathcal{G}_{ij}(E+{\rm i}\alpha)|^2}{\sum_i {\rm Im} \, \mathcal{G}_{ii}(E+{\rm i}\alpha)} \Bigg]
    \; .
\end{equation}
It represents the average correlation between the amplitude of the eigenstates on two sites at distance $|i-j|\delta$, with $\delta$ the 
lattice spacing. 
The algebraic steps leading to the final expression in Eq.~(\ref{eq:IPR}) are detailed in Ref.~\cite{rizzo2024localized}, and the last term in Eq.~(\ref{eq:corrfuncAL}) is derived in an analogous way. Since the delocalization-localization transition is 
second-order, from the correlation function one extracts a correlation length, $\xi_{\rm loc}$, which diverges 
algebraically $\xi_{\rm loc} \sim |E-E_c|^{-\nu}$. The critical exponent $\nu$ depends on the lattice dimension, 
and the following values are reported in the literature: 
$\nu =1.57 \pm 0.02$ in $d=3$, $\nu =1.1 \pm 0.1$ in $d=4$, and further decreasing for increasing $d$, until
$\nu=1/2$ in $d \to \infty$~\cite{abou1973selfconsistent,tarquini2017critical}. 

From Eqs.~(\ref{eq:rhoi})-(\ref{eq:corrfuncAL}),  it is clear  that the critical 
properties of Anderson localization on a generic lattice are fully determined by $\hat{\mathcal{G}}(z)$.

\subsection{The Green's function on the Bethe Lattice}
\label{sec:Anderson-Bethe}

The advantage of working with the Bethe lattice is that, thanks to the hierarchical structure of the graph, the computation of the components 
$\mathcal{G}_{ij}(z)$ can be performed analytically. With methods explained in~\cite{abou1973selfconsistent,biroli2010anderson}, one obtains that  the diagonal elements  are given by
\begin{equation}
\label{eq:gAL}
    \mathcal{G}_{ii}(z)=\frac{1}{\epsilon_i-z-t^2\sum\limits_{k\in \partial i}\mathcal{G}_{k\rightarrow i}(z)}
    \; ,
\end{equation}
where the symbol $\partial i$ represents the set of nearest neighbors of site $i$, and the terms $\mathcal{G}_{k\rightarrow i}(z)$ are the cavity Green's functions on site $k$ in the absence of site $i$. The latter satisfy an analogous recursion relation, 
\begin{equation}
\label{eq:gcavAL}
    \mathcal{G}_{k\rightarrow i}(z)=\frac{1}{\epsilon_k-z-t^2\sum\limits_{l\in {\partial k \setminus l}} \mathcal{G}_{l\rightarrow k}(z)}
    \; ,
\end{equation}
where $\partial  {k \! \setminus \! i}$ is a shorthand notation for $\partial {k \! \setminus \! \{i\}}$ and  ``$\setminus$" is the symbol of difference between sets.

The imaginary part of the Green's function appears as the 
most relevant object since, according to Eq.~(\ref{eq:rhoi}),  its probability distribution determines the mobility edge through the 
LDoS. Indeed,  the mobility edge is the curve in the $(E,W)$-plane 
where the Green's functions transition from vanishing to non-vanishing typical value of the imaginary part.
The interpretation of the real part of the Green's function has remained elusive so far.
The Localization Landscape Theory could give an interpretation for it. In fact, as we will show in Sec.~\ref{sec:EQGU}, the Localization Landscape $\mathbf{u}$ is directly related to the real part of $\hat{\mathcal{G}}(z)$. 

\subsection{Self-consistent distributional equation}
\label{sec:SCanderson}

The derivation that we present in this Section is a summary of the one in Ref.~\cite{rizzo2024localized}, with the only difference that we consider here the case with finite energy $E$.
Let us focus on the cavity Green's functions $\mathcal{G}_{k\to i}(z)$, with $z=E+{\rm i} \alpha$ where $\alpha \ll 1$ serves as a regularization parameter. From now on, and to lighten the notation, we only 
write $\mathcal{G}_{k\to i}$ without the explicit $z$-dependence. 
We decompose the cavity Green's function in the localized phase as
\begin{equation}
    \mathcal{G}_{k\to i} = \mathcal{G}^R_{k\to i} + {\rm i} \alpha\mathcal{G}^I_{k\to i}
    \; .
\end{equation}
Since in the localized phase the imaginary part of the Green's functions 
is linear in $\alpha$ for $\alpha\ll1$, we have factorized an $\alpha$ from the imaginary part.

The cavity equation (\ref{eq:gcavAL}) can be separated into real and imaginary parts. By expanding for small $\alpha$, and retaining the leading order, we obtain the two cavity equations governing the critical properties:
\begin{align}
\label{eq:ReGcavcrit}
    (\mathcal{G}_{k\to i}^R)^{-1} &= \epsilon_k- E - t^2 \sum_{l \in \partial k \setminus i} \mathcal{G}_{l\to k}^R \; ,  \\
    \label{eq:ImGcavcrit}
    \mathcal{G}_{k\to i}^I &= t^2 (\mathcal{G}_{k\to i}^R)^2 \sum_{l \in \partial k \setminus i} \mathcal{G}_{l\to k}^I \; .
\end{align}
We can now exploit the statistical independence of the cavity variables by noting that, after averaging over disorder, the Hamiltonian becomes translationally invariant. Consequently, the joint probability distribution of the on-site variables is identical on every site, and the same applies to the cavity variables. This allows us to write a recursive equation that relates the joint probability distribution of their real and imaginary parts [Eqs.~(\ref{eq:ReGcavcrit}) and (\ref{eq:ImGcavcrit})] to $K$ copies of itself evaluated on the nearest neighboring cavity sites~\cite{rizzo2024localized,abou1973selfconsistent,tikhonov2019critical}:
\begin{equation}
\label{eq:SCanderson}
    P(g, \hat g) = \int d\epsilon \, \gamma(\epsilon) \int \prod_{l=1}^K \left[dg_l d\hat g_l \, 
    P( g_l, \hat g_l )\right]  \; \delta \bigg( g - \frac{1}{\epsilon-E - t^2 \sum_l g_l} \bigg) \delta \bigg( \hat g - t^2 g^2 \sum_l \hat g_l \bigg),
\end{equation}
where the disorder distribution $\gamma$ is the one in Eq.~(\ref{eq:gammas}) and the functions $P$ in the left-hand-side and 
within the integrals in the right-hand-side are the same. This follows from the fact that the cavity variables are statistically independent, thus the joint probability distribution of $K$ cavity variables on the right-hand-side have been decomposed as
\begin{equation}
    P(\{ g_l, \hat g_l \}_{l=1,\dots,K})=\prod_{l=1}^K P( g_l, \hat g_l ).
\end{equation}

 This equation can be compared with Eq.~(\ref{eq:stochSC}), which represents its counterpart in the Localization Landscape percolation problem. This comparison will let us highlight the difference between the Anderson localization and the Localization Landscape percolation transitions on the Bethe lattice. 

\subsection{Linear stability analysis}
\label{sec:ALeigenval}

Equation (\ref{eq:SCanderson}) can be rewritten as an integral eigenvalue equation using the following procedure.  
Starting from Eq.~(\ref{eq:SCanderson}), we use the integral representation of the delta function for the variable $\hat g$, 
\begin{equation}
    \delta\bigg( \hat g - t^2 g^2\sum_l \hat g_l \bigg)
    = \int_{-\infty}^{\infty} 
    \frac{d\lambda'}{2\pi} \,e^{ - {\rm i} \lambda' \left(\hat g - t^2 g^2\sum_l \hat g_l \right)} \, ,
\end{equation}
to  integrate explicitly over all the $\hat g_l$'s in the right-hand-side of Eq.~(\ref{eq:SCanderson}). The result is
\begin{equation}
    P(g,\hat g)= \int \frac{d\lambda'}{2\pi}  \, e^{ -i \lambda' \hat g}\int d\epsilon \, \gamma(\epsilon) 
\int \prod_{l=1}^K \left[dg_l  \, \hat P( g_l , \lambda' t^2g^2)\right] \; 
    \delta \bigg( g - \frac{1}{\epsilon-E - t^2 \sum_l g_l} \bigg)\,.
    \label{eq:Pgghat1}
\end{equation}
with 
\begin{equation}
 \hat P( g_l, \lambda' t^2g^2) \equiv 
 \int  d\hat g_l \; P( g_l, \hat g_l) \; e^{ {\rm i} \lambda' t^2 g^2 \hat g_l} 
 \; .
\end{equation}
Then, Fourier transforming Eq.~(\ref{eq:Pgghat1}) 
over $\hat g$ on both sides, and recognizing the integral representation of $\delta(\lambda-\lambda')$, 
we obtain 
\begin{equation}  
    \hat P(g,\lambda)=\int d\epsilon \, \gamma(\epsilon) \int \prod_{l=1}^K \left[dg_l \; \hat P( g_l ,\lambda t^2g^2)\right]
    \; \delta \bigg( g - \frac{1}{\epsilon -E - t^2 \sum_l g_l} \bigg)\,.  
\end{equation}  
Next, we expand this characteristic function around the solution with zero imaginary part to lowest order in $\lambda$, assuming an algebraic form for the distribution~\cite{rizzo2024localized,abou1973selfconsistent,tikhonov2019critical},  
\begin{equation}  \label{eq:ansatz}
    \hat P(g,\lambda) \approx P(g)+f(g)|\lambda|^\beta\,,  
\end{equation}  
which gives  
\begin{equation} 
\label{eq:fAL}
    f(g)= \int d g'\,\mathcal{K}_{\rm loc}^\beta(g,g')f(g')\,.  
\end{equation}  
with  the kernel 
\begin{equation}
\label{eq:kernelAL}
    \mathcal{K}_{\rm loc}^\beta(g,g')=
    K|tg|^{2\beta} \int d\epsilon \, \gamma(\epsilon) \; \int d \tilde g \;  R_{\rm loc}(\tilde g) 
    \; \delta \bigg (g- \frac{1}{\epsilon-E-t^2 (\tilde g+g')}\bigg)\,,  
\end{equation}  
where  
\begin{equation}  
    R_{\rm loc}(\tilde{g})=\int \prod_{l=1}^{K-1}\left[dg_l\,P(g_l )\right] \; \delta\bigg(\sum_{l=1}^{K-1}g_l-\tilde{g}\bigg).  
\end{equation}  
The function $R_{\rm loc}$ represent the distribution of the sum of the $K-1$ real parts of the cavity Green functions. 
It can be evaluated numerically by computing the marginal distribution of the $\mathcal{G}_{k\to i}^R$'s through the population dynamics algorithm
explained in Sec.~\ref{subsec:population} and sampling from it sums of $K-1$ variables.

Given this equation, the critical line can be identified as the curve in the 
$(E,W)$ plane where the largest eigenvalue of the kernel equals one. In the localized phase, the only stable solution of the self-consistent distributional equation~\eqref{eq:SCanderson} is the one with a vanishing imaginary part of the Green's function; therefore, any perturbation $f(g)$ of this solution must decay under iteration of the integral operator. 
Conversely, in the delocalized phase, such perturbations grow under iteration. If the leading eigenvalue of 
${\mathcal K}_{\rm loc}$ is smaller than one, all components of $f$, when decomposed on the eigenbasis of 
${\mathcal K}_{\rm loc}$, decay upon iteration. If it exceeds one, any component not orthogonal to the corresponding eigenfunctions diverges. Hence, the boundary separating these two regimes is precisely the curve where the largest eigenvalue of the kernel equals one.

This integral operator can be diagonalized numerically with high precision for all $E$, as it has been done in Refs.~\cite{parisi2019anderson} and~\cite{tikhonov2019critical} for $E=0$. Because of the high computational cost of this procedure, in order to obtain the full curve (sacrificing accuracy) we used the population dynamic approach. 
This method will be explained in detail in Secs.~\ref{sec:numerics}, and the resulting phase diagram is discussed in the main text.

The integral eigenvalue equation \eqref{eq:fAL} is fundamentally different from its counterpart in the Localization Landscape percolation problem. In Sec. \ref{subsubsec:linstab}, we will show this explicitly in the high-connectivity limit. This difference suggests that the critical behavior in Anderson localization may differ from the one of Localization Landscape percolation and, as we show in the main text, 
this is indeed the case. 

\section{The Localization Landscape Theory on the Bethe Lattice}
\label{sec:EQGU}

In this Section we derive the relevant equations to study the critical behavior of the Localization Landscape percolation problem.
We begin by recalling the definition of the Localization Landscape, and then derive the set of coupled equations that let us compute the Localization Landscape on a generic site $i$ of the Bethe lattice as a function of independent cavity quantities defined on the cavity lattices rooted at the nearest neighbors of $i$.
Here we present an alternative derivation to that in the End Matter of the paper, employing known results for the Green's function of Anderson localization on the Bethe lattice~\cite{biroli2010anderson,rizzo2024localized}, Sec.~\ref{subsec:cavity-derivation}.
Next, in Sec. \ref{sec:EQP} we obtain the equation governing the probability that a generic site belongs to an infinite cluster. Then, in Sec.~\ref{sec:SCP} we show how this equation and the cavity equations of Sec.~\ref{subsec:cavity-derivation} can be recast into a self-consistent distributional equation.
Finally, in Sec.~\ref{sec:CFP} we derive the expressions for the percolation correlation function and average cluster size.

\subsection{The Localization Landscape}
\label{subsec:localization-landscape}

The Localization Landscape Theory (LLT)~\cite{filoche2012universal}, here treated in its discrete formulation as in Ref.~\cite{filoche2024anderson}, introduces an $N$-dimensional vector ${\mathbf u}$ as the solution of the equation
\begin{equation}
\hat {\mathcal H} {\mathbf u} = {\mathbf 1}
\end{equation}
with ${\mathbf 1}$ the vector with identical components equal to one, and defines a real-space effective potential, $1/u_i$.
Assuming a positive definite Hamiltonian—or else translating it to ensure that its eigenvalues are non-negative—we define the set of nearest-neighboring sites on the lattice with lower potential than the electron's energy, $\Omega_E = \{ i \; { {\rm such \; that}} \; 1/u_i \leq E \}$. These are the spatial regions where a particle is classically confined.
The LLT proposes that, when a macroscopic (``giant'') cluster of this kind exists, that is, when the set $\Omega_E$ forms a connected path spanning the system, the quantum particle delocalizes.

In principle the spectral shift of $\hat{\cal H}$ to ensure that it is positive definite is arbitrary. However, as discussed in Ref.~\cite{filoche2017localization}, the minimal shift, equal to minus the minimal eigenvalue of $\hat{\cal H}$, is also the optimal choice. We therefore follow Refs.~\cite{filoche2012universal,arnold2016effective,arnold2019localization,arnold2019computing,david2021landscape} and consider
\begin{equation}
\hat{\cal H}_+\mathbf{u}=(\hat{\cal H} - E_{\rm min} \hat{\cal I}) \mathbf{u} = \mathbf{1} \, .
\end{equation}
From Eqs.~\eqref{eq:g} and~\eqref{eq:pcavlin}-\eqref{eq:uk} below, we determine the critical percolation parameter $E_{c}^{\rm perc}(W)$ for the positive definite Hamiltonian $\hat{\cal{H}}_+$, as well as the exact position of the mobility edge $E_c^{\rm loc}(W)$ for the Anderson statistically symmetric Hamiltonian $\hat{\cal{H}}$~\cite{abou1973selfconsistent,biroli2010anderson}.
To compare the percolation threshold within the effective landscape with the mobility edge, we must subtract the spectral shift $-E_{\rm min}$. This subtraction yields a negative percolation threshold. However, exploiting the statistical symmetry of the spectrum, one can equivalently express the percolation threshold on the positive-energy side mapping ${E_c^{\rm perc}}(W) \mapsto -E_{\rm min} - E^{\rm perc}_c(W)$.

\subsection{Recursive derivation of the cavity equations}
\label{subsec:cavity-derivation}
As mentioned in the End Matter of the paper, for the Anderson model on the Bethe lattice the minimum possible eigenvalue of $\hat{\mathcal{H}}$ depends on $W$, and it is given by
\begin{equation}
E_{\rm min}(W) =
    \begin{cases}
        -2t\sqrt{K}-W/2 &\text{if} \quad W>W_{\rm min}\,,\\
        E_{\rm iso}(W) &\text{if} \quad W\leq W_{\rm min}\,.
    \end{cases}
\end{equation}
The resulting Hamiltonian $\hat{\mathcal{H}}_+$ has the same functional form of the Hamiltonian in Eq.~(\ref{eq:ham}), except that the on-site energies are redefined as $\varepsilon_i = \epsilon_i -E_{\rm min}(W)$, i.e., they are drawn from
\begin{equation}
\label{eq:gamma}
    \gamma_+(\varepsilon) =\gamma(\varepsilon+E_{\rm min}(W)) =
    \begin{cases}
        1/W \,\,\qquad &\text{if }\,\, \varepsilon \in [-W/2 -E_{\rm min}(W),\,W/2-E_{\rm min}(W)]
        \; ,\\
        0 \,\, \qquad &\text{else}
        \; ,
    \end{cases}
\end{equation}
where $\gamma$ is the one in Eq.~(\ref{eq:gammas}).
Henceforth, we omit the explicit $W$-dependence in $E_{\rm min}(W)$.

From the inversion of the Green's function definition, $( \hat {\mathcal H}-z \hat {\mathcal I} ) \hat {\mathcal G}(z) = \hat {\mathcal I}$, one obtains that the components of the Localization Landscape $u_i$ are all non-negative, and are computed as
\begin{equation}
\label{eq:ui}
\hat{\mathcal H}_+ {\mathbf u} =\left( \hat {\mathcal H}-E_{\rm min} \hat {\mathcal I}\right)\mathbf{u}= \hat {\mathcal G}^{-1}(E_{\rm min})  {\mathbf u}  =  {\mathbf 1} \implies u_i= \sum_{j=1}^N \mathcal{G}_{ij}(E_{\rm min})
\; .
\end{equation}
Since, as mentioned in the End Matter of the paper, for $E=E_{\rm min}$ the imaginary part of $\mathcal{G}_{ij}(E)$ vanishes, Eq.~\eqref{eq:ui} can be 
rewritten in the precise form adopted in the main text
\begin{equation}
    u_i= \sum_{j=1}^N \mathrm{Re} \, \mathcal{G}_{ij}(E_{\rm min})\,.
\end{equation}
Henceforth, we omit the explicit $E_{\rm min}$-dependence and the $\rm Re$ symbol.

In the End Matter of the paper we showed that
\begin{equation}
    \label{eq:etau}
    u_i = \mathcal{G}_{ii}\eta_i 
\end{equation}
where
\begin{minipage}[t]{0.48\textwidth}
\begin{align}
    \label{eq:g}
    \mathcal{G}_{ii}^{-1} & =  \varepsilon_i - t^2 \sum_{k \in \partial i} \mathcal{G}_{k \to i} \, ,\\
    \label{eq:eta}
    \eta_{i} &= 1 + t \sum_{k \in \partial i} \mathcal{G}_{k \to i} \eta_{k \to i}
    \; ,
\end{align}
\end{minipage}
\hfill
\begin{minipage}[t]{0.48\textwidth}
\begin{align}
    \label{eq:cavg}
    \mathcal{G}_{k \to i}^{-1} & = \varepsilon_k - t^2 \sum_{l \in \partial k \setminus i} \mathcal{G}_{l \to k} \,  ,\\
    \label{eq:caveta}
    \eta_{k \to i}  &= 1 + t \sum_{l \in \partial k \setminus i}
    \mathcal{G}_{l \to k} \eta_{l \to k}   \,.
\end{align}
\end{minipage}
We obtained this result by representing the Green's functions of the system as
\begin{equation}
    \label{eq:gGaussian}
    \mathcal{G}_{ij} =  \frac{1}{Z_0}\int {\cal D}{\mathbf x} \;  x_i x_j \; e^{-S_0[{\mathbf x}]}
    \equiv \langle x_i x_j\rangle_{0}
    \, ,
\end{equation}
where
\begin{eqnarray}
    S_J[{\mathbf x}]  &=& \frac{1}{2}  \sum_i \varepsilon_i x_i^2 - t \sum_{\langle i,j \rangle} x_i x_j -J\sum_j x_j =\frac{1}{2}\mathbf{x^t}{\mathcal H}_+\mathbf{x}-J\sum_j x_j
    \; ,
\\
    Z_J  &=& \int {\cal D} x \, e^{-S_J[{\mathbf x}]}
    \; ,
\end{eqnarray}
which is justified by the fact that $\mathcal{H}_+$ is a positive definite matrix, thus $e^{-S_0[{\mathbf x}]}/Z_0$ is a properly normalized Gaussian measure. In Eq.~\eqref{eq:gGaussian} the symbol $\langle \,\cdot \, \rangle_J$ indicates the expectation value defined by the Gaussian measure with action $S_J[\mathbf{x}]$.
The component $u_i$ is given by the Gaussian expectation
\begin{equation}
    u_i   = \sum_j \mathcal{G}_{ij} = \langle x_i \sum_j x_j \rangle_0 = \frac{\partial \langle x_i \rangle_J}{\partial J} \Bigg\vert_{J=0},
\end{equation}
parametrizing the on-site marginal distributions on the normal and cavity lattice as
\begin{equation}
    \mu_i(x_i) \propto e^{-\frac{x_i^2}{2 \, \mathcal{G}_{ii}} + J\eta_i x_i} \,,
    \qquad\qquad \mu_{k \to i}(x_k) \propto e^{-\frac{x_k^2}{2 \, \mathcal{G}_{k \to i}} + J\eta_{k \to i} x_k} \,,
\end{equation}
and finding the self-consistent recursive equations~\eqref{eq:g}-\eqref{eq:caveta}.

To validate this result, we compute $u_i$ recursively by 
using the standard cavity method on a tree, see Ref.~\cite{mezard2009information}. 
The $i$-th component of the Localization Landscape can be expressed as
\begin{equation}
\label{eq:ubari}
    u_i = \sum_j \mathcal{G}_{ij} = \mathcal{G}_{ii}+\sum_{k\in \partial i}\sum_{j\in B_{k\rightarrow i}}\mathcal{G}_{ij},
\end{equation}
where $B_{k \rightarrow i} \equiv \left\{ j \,\middle|\, j \text{ belongs to the cavity lattice rooted at } k\in \partial i \text{ after the removal of site } i \right\}$.
The second sum in the second term can be rewritten as
\begin{equation}
\label{eq:sumBranch}
    \sum_{j\in B_{k\rightarrow i}}\mathcal{G}_{ij} = \mathcal{G}_{ik}+\sum_{j\in B_{k\rightarrow i}\setminus k}\mathcal{G}_{ij}.
\end{equation}

Take two sites on the Bethe lattice, say $0$ and $r$, connected by a non-intersecting path of length $r$.
Using the Gaussian representation of the Green's function, $\mathcal{G}_{0r}$ reads
\begin{equation}
    \mathcal{G}_{0r}= \frac{1}{Z_0} \int \mathcal{D}x \, x_0x_r \; e^{-S_0[{\mathbf x}]} \,.
\end{equation}
After integrating out all the variables that are not on the chosen path which connects $0$ and $r$, the integral can be written in terms of the normal and cavity marginals (introduced also in the End Matter of the paper), which in absence of the source $J$ read
\begin{equation}
\label{eq:mus}
    \mu_i(x_i) \propto e^{-\frac{x_i^2}{2 \mathcal{G}_{ii}}}
    \; ,
    \qquad\qquad
    \mu_{k\to i}(x_k) \propto e^{-\frac{x_k^2}{2 \mathcal{G}_{k\to i}}}
    \; .
\end{equation}
Thus,
\begin{eqnarray}
         \mathcal{G}_{0r}&= & \int \prod_{s = 0}^{r-1}  \left[ {\rm d} x_s \, \mu_{s \to s+1} (x_s) \right] dx_r\,\mu_r(x_r) \;
         x_0x_r
         \; e^{t \sum_{s=0}^{r-1}x_sx_{s+1}}
         \nonumber\\
         &= & \int \prod_{s = 1}^{r-1}  \left[ {\rm d} x_s \, \mu_{s \to s+1} (x_s) \right] dx_r\,\mu_r(x_r) \;
         (t\mathcal{G}_{0 \to 1}x_1)x_r \; e^{t \sum_{s=1}^{r-1}x_sx_{s+1}}
         \nonumber\\
         &=& \mathcal{G}_{rr}
         \prod_{s=0}^{r-1} (t\mathcal{G}_{s\rightarrow s+1}) \; .
         \label{eq:chain-rule}
\end{eqnarray}
In the second line we have explicitly computed the Gaussian integral
\begin{equation}
    \int dx_0 \mu_{0\to 1}(x_0) \; x_0 \; e^{tx_0x_1}= t\mathcal{G}_{0 \to 1}x_1\,,
\end{equation}
and in the last line we have just repeated the same integration procedure for all $s$.

Another version of the chain rule of Eq.~(\ref{eq:chain-rule}), can be obtained in an analogous way by using the latter formula to compute the matrix element $\mathcal{G}_{r0}$, and by noticing that $\mathcal{G}_{ij}=\mathcal{G}_{ji}$.
Therefore,
\begin{equation}
\label{eq:chain1}
\mathcal{G}_{0r}=\mathcal{G}_{00}\prod_{s=1}^r (t\mathcal{G}_{s\rightarrow s-1}) =\\
\mathcal{G}_{rr}\prod_{s=0}^{r-1} (t\mathcal{G}_{s\rightarrow s+1})
\; .
\end{equation}
Here, the cavity Green's functions in the product are the ones connecting the sites along the path going from site $r$ to site $0$ (in the first equality) or vice versa (in the second). Note that the recursive equations defining these normal and cavity Green's functions are the same as those for the real part of the normal and cavity Green's functions in the Anderson localization context, after substituting $E=E_{\rm min}$ [Eqs.~\eqref{eq:gAL} and \eqref{eq:gcavAL}], or can be derived following the same procedure as in the End Matter, starting from an action with $J=0$, and using the Ansatz for the normal and cavity marginals of Eq.~\eqref{eq:mus}.

Inserting Eq.~(\ref{eq:chain1}) in Eq.~(\ref{eq:sumBranch}) we obtain
\begin{eqnarray}
    \mathcal{G}_{ij} &= & \mathcal{G}_{jj}\prod_{(s,s')\in p_{ij}}t\mathcal{G}_{s \rightarrow s'}=t\mathcal{G}_{i \rightarrow k}\underbrace{\mathcal{G}_{jj}\prod_{(s,s')\in p_{kj}}t\mathcal{G}_{s \rightarrow s'}}_{\mathcal{G}_{kj}}=t\mathcal{G}_{i \rightarrow k}\mathcal{G}_{kj},
\end{eqnarray}
where $p_{ij}$ represents the directed path connecting site $i$ and site $j$.
Plugging this expression and $\mathcal{G}_{ik} = t\mathcal{G}_{i \rightarrow k}\mathcal{G}_{kk}$ in Eq.~(\ref{eq:ubari}) we immediately obtain
\begin{equation}
\label{eq:u}
u_i = \mathcal{G}_{ii} + t \sum_{k \in \partial i} \mathcal{G}_{i \rightarrow k} u_{k \rightarrow i} \,,
\qquad\qquad
u_{k \rightarrow i} = \mathcal{G}_{kk} + t \sum_{l \in \partial k \setminus i} \mathcal{G}_{k \rightarrow l} u_{l \rightarrow k} \, ,
\end{equation}
where we have defined
\begin{equation}
    u_{k \rightarrow i}\equiv \sum_{j\in B_{k\rightarrow i}}\mathcal{G}_{kj}
    \; .
\end{equation}

These equations are equivalent to Eqs.~(\ref{eq:eta}). In fact, if we replace
\begin{equation}
    u_{k\to i}=\mathcal{G}_{kk}\eta_{k\to i}
\end{equation}
in Eqs.~(\ref{eq:eta}), we recover the recursive Eqs.~(\ref{eq:u}):
\begin{align}
    u_i &= \mathcal{G}_{ii}\eta_{i}= \mathcal{G}_{ii}+t\sum_{k \in \partial i}\mathcal{G}_{ii} \eta_{k \to i}  \mathcal{G}_{k \to i}=\mathcal{G}_{ii}+t\sum_{k\in \partial i}\mathcal{G}_{i \rightarrow k}u_{k \rightarrow i}\,,\\
    u_{k \to i} &= \mathcal{G}_{kk}\eta_{k \to i}= \mathcal{G}_{kk}+t\sum_{l \in \partial k \setminus i}\mathcal{G}_{kk} \eta_{l \to k}  \mathcal{G}_{l \to k}=\mathcal{G}_{kk}+t\sum_{l\in \partial k \setminus i}\mathcal{G}_{k \rightarrow l}u_{l \rightarrow k}\,.
\end{align}
The second equalities in the two equations follow from Eqs.~(\ref{eq:chain1}). Indeed,
\begin{equation}
    \mathcal{G}_{kk} t \mathcal{G}_{l \to k} \eta_{l \to k} = \mathcal{G}_{kl}  \eta_{l \to k}=\mathcal{G}_{ll}t\mathcal{G}_{k \to l}\eta_{l \to k}=t\mathcal{G}_{k \to l}u_{l \to k}\,.
\end{equation}
Conversely, it is easy to check that by defining $\eta_i =u_i/\mathcal{G}_{ii}$ and $\eta_{k\to i} =u_{k\to i}/\mathcal{G}_{kk}$ Eqs.~(\ref{eq:u}) boil down to Eqs.~(\ref{eq:eta}).

As we can see, Eqs.~(\ref{eq:u}) depend both on the normal and the cavity Green's functions. Therefore, Eqs.~(\ref{eq:g}), (\ref{eq:cavg}), (\ref{eq:u}) form a set of coupled recursive equations that can be solved self-consistently. However, for computing the percolation critical curve it is more convenient to work with the set of equations (\ref{eq:g})-(\ref{eq:caveta}).

\subsection{The percolation probability}
\label{sec:EQP}

In order to study the percolation problem for the Hamiltonian $\hat{\cal{H}}_+$ at energy $E$, one has to consider the standard cavity equation for the random site percolation on the Bethe lattice \cite{stauffer2018introduction}, where the probability of a node being occupied is replaced by the condition $u_i \geq 1/E$. 
In this setting, we define
\begin{equation}
    p_i\equiv \Pr\{\text{site $i$ belong to the giant cluster}\}.
\end{equation}

Since the value of the Localization Landscape $u_i$ at site $i$ is correlated to the ones of its nearest neighboring sites, 
$u_{k \in \partial i}$,
the probability of a site being occupied is site dependent. 

The recursive equations read
\begin{align}
\label{eq:pi}
    p_{i} &= \theta(u_i-1/E) \Big[ 1 - \prod_{k \in \partial i } (1 -  p_{k \to i}) \Big] , 
    \\
    \label{eq:cavpi}    
    p_{k\to i} &= \theta(u_k-1/E) \Big[ 1 - \prod_{l \in \partial k \setminus i } (1 -  p_{l \to k}) \Big] .
\end{align}
From now on we will refer to the $p_i$'s as the percolation probabilities.
An important observation is that $\{p_{k \to i}\}_{k\in \partial i}$ in Eq.~(\ref{eq:pi}) are not statistically independent, because for each of them we need to compute the $\theta(u_k-1/E)$'s, which also depend 
on quantities evaluated at site $i$. For this reason, it is useful to switch to a different set of cavity variables that we define as
\begin{equation}
    \bar p_{k \to i} \equiv \Pr\{\text{cavity site $k$ (in absence of $i$) belong to the giant cluster if occupied}\}\,,
\end{equation}
thus:
\begin{equation}
   p_{k \to i} =\theta(u_k-1/E)\,\bar p_{k \to i}\,.
\end{equation}
In terms of these variables, Eqs.~(\ref{eq:pi}) become 
\begin{align}
    \label{eq:pi2}
    p_{i} &= \theta(u_i-1/E) \Big[ 1 - \prod_{k \in \partial i } \Big(1 - \theta(u_k-1/E)\bar p_{k \to i} \Big ) \Big]\,, \\
    \label{eq:cavpbari}
    \bar p_{k \to i} &=1 - \prod_{l \in \partial k \setminus i} \Big(1 - \theta (u_l -1/E)\bar p_{l \to k} \Big).
\end{align}
Since in the non-percolating phase all these probabilities must be exactly zero, and the critical percolation
transition is expected to be continuous,  close to the percolation critical curve we can 
linearize the second equation above. Under this assumption it becomes
\begin{equation}
\label{eq:pcavlin}
    \bar p_{k \to i} = \sum_{l \in \partial k \setminus i} \theta (u_l -1/E)\bar p_{l \to k} \, .
\end{equation}
Here $u_l=\mathcal{G}_{ll}\eta_l$, see Eq.~(\ref{eq:etau}), and it can be computed in terms of independent cavity variables once we realize that
\begin{equation}
    \mathcal{G}_{ll}^{-1} = \mathcal{G}_{l \to k}^{-1} - t^2\mathcal{G}_{k \to l} , 
    \qquad\qquad
    \eta_l = \eta_{l \to k} + t\mathcal{G}_{k \to l} \eta_{k \to l} ,
\end{equation}
and 
\begin{equation}
    \mathcal{G}_{k \to l}^{-1} = \varepsilon_k - t^2\sum_{m \in \partial k \setminus l} \mathcal{G}_{m \to k} \,, 
    \qquad
    \eta_{k \to l}  = 1 +t \sum_{m \in \partial k \setminus l} \mathcal{G}_{m \to k} \eta_{m \to k} \,, 
\end{equation}
which follow from Eqs.~(\ref{eq:cavg}).
The final expression for the Localization Landscape for a site $l\in \partial k\setminus i$ is 
\begin{align}
        \label{eq:uk}
        u_l&= \left( \mathcal{G}_{l \to k}^{-1}-\frac{t^2}{\varepsilon_k-t^2\sum\limits_{m \in \partial_k \setminus l} \mathcal{G}_{m \to k}} \right)^{-1} \left(\eta_{l \to k}+t \; \frac{1+t\sum\limits_{m\in \partial_k \setminus l}\mathcal{G}_{m \to k}\eta_{m\to k}}{\varepsilon_k-t^2\sum\limits_{m \in \partial_k \setminus l} \mathcal{G}_{m \to k}}\right) 
        \equiv U_l(\{\mathcal{G}_{l\to k}, \eta_{l\to k}\}_{l \in \partial k}).
\end{align}

\subsection{Self-consistent distributional equation}
\label{sec:SCP}

We can now exploit the statistical independence of the cavity variables by observing that, after averaging over disorder, the Hamiltonian becomes translational invariant. Therefore, the joint probability distribution of on-site variables must be the same on every site (and the same must hold for the cavity variables). This means that we can write a recursive equation that relates the joint probability distribution of $\mathcal{G}_{k\to i}, \eta_{k\to i},\bar p_{k \to i} $ on a cavity site, to independent copies of itself. In other words, $\mathcal{G}_{k\to i}, \eta_{k\to i},\bar p_{k \to i} $ are written in terms of independent sets of cavity variables $\{\mathcal{G}_{l\to k}, \eta_{l\to k},\bar p_{l \to k}\}_{l \in \partial k}$ drawn from the same probability distribution $P(g,\eta,\bar p)$.
Because of translational invariance $P(g,\eta,\bar p)$ must also be the probability distribution of $\mathcal{G}_{k\to i}, \eta_{k\to i},\bar p_{k \to i} $ themselves. Therefore, using all the cavity equations which define the cavity variables, i.e. Eqs.~(\ref{eq:eta}), (\ref{eq:cavg}),  (\ref{eq:pcavlin}) and (\ref{eq:uk}), we can write the self-consistent distributional equation 
\begin{align}
\label{eq:stochSC}
    P(g,\eta,\bar p) =& \int d\varepsilon\, \gamma_+ (\varepsilon)
    \prod_{l=1}^{K+1} \left[dg_ld\eta_l d\bar p_l  \,P( g_l,\eta_l,\bar p_l ) \right] \; 
    \delta \bigg ( g-\frac{1}{\varepsilon-t^2\sum\limits_{l=1}^K g_l}\bigg ) \nonumber\\
    &\times \delta \bigg (  \eta- 1-t\sum\limits_{l=1}^K g_l \eta_l \bigg ) \delta \bigg(\bar p-\sum\limits_{l=1}^K\theta\Big(U_l\big(\{g_l,\eta_l\}_{l\leq K+1}\big)-1/E\Big)\bar p_l \bigg).
\end{align}
This equation holds close to criticality.  If one wants to obtain the distribution for a generic pair $(E,W)$, the equation must be solved exchanging the delta function enforcing Eq.~(\ref{eq:pcavlin}) with a delta enforcing the full relations in 
Eqs.~(\ref{eq:cavpbari}). 

The critical curve of this percolation problem is given by the points in the $(E,W)$-plane where $\bar p_{k \to i}$ passes from having zero to non-zero expectation value.

Equation~(\ref{eq:stochSC}) can be solved by using population dynamics, which is  a standard numerical technique used to solve self-consistent distributional equations, see Sec~\ref{subsec:population}.

\subsection{Correlation function and average cluster size}
\label{sec:CFP}

The correlation function $C_{p}(r)$ for a percolation problem  is defined 
as the probability that two sites at distance $r$ belong to the same cluster. 
Denoting as $0,\dots,r$ the sites on the lattice along a non-intersecting path of length $r$,  we have
\begin{equation}
\label{eq:corrfuncnobond}
     C_{\rm perc}(r) = \Pr \{O_0=1, \dots, O_{r}=1\}=\mathbb{E}\bigg[\prod_{i=0}^r O_i\bigg]\,.
\end{equation}
with the occupation variables $O_i$ defined as
\begin{equation}
    \label{eq:Oiperc}
    O_{i} = \begin{cases}
        1 & \text{if } u_{i} \geq 1/E \; ,  \\
        0 & \text{otherwise} \; 
    \end{cases} =\theta(u_i -1/E)
\end{equation}

Since the Bethe lattice is statistically translational invariant the occupation probability will be site-independent, therefore we can define
\begin{equation}
\label{eq:q}
    q \equiv \Pr\{O_i=1\} = \Pr \{u_i\geq 1/E\}
    \; . 
\end{equation}
Since the $O_i$ depend only on the set of independent cavity quantities $\big\{(\mathcal{G}_{k\to i},\eta_{k \to i})\,|\,i\in\{0,\dots, r\}\,,k\in \partial i \cap\partial \{0,\dots, r\}\}$, given the joint probability distribution $P(g,\eta,\bar p)$ we can compute $C_{\rm perc}(r)$. The expected scaling of the correlation function for large distances is $C_{\rm perc}(r)\sim e^{-r/\xi_{\rm perc}}/K^r$ and from this expression one extracts the 
correlation length $\xi_{\rm perc}$.  The procedure for the numerical computation of $C_{\rm perc}(r)$ and $\xi_{\rm perc}$ is explained in Sec.~\ref{app:numeval}. 

The average cluster size $S$ represents the average size of the connected component to which a generic occupied site belongs. This can be computed directly from the correlation function, by observing that the average number of sites at distance $r$ belonging to the same cluster as site $i$ is $(K+1)K^{r-1}C_{\rm perc}(r)/q$ (the factor $1/q$ is needed to condition the probability on the occupation of the reference site). Therefore, the average number of sites belonging to the same cluster of site $i$, regardless of their distance, will be just
\begin{equation}
\label{eq:avgclsize}
    S=1+\sum_{r=1}^\infty (K+1)K^{r-1}\frac{C_{\rm perc}(r)}{q}
    \; . 
\end{equation}
$q$ can be computed numerically from $P(g,\eta)$ as explained in Sec.~\ref{subsubsec:qqbar}.

\section{Numerical Methods}
\label{sec:numerics}

In this Section we explain the numerical methods that we used to solve the critical percolation properties of the 
Localization Landscape. The numerical procedures for computing the critical properties of the Anderson model on the Bethe lattice are described in detail in Refs.~\cite{rizzo2024localized,tikhonov2019critical}.

\subsection{Population dynamics}
\label{subsec:population}

The population dynamics algorithm is a method used to solve self-consistent distributional equations. It is accurately described in the context of message-passing algorithms on random graphical models in Ref.~\cite{mezard2009information}.
Consider the Bethe lattice with connectivity $K+1$ and the cavity equation that relates a cavity variable $X$ to $K+1$ independent copies of itself on the nearest neighboring sites $\{X_l\}_{l=1,\dots,K+1}$, 
along with a random variable $Y$ drawn from a distribution $\gamma_+$:
\begin{equation}  
\label{eq:cavgeneric}  
    X = \Psi\left(\{X_l\}_{l=1,\dots,K+1}\,; Y\right).  
\end{equation}  
It is important to note that the variables in this equation are not necessarily scalars. For example, in the set of cavity equations given by Eq.~(\ref{eq:cavg}), the equation (\ref{eq:eta}) in the right column, and Eq.~(\ref{eq:cavpbari}), we will have $X = (\mathcal{G}_{k \to i}, \eta_{k \to i}, \bar p_{k \to i})$, $Y=\varepsilon_k$,
and $\{X_l\}_{l=1,\dots,K+1} = \{\mathcal{G}_{l \to k}, \eta_{l \to k}, \bar p_{l \to k}\}_{l\in \partial k}$, 
and the function $\Psi$ relating these.

We aim to find the probability distribution $P$ such that, when $\{X_l\}_{l=1,\dots,K+1}$ are independent random variables drawn from $P$, the equality in Eq.~(\ref{eq:cavgeneric}) holds in distribution. This means that $P$ is the solution to the equation  
\begin{equation}  
\label{eq:popdyneq}  
    P(x) = \int dy \, \gamma_+(y) \int \prod_{l=1}^{K+1} \left[dx_l \, P( x_l )\right]  \delta \bigg( x - \Psi \left( \{x_l\}_{l=1,\dots,K+1} \,; y \right) \bigg),  
\end{equation}  
where the statistical independence of the $X_l$ variables has let us factorize the joint probability distribution within the integral of 
the right-hand-side
\begin{equation}
    P(\{x_l\}_{l=1,\dots,K+1}) = \prod_{l=1}^{K+1} P(x_l)
    \; .
\end{equation}
The algorithm is based on defining a ``pool'' (or ``population'') of $N$ variables $\{X_i\}_{i=1,\dots,N}$, which approximately represents the probability distribution:
\begin{equation}
    P(x) \simeq \frac{1}{N} \sum_{i=1}^N \delta(x - X_i) \, .
\end{equation}
Each element of the pool is initialized independently at random. 
We select an integer $T$ as the number of iterations. The empirical distribution of $\{X_i^{(\tau)}\}_{i=1,\dots,N}$ at iteration $\tau$ is denoted $\tilde P^{(\tau)}$. At each iteration step $t$, we sample a variable $y^{(\tau)}$ from $\gamma_+$ and randomly select $K+1$ indices $\{i_1, \dots, i_{K+1}\}$ from $\{1, \dots, N\}$. 
 We then replace $X_{i_1}^{(\tau-1)}$ in the population with  
\begin{equation}  
    X_{i_1}^{(\tau)} = \Psi \left( \{X_{i_l}^{(\tau-1)}\}_{l=1,\dots,K+1} \,; y^{(\tau)} \right),  
\end{equation}  
while keeping all other variables unchanged.

Assuming Eq.~(\ref{eq:popdyneq}) has a solution, it can be argued that for sufficiently large $T$ and $N$, the empirical distribution $\tilde P^{(\tau)}$ will be a good approximation of $P$.

\subsection{Numerical determination of the mobility edge}

Here we describe the numerical procedure used to determine the localization transition. The starting point is the set of recursion equations for the cavity Green’s functions, Eq.~\eqref{eq:gcavAL}. As discussed in Sec.~\ref{sec:SCanderson}, in the localized phase the imaginary part of the Green’s function vanishes proportionally to the regulator~$\alpha$. Anderson localization can therefore be investigated through the linear stability of Eqs.~\eqref{eq:gcavAL} with respect to a small imaginary component, which leads to the simplified recursion equations~\eqref{eq:ReGcavcrit}–\eqref{eq:ImGcavcrit}. Equation~\eqref{eq:ReGcavcrit} determines the distribution of the real parts alone, independently of the (small) imaginary components, while Eq.~\eqref{eq:ImGcavcrit} is linear in the imaginary parts. As a consequence, the typical value of the imaginary parts either grows exponentially (in the delocalized phase) or decreases exponentially (in the localized phase) under iteration. The corresponding growth (or decay) rate is given by the largest eigenvalue~$\lambda$ of the linear integral operator~\eqref{eq:fAL},
\begin{equation}
\mathcal{G}^I_{\rm typ} \propto \lambda^{n_{\rm iter}} \, ,
\end{equation}
and can be computed numerically using the population dynamics algorithm.

The numerical procedure proceeds as follows. We initialize a pool of $N$ pairs of cavity Green’s functions ${ \mathcal{G}_i^R, \mathcal{G}_i^I }$. We first iterate Eq.~\eqref{eq:ReGcavcrit} using population dynamics for the real parts only, until their distribution converges to a stationary form. We then start updating the entire pool, real and imaginary parts together, according to Eqs.~\eqref{eq:ReGcavcrit}–\eqref{eq:ImGcavcrit}. The only modification compared to the standard algorithm is the following: since our goal is to monitor the exponential growth or decay of the typical imaginary part, at each iteration step $n+1$ we generate an entirely new pool of $N$ elements from the pool at “time’’~$n$, replace the old pool, and iterate again. After a transient of about $200$ iterations, the typical value of $\mathcal{G}_i^I$ begins to grow or decrease exponentially, depending on whether the parameters $(E,W)$ lie in the delocalized or localized phase. We extract the growth exponent through a fit of the resulting data, and repeat the whole procedure several times to improve the statistical accuracy.

The main difficulty of this approach stems from the fact that the probability distributions of $(\mathcal{G}_i^R, \mathcal{G}_i^I)$ develop power-law tails at large arguments, as extensively discussed in the literature~\cite{abou1973selfconsistent,tikhonov2019critical,rizzo2024localized}. Consequently, finite-size effects associated with the finite pool size $N$ used to represent these distributions are very strong. These finite-size effects have been investigated in detail and with high precision in Ref.~\cite{tikhonov2019critical} for $k=2$ and $E=0$, where the localization transition is driven by increasing $W$. In that specific case, the operator~\eqref{eq:fAL} has been diagonalized numerically with very high accuracy in Refs.~\cite{parisi2019anderson,tikhonov2019critical}, allowing for a careful characterization and control of the finite-$N$ corrections, since the $N \to \infty$ asymptotic result is known exactly. We take advantage of this analysis to carry out an accurate determination of the mobility edge at $E>0$.

\begin{figure}[h!]
    \begin{center}
         \includegraphics[width=0.46\textwidth]{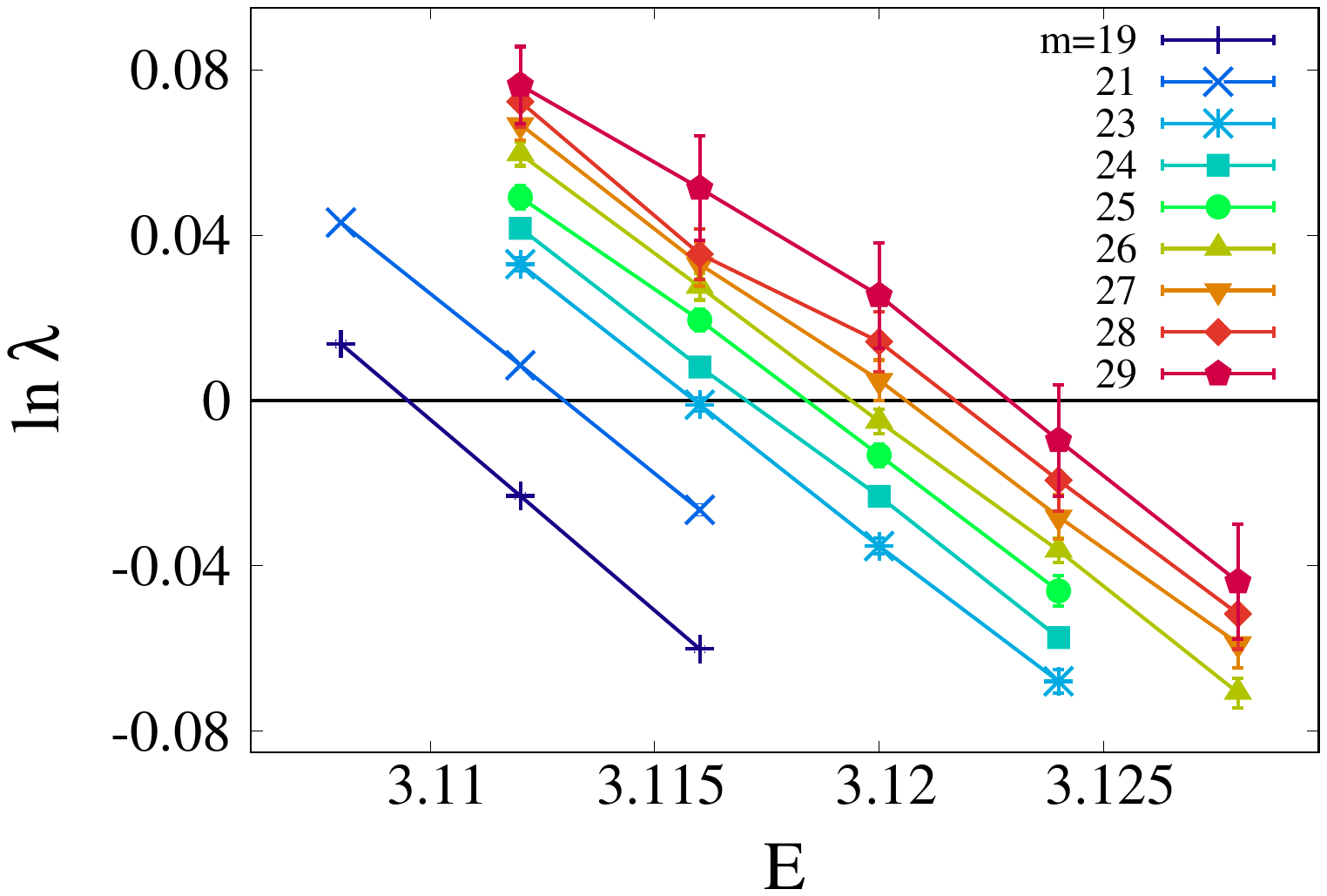}
    \end{center}
       \vspace{-0.5cm}
    \caption{\small{Exponent governing the exponential growth or decay of the typical imaginary part of the Green’s function in the linearized regime, $\ln \lambda$, as a function of the energy $E$ for $W = 1.5$ and $K = 2$, shown for several population sizes $N = 2^m$ with $m$ ranging from $19$ to $29$. The finite-$N$ estimate of the mobility edge, $E_c(N)$, is obtained from a linear fit of the data in the region where $\ln \lambda$ is small.}}
    \label{fig:critical-W1}
\end{figure}

\begin{figure}[h!]
    \begin{center}
         \includegraphics[width=0.332\textwidth]{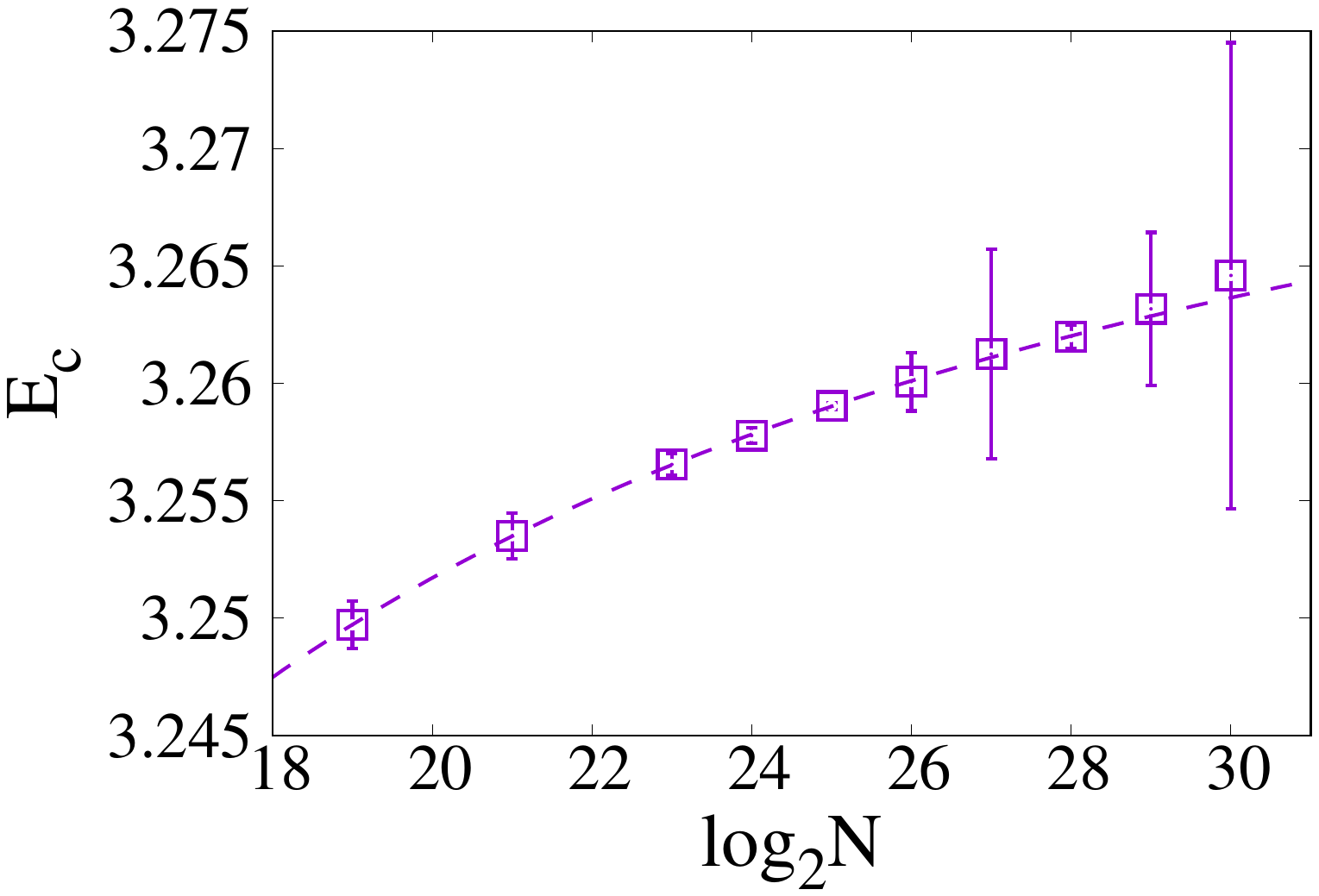} \put(-127,91){\small (a)}
         \includegraphics[width=0.332\textwidth]{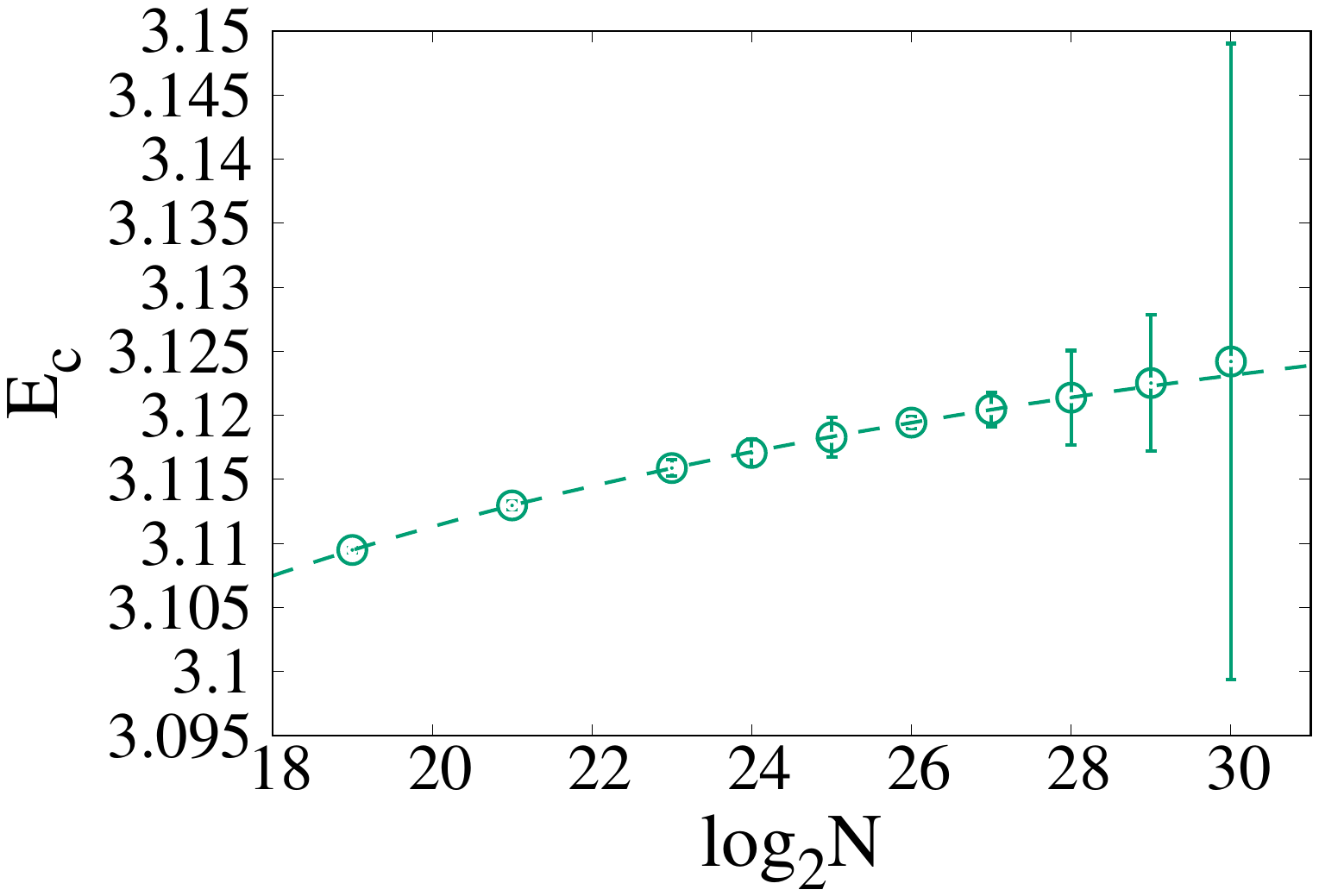} \put(-127,91){\small (b)}
         \includegraphics[width=0.332\textwidth]{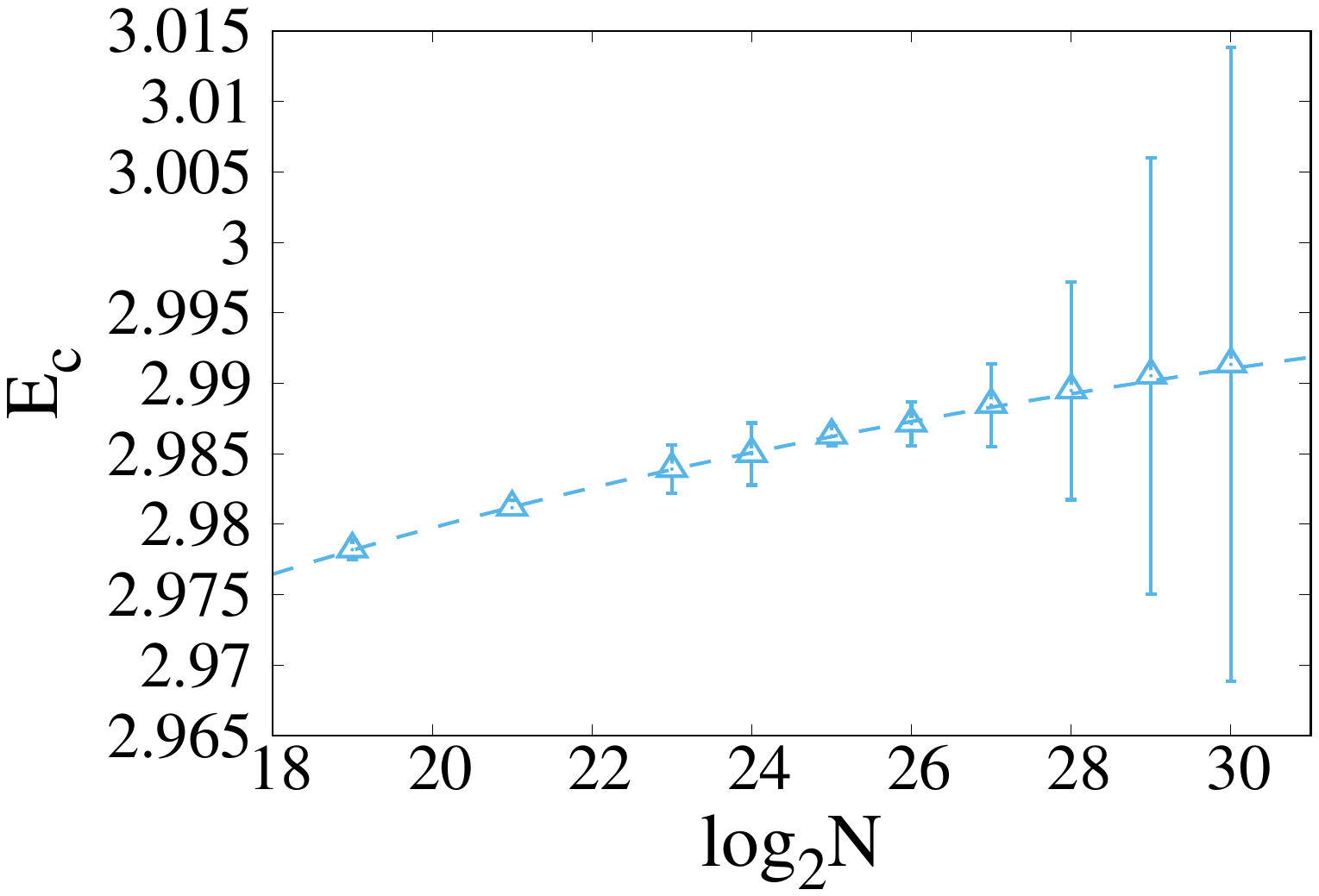} \put(-127,91){\small (c)}
         
         \includegraphics[width=0.332\textwidth]{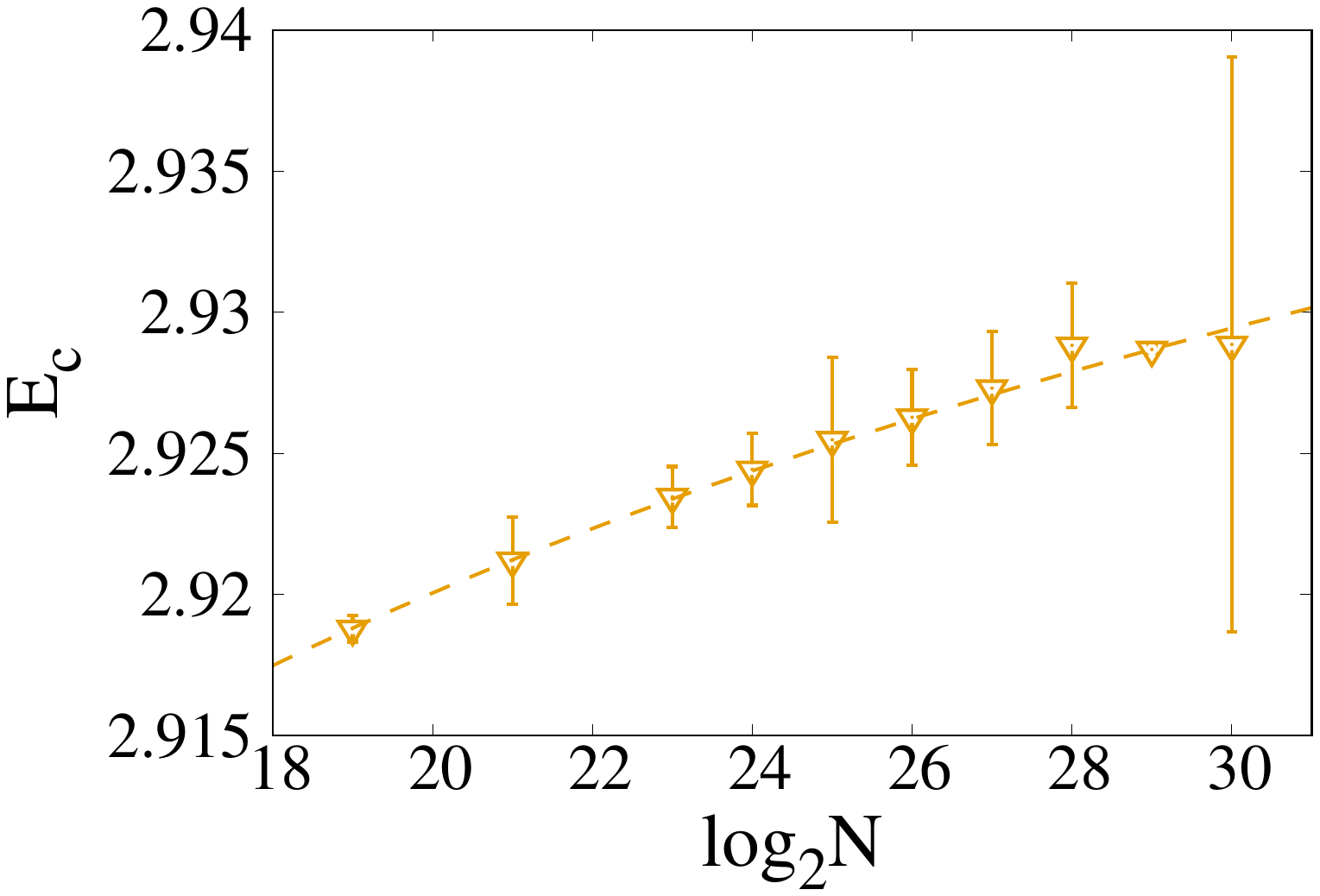} \put(-127,91){\small (d)}
         \includegraphics[width=0.332\textwidth]{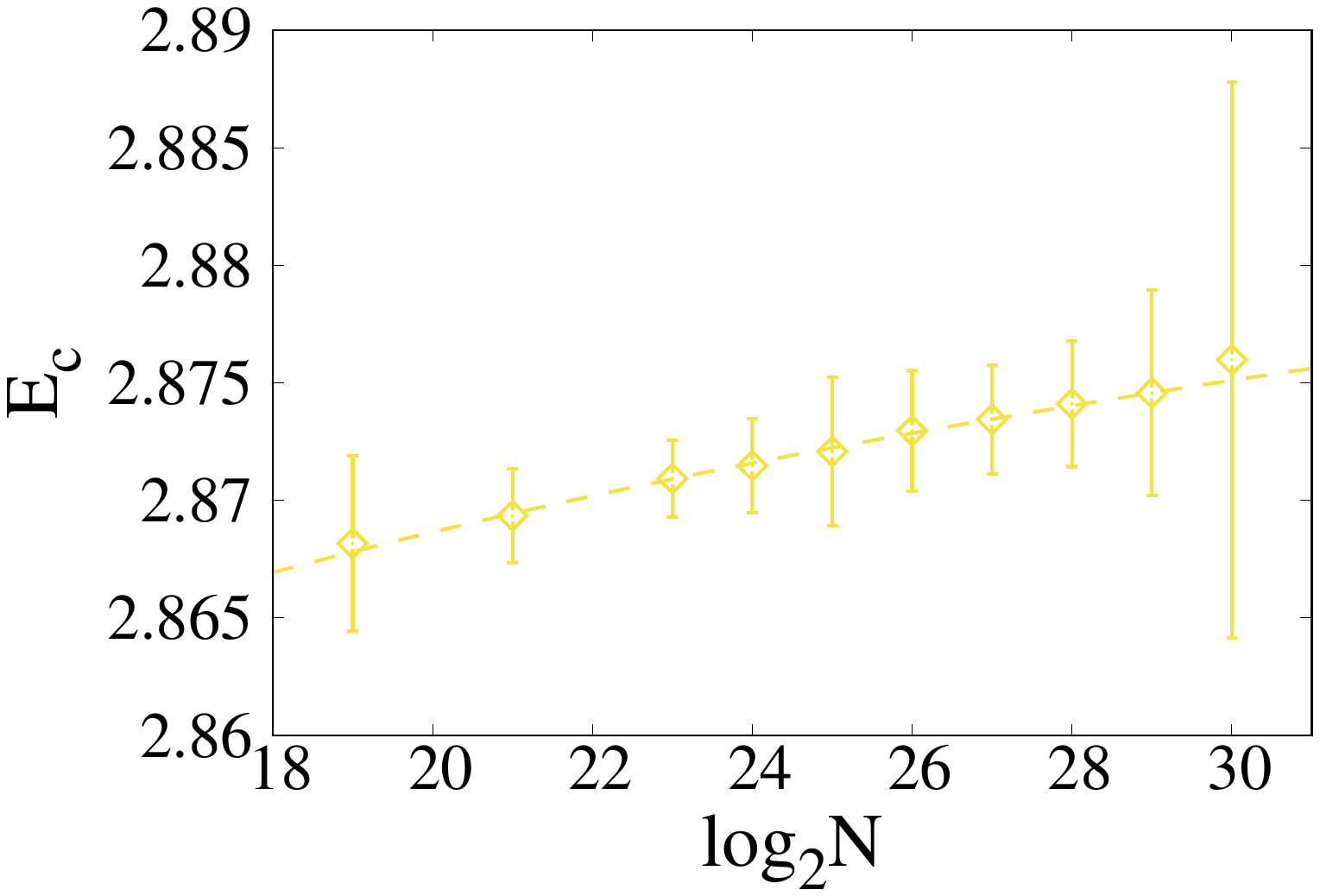} \put(-127,91){\small (e)}
         \includegraphics[width=0.32\textwidth]{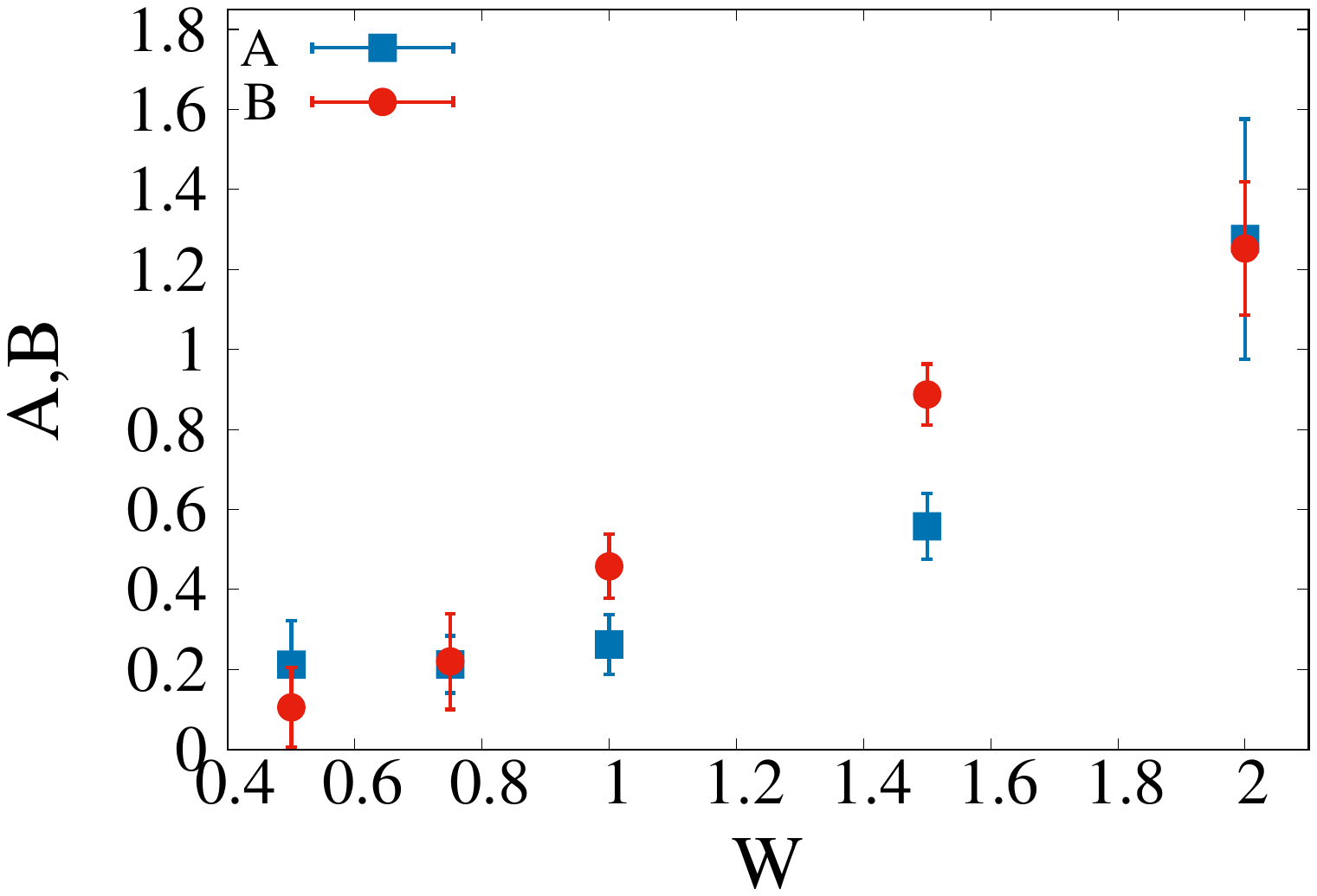} \put(-78,88){\small (f)}
         
    \end{center}
       \vspace{-0.5cm}
    \caption{\small{Fitting procedure used to determine the asymptotic critical energy $E_c$ according to Eq.~\eqref{eq:fit_Ec}, for $W = 2$ (a), $W = 1.5$ (b), $W = 1$ (c), $W = 0.75$ (d), and $W = 0.5$ (e). In panel (f) we plot the $W$-denepndence of the fitting parameters $A$ and $B$.}}
    \label{fig:fits_Ec}
\end{figure}

In Fig.~\ref{fig:critical-W1} we show $\ln \lambda$ as a function of $E$ for $W = 1.5$ (with $K = 2$) for several pool sizes $N = 2^m$, with $m$ ranging from $19$ to $29$. The figure clearly demonstrates that the energy at which $\ln \lambda$ crosses zero, $E_c(N)$, obtained from a linear fit of the data and corresponding to the estimated position of the mobility edge, drifts slowly but systematically to larger energies as the pool size increases. At $E=0$, where the exact $N \to \infty$ asymptotic value of the localization transition is known with high precision, the analysis of Ref.~\cite{tikhonov2019critical} shows that convergence to the thermodynamic limit is logarithmically slow in $N$. Motivated by this, we extrapolate the asymptotic mobility edge assuming the same type of finite-size corrections, and fit $E_c(N)$ with
\begin{equation} \label{eq:fit_Ec}
E_c(N) = E_c - \frac{A}{(\ln N)^B} \, ,
\end{equation}
where $A$ and $B$ are disorder-dependent fitting parameters.
In Figs.~\ref{fig:fits_Ec}(a)–(e) we show the results of these fits for several disorder strengths. As $W$ decreases, the concavity of $E_c(N)$ becomes flatter, corresponding to a decrease of the exponent $B$. This indicates that finite-size effects become increasingly severe at smaller disorder. Combined with the limited accessible range of $\ln N$-due to increasing computational cost-and with the increasing statistical uncertainty for the largest $N$, this results in rather large error bars in the extrapolated asymptotic value of the mobility edge, especially for small $W$. Consequently, we are unable to obtain reliable extrapolations for $W < 0.5$.

A further important point is that, because the accessible range of $\ln N$ is limited and the error bars for the largest $N$ are substantial, a relatively broad range of values of the fitting parameters $E_c$, $A$, and $B$ provides fits of comparable quality for each disorder strength. Within this range, we select the fitting parameters such that their dependence on $W$ is smooth (for instance, by imposing that $E_c$, $A$, and $B$ are decreasing functions of $W$), as illustrated in Fig.~\ref{fig:fits_Ec}(f).

The result of this procedure finally yields the estimation of the asymptotic position of the mobility edge which is plotted in the phase diagram of the main text.

\subsection{Numerical algorithm to compute the relevant observables in the localized phase}

In this section, we show how the linearized cavity recursion equations~\eqref{eq:ReGcavcrit}-\eqref{eq:ImGcavcrit} can be efficiently used to compute relevant observables in the localized phase. For clarity, we illustrate the method using the IPR. The extension to the computation of the two-point correlation function is straightforward and will not be discussed in detail. The approach is a direct generalization of the algorithm introduced in Ref.~\cite{rizzo2024localized} to the case $E \neq 0$.

The IPR is related to the second moment of $|\mathcal{G}_{ii}|$, which is broadly distributed according to Eq.~\eqref{eq:ansatz}. The power-law tails of its probability distribution would lead, in principle, to divergent expressions for $\langle |\mathcal{G}_{ii}|^2 \rangle$. In practice, this divergence is avoided because the power-law behavior is cut off at large values of the imaginary part around $\alpha^{-1}$, which marks the limit of validity of the linearized equations. Nevertheless, computing $\langle |\mathcal{G}_{ii}|^2 \rangle$ requires accessing the region of large imaginary parts of the Green's function, rather than the region of typical finite values. At first sight, this seems to suggest that the linearized equations might not be useful.

Fortunately, this is not the case. To see this, we define:
\begin{equation} \label{eq:mii1}
{\cal M}_{ii} \equiv \frac{1}{{\cal G}_{ii}} = \epsilon_{i} - E - {\rm i} \alpha - \!\! \sum_{m \in \partial i} {\cal G}_{m \to i} = m_{ii} - {\rm i} \alpha \, \hat{m}_{ii} \, , 
\end{equation}
from which  one immediately obtains that 
\[
\langle |{\cal G}_{ii}|^2 \rangle  = \int  Q(m,\hat{m}) \frac{1}{m^2+\hat{m}^2 \alpha^2}   \, \de m \, \de \hat{m} \, .
\]
Similarly, using the fact that ${\cal G}_{ii}^I = \alpha \hat{m}_{ii}/(m_{ii}^2 + \alpha^2 \hat{m}_{ii}^2)$, $\langle {\rm Im} {\cal G}_{ii} \rangle$ is expressed as:
\[
\langle {\rm Im} {\cal G}_{ii} \rangle = \int  Q(m,\hat{m}) \frac{\alpha {\hat{m}}}{m^2+\hat{m}^2 \alpha^2}   \, \de m\, \de \hat{m} \, ,
\]
Given that $\hat{m}$ is strictly positive we can make the change of variables $m = \alpha \hat{m} x$ that leads to
\begin{eqnarray} 
\langle |{\cal G}_{ii}|^2 \rangle &=& \int  Q(\alpha \hat{m} x  ,\hat{m}) 
\frac{(\alpha \, \hat{m})^{-1}}{1 + x^2}   \, \de x \, \de \hat{m} \, , \label{eq:gii} \\
\label{eq:img} \langle {\rm Im} {\cal G}_{ii} \rangle &=&  \int  Q(\alpha \hat{m} x  ,\hat{m}) \frac{1}{1 + x^2}   \, \de x \, \de \hat{m} \, .
\end{eqnarray}
In the $\alpha \to 0$ limit we can approximate $Q(\alpha \hat{m}  x,\hat{m}) \approx Q(0,\hat{m})$ and perform the integration over $x$ explicitly. Plugging Eqs.~\eqref{eq:gii} and \eqref{eq:img} into Eq.~\eqref{eq:IPR}, one finally obtains~\cite{mirlin1994statistical,rizzo2024localized}:
\begin{equation} \label{eq:IpQ}
    I_2 = 
    \frac{ \int  Q(0  ,\hat{m}) \, \hat{m}^{-1} \,\de \hat{m}}
    {\int  Q(0 ,\hat{m})  \, \de \hat{m}} \, .
\end{equation}
To sum up, although the moments of the local Green's functions are controlled by the fact that  $|{\cal G}_{ii}|$ is $O(1/\alpha)$ with probability $O(\alpha)$, they can be computed in terms of the {\it typical} values of ${\cal M}_{ii}$, whose real part that is typically $O(1)$ and whose imaginary part that is typically $O(\alpha)$. The fact that in the localized phase one can use the linearized equations to compute the relevant observables, such as the (generalized) IPR, facilitates the adoption of highly efficient computational methods that strongly reduces the effect of the finite size of the population.

We thus introduce a modification of the population dynamics algorithm (see Ref.~\cite{rizzo2024localized} for more details) which allows us performing the extrapolation of $Q(m,\hat{m})$ to $m=0$ very efficiently, thereby allowing one to evaluate the numerator and the denominator of Eq.~\eqref{eq:IpQ} with arbitrary accuracy.  In fact, from Eq.~\eqref{eq:mii1} we have that $m_{ii} = \epsilon_{i} - E - \sum_{m \in \partial i} {\cal G}_{m \to i}^R$. Hence, the probability that $m_{ii}=0$ is equal to the probability that $\epsilon_i = E + \sum_{m \in \partial i} g_{m \to i}$. This occurs with probability density $1/W$  if $|E + \sum_{m \in \partial i} g_{m \to i}|<W/2$, and with zero probability otherwise. Based on this observation, we thus proceed in the following way:
For a given pair of values $(E,W)$ in the localized phase, we implement the standard population dynamics algorithm described in Sec.~\ref{subsec:population} and obtain the stationary probability distribution of the cavity Green's function $P({\cal G}^R,{\cal G}^I)$ in the linearized regime, corresponding to the solution of Eqs.~\eqref{eq:ReGcavcrit}–\eqref{eq:ImGcavcrit}; We extract $K+1$ elements from the pool and compute $m$ and $\hat{m}$ from Eq.~\eqref{eq:mii1}. We define $S = E + \sum_{i=1}^{K+1} {\cal G}^R_{i}$; If (and only if) $|S|<W/2$ we add $\hat{m}^{-1}/W$ to the numerator and $1/W$ to the denominator of $I_2$; We repeat this process several times and divide the numerator and the denominator by the total number of attempts; We renew the elements of the pool of the cavity Green's function by performing a few steps of the standard population dynamics algorithm and repeat the whole process several times until the desired accuracy on $I_2$ is reached. The algorithm described here can be straightforwardly extended to the computation of generic two-points correlation functions.

\subsection{Montecarlo sampling of $P(g,\eta)$}
\label{app:numeval}

We now summarize the method that we have used to compute numerically 
many of the quantities related to the Localization Landscape percolation. 
All the following procedures have in common that they involve the sampling of pairs of 
cavity Green's functions and cavity rescaled fields from the $P(g,\eta)$ that solves
\begin{equation}
    P(g,\eta)=\int d\varepsilon\, \gamma_+(\varepsilon) \int \prod_{k=1}^K\left[dg_k d\eta_k \,P( g_k,\eta_k ) \right]\; \delta\bigg(g-\frac{1}{\varepsilon-t^2\sum_k g_k}\bigg)\delta \bigg(\eta -1-t\sum_kg_k\eta_k\bigg)\,.
\end{equation}
 $P(g,\eta)$ can be  computed from this equation via population dynamics, see Sec.~\ref{subsec:population}. 

 As argued in Sec.~\ref{subsubsec:joint}, with a negligible error we can substitute $P(g,\eta)$ 
with the product $P_g(g)P_\eta(\eta)$ where the marginal distributions $P_g(g)$ and $P_\eta(\eta)$ are the ones derived in Sec.~\ref{subsubsec:marginals}.  This avoids repeated population dynamics simulations when varying $W$. Numerically, we found that the Pearson correlation coefficient $r \sim O(10^{-2})$ and the mutual information is of $O(10^{-2} \, \text{bits})$ for the parameter ranges that we used, confirming that the factorization approximation proposed in Eq.~\eqref{eq:MF} introduces negligible errors.

 \subsubsection{The occupation probabilities $q$ and $\bar q$}
\label{subsubsec:qqbar}

The occupation probability $q$ for the Localization Landscape percolation problem (see Eqs.~\eqref{eq:Oiperc} and \eqref{eq:q}), 
can be written as
\begin{equation}
    q \equiv \Pr\{O_i=1\}= \mathbb{E}[O_i]
\end{equation}
and it can be computed explicitly through Montecarlo sampling:
\begin{equation}
    q = \frac{1}{M}\sum_{m=1}^M O_i^{(m)}\,,
\end{equation}
where for each sample $m$ we draw $\{\mathcal{G}_{k\to i}^{(m)}, \eta_{k \to i}^{(m)}\}_{l \in \partial k}$ from $P(g,\eta)$ to compute the occupation variable $O_{i}^{(m)}$. Alternatively, in the high-connectivity limit, we have an analytical expression of 
the marginal distribution of $u_{i}$ (see Eq.~\eqref{eq:P_u}) and we can just calculate $q= \int du \;P_u(u) \, \theta(u-1/E)$.

The conditional probability $\bar{q}$ is defined in Eq.~\eqref{eq:qbar}. It reads
\begin{equation}
  \bar{q} \equiv \Pr \{ O_{k\to i} =1 \mid O_{i\to j}=1 \} \quad \forall i, \; \forall j \in \partial i, \; \forall k \in \partial i \setminus j \; , 
\end{equation}
with the cavity occupation variables defined in Eq.~\eqref{eq:cavOi}, 
and governs the high-connectivity transition. It can computed numerically by rewriting it as
\begin{equation}
\label{eq:ProbOsol}
    \bar{q} = \frac{\Pr \{ O_{k \to i} =1, O_{i \to j}=1 \}}{\Pr \{ O_{i \to j} =1 \}} =  \frac{\mathbb{E}[O_{k\to i}O_{i\to j}]}{\mathbb{E}[O_{i\to j}]}\,.
\end{equation}
Again, the expectation values of this expression can be evaluated using Montecarlo sampling, i.e.
\begin{equation}
\label{eq:qbarmonte}
    \bar{q} = \frac{\sum_{m=1}^M O_{i \to j}^{(m)} O_{k \to i}^{(m)}}{\sum_{m=1}^M O_{i \to j}^{(m)}} \overset{d}{=} \frac{1}{K} \frac{\sum_{m=1}^M O_{i \to j}^{(m)} \sum_{k \in \partial i \setminus j} O_{k \to i}^{(m)}}{\sum_{m=1}^M O_{i \to j}^{(m)}}\,,
\end{equation}
where for each of the $M$ samples, $\{\mathcal{G}_{k\to i}^{(m)}, \eta_{k \to i}^{(m)}\}_{k \in \partial i \setminus j}$ are drawn from $P(g,\eta)$ to compute the occupation variables $O_{i \to j}^{(m)}$ and $\{O_{k \to i}^{(m)}\}_{k \in \partial i}$. The last equality in Eq.~(\ref{eq:qbarmonte}) follows from the statistical equivalence of the variables $\{O_{k \to i}\}_{k \in \partial i}$.

\subsubsection{The percolation correlation function and the average cluster size}

From its definition in Eq.~(\ref{eq:corrfuncnobond}), we compute the correlation function numerically by approximating the expectation value with the average over $M$ samples as

\begin{equation}
    C_{\rm perc}(r)=\frac{1}{M}\sum_{m=1}^M\prod_{s=0}^r O_s^{(m)}\,,
\end{equation}
where the $O_s$ are the ones defined in Eq.~(\ref{eq:Oiperc}).  
Each $u_s$ is computed in terms of the on-site Green's function  and the  rescaled field  as $u_s=\mathcal{G}_{ss}\eta_s$, where $\mathcal{G}_{ss}$ and $\eta_s$ are computed as functions of their cavity counterparts on the nearest-neighboring sites of the path between site $0$ and site $r$ (i.e. $\big\{(\mathcal{G}_{k\to i},\eta_{k \to i})\,|\,i\in\{0,\dots, r\}\,,k\in \partial i \cap\partial \{0,\dots, r\}\}$)  according to Eq.~(\ref{eq:g}) and the equation on the left in (\ref{eq:eta}). 
Thus, in order to compute a sample of $\prod_{s=0}^r O_s^{(m)}$, we just need to draw $K(r+1)+2$ independent identically distributed pairs of cavity Green's functions and cavity rescaled fields from $P(g,\eta)$.

The correlation length of $C_{\rm perc}(r)$ is evaluated by fitting the functional form $e^{-r/\xi_{\rm perc}}/K^r$ to the curve $C_{\rm perc}(r)$ in the non-percolating phase, for $r\gg 1$.

For the computation of $S$ we just use the definition in Eq.~(\ref{eq:avgclsize}). Therefore, we need to compute the correlation function $C_{\rm perc}(r)$ up to a cutoff distance $r_{max}$ and perform explicitly the sum in Eq.~(\ref{eq:avgclsize}). 

\section{Analytic results}
\label{sec:analytic}

In this Section we derive some analytic results. In Sec.~\ref{subsec:indsites} we present the equation determining the parameter dependence of the percolation critical curve in the independent-site approximation. Next, in Sec.~\ref{subsec:hicon} we define the high-connectivity limit, we write the equations governing the system in this regime, 
we derive many exact and approximate results which provide very good approximations for finite and not too small 
$K$, and we obtain the exact solution in the high connectivity limit for the critical properties of the Localization Landscape percolation problem.

\subsection{The independent-site approximation}
\label{subsec:indsites}

The uncorrelated site percolation model on a lattice assumes that each site is independently occupied with probability $q$ (and not occupied with probability $1-q$). On the Bethe lattice with connectivity $K+1$ the critical value of the occupation probability $q_c$ above which the system is in the percolating phase is $1/K$ \cite{stauffer2018introduction}. The independent-site approximation for the Localization Landscape percolation problem consists of treating the components of the Localization Landscape $u_i$ on different sites as statistically independent. Under this assumption, the problem maps directly onto an uncorrelated percolation problem where the probability $q$ that a generic site $i$ is occupied, is given by the probability of $u_i$ being greater or equal than $1/E$ (see Eq.~\eqref{eq:q}). Thus, the equation determining the critical curve $E^{\rm perc}_c(W)$ in the $(E,W)$-plane in the independent-site approximation reads 
\begin{equation} 
\label{eq:indsitescrit}
    \Pr\{u_i\geq1/E^{\rm perc}_c(W)\}\equiv q_c=1/K.
\end{equation}

The probability $q$ can be computed numerically as explained in Sec.~\ref{subsubsec:qqbar}.
The critical curve in the high-connectivity limit has been plotted in Fig.~\ref{fig:critcurve} with a solid red line. The independent-site approximation 
curve deviates significantly from the exact one in the high-connectivity limit, indicating that correlations are not negligible 
while determining the critical behavior of the system.

\subsection{The high-connectivity limit}
\label{subsec:hicon}

 In the high-connectivity limit  $K \gg1$, many of the  quantities describing the transition 
 can be obtained either analytically, under a few approximations, or with little numerical effort. 
 Moreover, many of these results remain accurate even for low values of $K$, 

 making the high-connectivity limit a very convenient 
 regime to study the main features of 
 the transition. 
 
Here, we derive the equations that describe the critical curve in the high-connectivity limit in two ways. 
The first one is based on defining an integral eigenvalue equation analogous to the one that we derived 
in Sec.~\ref{sec:ALeigenval} for Anderson localization. This derivation lets us 
understand that the equations describing the percolation transition are different from the ones describing the localization transition. 
The second method is more straightforward, and it is based on enforcing that the average cluster size diverges at the transition.

The main idea behind the high-connectivity limit is that normal and cavity variables can be considered to be 
statistically equivalent, i.e.
\begin{align}
    \mathcal{G}_{ii}&\overset{d}{=} \mathcal{G}_{i\to j} \,,\,\,\,\,
    \eta_i\overset{d}{=} \eta_{i\to j} \,,\,\,\,\,
    u_i\overset{d}{=} u_{i\to j}\overset{d}{=}\mathcal{G}_{i\to j}\eta_{i\to j} \,,\,\,\,\,
    p_i\overset{d}{=} p_{i\to j} \; , \,\,\,\qquad\qquad \forall i, \,\,\, \forall j \in \partial i\,,
\end{align}
for $K\gg 1$, 
where the symbol $\overset{d}{=}$ represents equality in distribution.
 This is because recursive equations for normal and cavity variables differ by just one of the $O(K)$ terms inside sums of the type $\sum_{k\in \partial i}$ or $\sum_{k\in \partial i \setminus j}$, and according to large deviation theory, under the hypothesis that cavity variables are random variables with finite mean and variance, we can safely neglect one of the terms in the sums. 

 The sets of cavity variables $\{\mathcal{G}_{k\to i}, \eta_{k\to i},p_{k\to i}\}_{k \in \partial i \setminus j}$ are not independent in the general case, because for each of the $p_{k \to i}$ one has to compute $u_k$, and all the $u_k$'s are statistically dependent, since they depend on quantities evaluated at their common neighbor $i$. 
 Now, since $u_k\overset{d}{=}\mathcal{G}_{k \to i}\eta_{k \to i}$, the cavity percolation probabilities become independent cavity variables. For this reason, the three cavity equations that determine the transition in the high-connectivity limit are
 \begin{align}
     \label{eq:gsys}
    \mathcal{G}_{i\to j}^{-1} & =  \varepsilon_i - t^2 \sum_{k \in \partial i \setminus j} \mathcal{G}_{k \to i} \, ,\\
    \label{eq:etasys}
    \eta_{i \to j} & = 1 + t \sum_{k \in \partial i \setminus j} \mathcal{G}_{k \to i } \eta_{k \to i}   \, ,\\
    \label{eq:psys}
    p_{i \to j} &= \theta(\mathcal{G}_{i\to j}\eta_{i\to j} -1/E)\sum_{k \in \partial i \setminus j} p_{k \to i} \, ,
 \end{align}
 where the last one has been obtained by expanding Eq.~\eqref{eq:cavpi} for small cavity percolation probabilities close to the critical curve. The self-consistent distributional equation that we have to solve is much simpler compared to Eq.~(\ref{eq:stochSC}), and reads
 \begin{align}
    P(g,\eta,p)&=\int d\varepsilon \,\gamma_+(\varepsilon)
    \int \prod_{k=1}^K \left[dg_kd\eta_kdp_k \,P( g_k,\eta_k,p_k ) \right]
    \; 
    \delta \bigg ( g-\frac{1}{\varepsilon-t^2\sum_kg_k}\bigg ) \nonumber\\ 
    &\qquad \qquad\qquad \qquad \times \delta \bigg (  \eta- 1-t\sum_k g_k \eta_k \bigg ) \delta \bigg (  p -\theta(g\eta-1/E)\sum_k p_k \bigg ). 
\end{align}
 
 In Sec.~\ref{subsubsec:marginals} we show that the marginal distributions of $\mathcal{G}_{i \to j}$, $\eta_{i \to j}$ and $u_{i\to j}$ admit thin-tailed solutions, 
 and that these solutions are very close to the ones obtained with population dynamics simulations also for $K \sim O(1)$. 
 In Sec.~\ref{subsubsec:joint} we show numerically
  that the joint distribution $P(g,\eta)$ can effectively be approximated with the product of their marginals. Section~\ref{subsubsec:lowW} determines the position of the isolated eigenvalue as a function of disorder and minimal disorder needed to have a percolation transition.
In Sec.~\ref{subsubsec:linstab} we perform a linear stability analysis and in Sec.~\ref{subsubsec:divergence-cluster-size}
we study the  average cluster size to locate the critical percolation curve.

\subsubsection{Marginal distributions}
\label{subsubsec:marginals}
 In order to obtain a self-consistent solution for the marginal distribution of the cavity Green's function $P_g(g)= \int d\eta dp \, P(g,\eta,p)$, we observe that Eq.~(\ref{eq:gsys}) 
 is closed, therefore it can be solved independently from the other cavity equations.
If $K$ is large enough, assuming that $P_g$ has finite mean and variance,  we can approximate the sum over the 
nearest neighbors of site $i$ in absence of $j$ as  
\begin{equation}
\label{eq:CLTG}
\sum_{k\in \partial i \setminus j} \mathcal{G}_{k\to i} \approx K \mu_g +\sqrt{K\sigma_g^2}Y
\; ,
\end{equation}
where
\begin{equation}
    Y  \sim \mathcal{N}(0,1) \; , \quad
    \mu_g=\mathbb{E}[\mathcal{G}_{k\to i}] \; , \quad   \sigma_g^2 = \mathbb{V}[\mathcal{G}_{k\to i}] \; ,
\end{equation}
and $\mathbb{V}[ \dots ]$ is the variance of the random variable. 
This follows directly from the central limit theorem. Now, the roughest approximation that we can make is to neglect Gaussian fluctuations.
In this limit the probability distribution of the cavity Green's functions is simply obtained by changing variables in the probability distribution of $\varepsilon_i$.
Thus, the probability distribution of the cavity Green's functions becomes
\begin{equation}
\label{eq:roughappr}
    Q_g(g;\mu_g) = 
    \begin{cases}
        \displaystyle \frac{1}{W g^2} & \qquad \text{if } g \in D_g \equiv\left[\frac{1}{W/2-E_{\rm min}-t^2 K \mu_g },\; \frac{1}{ -W/2-E_{\rm min}-t^2 K \mu_g}\right] 
        \; , \\
        0 & \qquad \text{else}
        \; . 
    \end{cases}
\end{equation}
From this expression, $\mu_g$ and $\sigma^2_g$ are easily obtained self-consistently by solving
\begin{equation}
\label{eq:gavgapprox}
\mu_g = \frac{1}{W} \ln\left| 1-\frac{W}{W/2+E_{\rm min} +t^2 K \mu_g} \right|
\; ,
\end{equation}
and computing
\begin{equation}
\label{eq:gvarapprox}
    \sigma^2_g = \frac{1}{(t^2 K \mu_g+E_{\rm min})^2-W^2/4}-\mu_g^2
    \; .
\end{equation}
From population dynamics calculations we have observed that these approximations of the mean and variance of the distribution are quite accurate 
already for $K>4$ and any $W$. 

A more precise expression for the full probability distribution of the cavity Green's functions can be found keeping the Gaussian fluctuations. Starting from
\begin{equation}
    \mathcal{G}_{i\to j}^{-1} \approx \varepsilon-t^2\Big( K \mu_g +\sqrt{K\sigma_g^2}Y\Big)
    \; ,
\end{equation}
we obtain
\begin{equation}
\label{eq:Pgacc}
    P_g(g;\mu_g,\sigma_g^2) = \frac{1}{W g^2} \int_{-W/2-E_{\rm min}}^{W/2-E_{\rm min}} \!\! d\varepsilon \, \mathcal{N} \Big( \varepsilon \,; Kt^2 \mu_g +1/g \,,\, Kt^4\sigma_g^2\Big) 
    \; ,
\end{equation}
where  mean and variance have to be obtained self-consistently from 
\begin{equation}
\label{eq:SCgind}
\mu_g = \int dg\,P_g(g\,;\mu_g,\sigma_g^2)g\,,\qquad\qquad
\sigma^2_g = \int dg\,P_g(g\,;\mu_g,\sigma_g^2)(g-\mu_g)^2 \,.
\end{equation}

We compute now the marginal distribution of the cavity rescaled fields by employing a similar approximation. We assume that
\begin{equation}
\label{eq:approxsumeta}
    \eta_{i\to j} \approx 1+t\Big(  K\mathbb{E}[\mathcal{G}_{k \to i }\eta_{k \rightarrow i}] + \sqrt{K\mathbb{V}[\mathcal{G}_{k \to i }\eta_{k \rightarrow i}]}Y\Big)\,, \quad \text{with } \, Y  \sim \mathcal{N}(y\,;0,1)
    \; . 
\end{equation}
We will argue later that this Gaussian approximation does not hold, even at high connectivity, when the disorder is too strong. Nevertheless, as we will see, it remains sufficiently accurate to provide an excellent estimate of the marginal distribution of the cavity Localization Landscapes, even at low connectivity. This is because the distribution of the cavity Green's function becomes highly asymmetric in the strong-disorder regime. Consequently, this asymmetry propagates to the distribution of the rescaled cavity fields, which thereby loses its Gaussian form.

The variables $\eta_{k \to i}$ and $\mathcal{G}_{k \to i }$ are not independent. Therefore, computing the expectation value and the variance of their product in Eq.~(\ref{eq:approxsumeta}) is not trivial. However, for high-connectivity, we can argue that they are weakly correlated. 
Therefore, we will  consider them as independent, i.e.
\begin{equation}
\label{eq:MF}
    P(g,\eta) = \int dp \; P(g,\eta,p)\approx P_g(g)P_{\eta}(\eta) \; .
\end{equation}

Under this approximation,  the mean and variance in Eq.~(\ref{eq:approxsumeta}) now read
\begin{equation}
\label{eq:meanvaretag}
    \mathbb{E}[\mathcal{G}_{k \to i }\eta_{k \rightarrow i}] =\mu_g\mu_{\eta}\,, \qquad
    \mathbb{V}[\mathcal{G}_{k \to i }\eta_{k \rightarrow i}]=\sigma_g^2\sigma_{\eta}^2+\sigma_g^2\mu_{\eta}^2+\sigma_{\eta}^2\mu_g^2
    \; , 
\end{equation}
where
\begin{equation}
    \mu_\eta=\mathbb{E}[\eta_{k\to i}]\,, \qquad  \sigma_{\eta}^2 = \mathbb{V}[\eta_{k\to i}]\,.
\end{equation}
This implies that the probability distribution of the cavity rescaled fields is
\begin{equation}
\label{eq:Peta}
    P_{\eta}(\eta;\mu_{\eta},\sigma_{\eta}^2) = \mathcal{N}(\eta \,;\mu_{\eta},\sigma_{\eta}^2) \,, 
\end{equation}
with
\begin{equation}
\label{eq:etastats}
\mu_{\eta} = \frac{1}{1-K t \mu_g }\,, \qquad \sigma_{\eta}^2 = \frac{Kt^2\sigma^2_g}{[1 - Kt^2 (\sigma^2_g+\mu_g^2)](1-Kt\mu_g)^2}\,.   
\end{equation}
These expressions have been obtained by enforcing that both sides of Eqs.~(\ref{eq:approxsumeta}) have the same mean and variance.

The factorization of $P(g, \eta)$ in Eq.~(\ref{eq:MF})  also simplifies the calculation of the marginal distribution of the cavity
 Localization Landscapes, 
which in distribution are given by 
 $u_{i \to j} \overset{d}{=} \mathcal{G}_{i\to j} \eta_{i \to j}$. The result is
\begin{align}
\label{eq:P_u}
    P_u(u) &= \int dg \,P(g\, ,\eta =u/g )\left | \frac{d(u/g)}{du}\right| = \int \frac{dg}{|g|} \, P_g(g)P_{\eta} (u/g)\nonumber\\&= \int dg \, P_g(g) \, \mathcal{N}(u \, ; g\mu_{\eta} \, ,g^2\sigma^2_{\eta})
    \; .
\end{align}

\begin{figure}[!ht]
\hspace{-6cm} (a) \hspace{8cm} (b) 
 \\
    \begin{center}
         \includegraphics[width=0.46\textwidth]{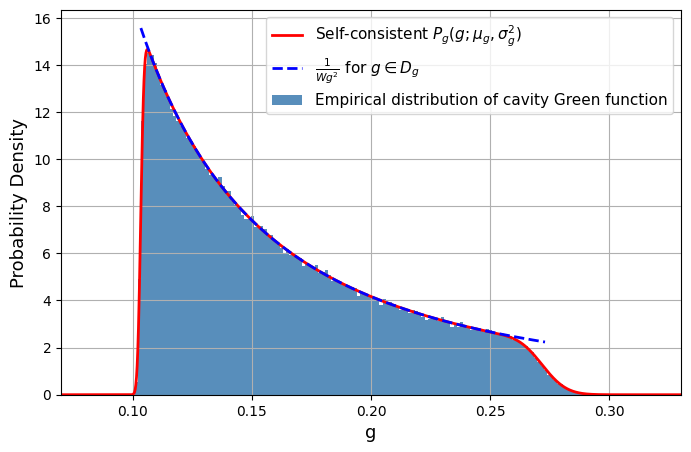}
        \includegraphics[width=0.4\textwidth]{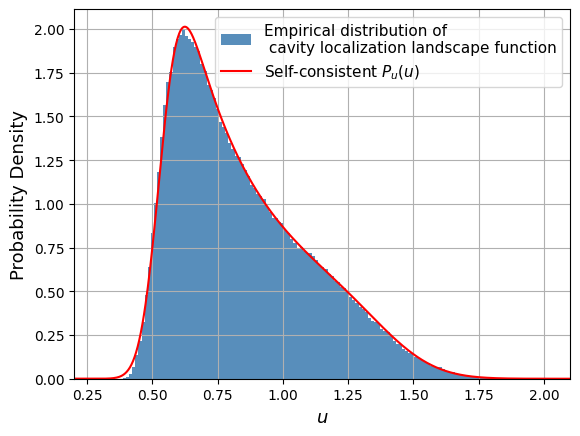}
    \end{center}
       \vspace{-0.5cm}
    \caption{\small{Marginal distributions of cavity Green's functions $\mathcal{G}_{i \to j}$ (a) and cavity Localization Landscapes $u_{i\to j}$ (b) for  $K=5$, $t=1$, and $W=6$. 
      Light blue histograms: empirical distributions obtained with population dynamics with $N=5\times 10^5$. 
    (a) Red curve: Probability distribution in Eq.~(\ref{eq:Pgacc}) with parameters $\mu_g$ and $\sigma_g^2$ computed solving by iteration the self-consistent 
    Eqs.~(\ref{eq:SCgind}). Blue curve: rougher approximation for the probability distribution given in 
    Eq.~(\ref{eq:roughappr}) with  $\mu_g$ computed self-consistently by iteration of Eq.~(\ref{eq:gavgapprox}).
     (b) Red curve: probability distribution $P_u$ in Eq.~(\ref{eq:P_u}), with parameters $\mu_u$ and $\sigma^2_u$ obtained self-consistently from 
     Eqs.~(\ref{eq:ustats}).}}
    \label{fig:PgPu}
\end{figure}

We can observe that $u_{i\to j}$ is the quantity with mean and variance 
 in Eqs.~(\ref{eq:meanvaretag}). Thus, using the results in Eqs.~(\ref{eq:etastats}) we have
\begin{equation}
    \label{eq:ustats}
    \mu_u =\mu_{\eta}\mu_g = \frac{\mu_g}{1-Kt\mu_g}\,,\qquad\qquad
    \sigma^2_u = \frac{\sigma^2_{\eta}}{Kt^2}=\frac{\sigma^2_g}{[1 - Kt^2 (\sigma^2_g+\mu_g^2)](1-Kt\mu_g)^2}\,.
\end{equation}

In Fig.~\ref{fig:PgPu} we plotted the distribution of the cavity Green's functions  $\mathcal{G}_{i \to j}$ (a) 
and the distribution of $u_{i \to j}=\mathcal{G}_{i\to j}\eta_{i \to j}$ (b) from population dynamics simulations
in a model with $K=5$, $t=1$, and $W=6$. 
We compare the histograms (light blue) to the large $K$ approximations (red curves) and found that the agreement is very good.

\subsubsection{Quality of the factorization approximation $P(g,\eta) \approx P_g(g) P_\eta(\eta)$}
\label{subsubsec:joint}

The joint probability distribution of the cavity Green's function and the cavity auxiliary fields $P(g,\eta)$ is the one satisfying the self-consistent distributional equation
\begin{equation}
\label{eq:PgetaSC}
     P(g,\eta)=\int d\varepsilon\, \gamma_+(\varepsilon) \int \prod_{k=1}^K\left[dg_k d\eta_k \; P( g_k,\eta_k )\right]\delta\bigg(g-\frac{1}{\varepsilon-t^2\sum_k g_k}\bigg)\delta \bigg(\eta -1-t\sum_kg_k\eta_k\bigg)\,.
\end{equation}
As we have anticipated in the previous section,  $P(g,\eta)$ is well approximated by the product of the two marginals $P_g(g)$ and $P_\eta(\eta)$ as in Eq.~(\ref{eq:MF}), and checked numerically. The quality of the approximation can be further quantified by computing the mutual information.

The mutual information of two continuous random variables $X,Y$ distributed with $P$ is defined as
\begin{equation}
    I(X,Y) \equiv \int dxdy\; P(x,y) \, \log_2 \frac{P(x,y)}{P(x)P(y)}\,,
\end{equation}
and it represents the information theoretical measure of the dependence of two random variables. More precisely, it is the information about the variable $X$ that one obtains after measuring $Y$, quantified in terms of Shannon's entropy.
The mutual information of two random variables is always greater or equal than zero. $I(X,Y)=0$ corresponds to the case in which $X$ and $Y$ are statistically independent.

\begin{figure}[!ht]
\vspace{0.5cm}
    \begin{center}
    \includegraphics[width=0.5\linewidth]{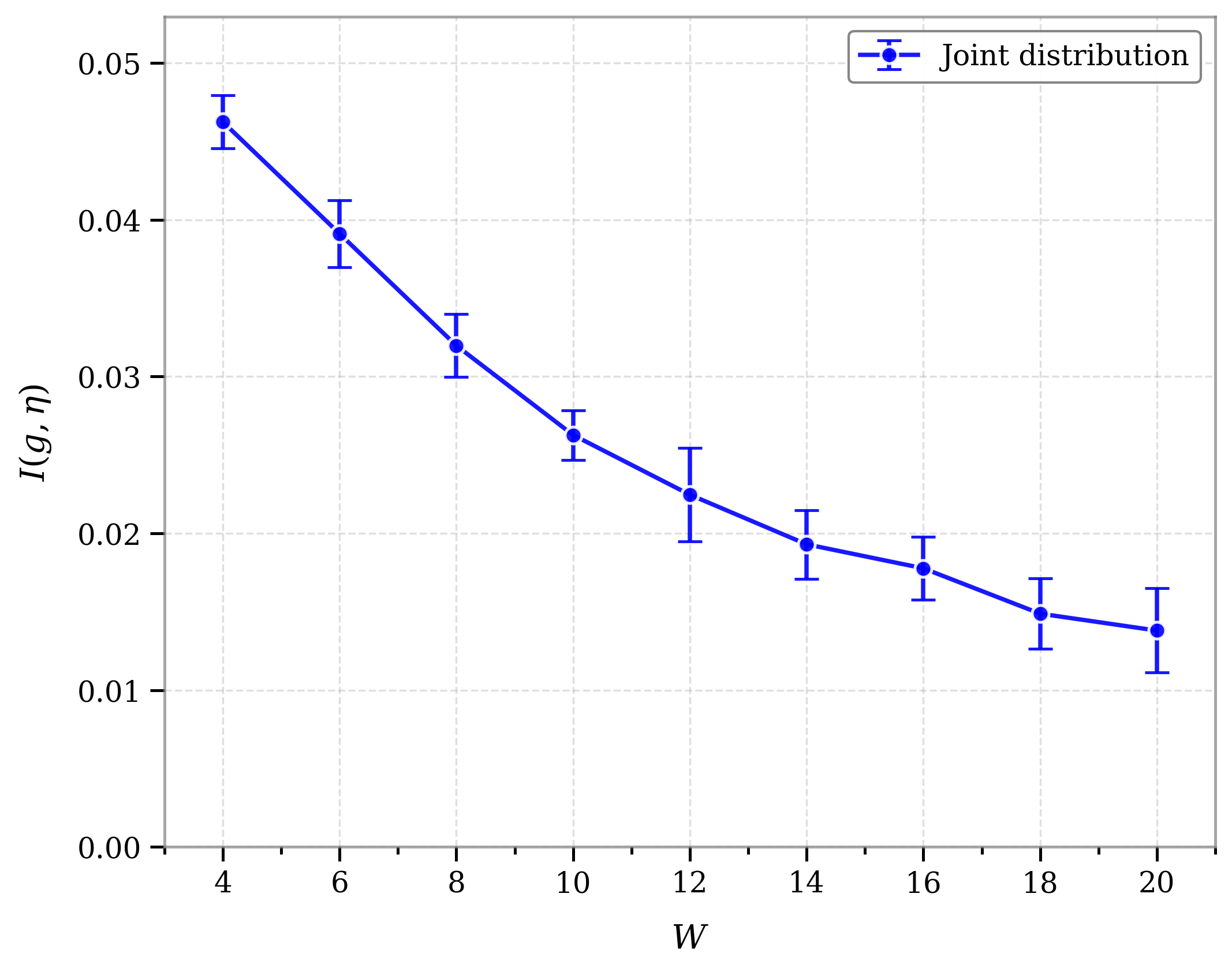}
    \end{center}
    \vspace{-0.5cm}
    \caption{\small{Mutual information of cavity Green's functions and cavity auxiliary fields. The mutual information, computed numerically following the procedure devised in Ref.~\cite{kraskov2004estimating} from the population dynamics distribution with a population of $N=5 \times10^5$ cavity variable pairs, is plotted for many values of the disorder strength above the lower critical disorder $W_{\rm min}$ (see Sec. \ref{subsubsec:lowW}}) for $K=5,\, t=1$. As it can be seen from the plot, the mutual information decreases monotonically with $W$, remaining always below $5\cdot 10^{-2}\, \mbox{\rm bits}$. This indicates that the factorization approximation of Eq.~\eqref{eq:MF} represents the exact joint distribution with good accuracy.}
    \label{fig:MI}   
\end{figure}

The estimation of the mutual information can be done using population dynamics. 
Assuming that the population of $N$ pairs of cavity Green's functions and cavity rescaled fields represents a set of $N$ samples drawn from the exact distribution, we can compute $I(\mathcal{G}_{i\to j},\eta_{i\to j})$ empirically through the procedure devised in Ref.~\cite{kraskov2004estimating}.
We computed the mutual information with $K=5,\,t=1$, and  many values of $W$ in the range that will be used later to compute the critical curve. The results are plotted in Fig.~\ref{fig:MI}. As we can see from the plot, the mutual information is of $O(10^{-2}\,bits)$ or less for the analyzed range of disorder, indicating that the factorization approximation describes the real distribution with good accuracy for this set of parameters. Moreover, it has been checked that for a fixed value of $W$, the mutual information decreases as $K$ increases, signaling that the factorization approximation improves for increasing $K$.

The mutual information analysis indicates that the quantities $\mu_\eta$ and $\sigma^2_\eta$ can be computed self-consistently with minimal error through Eqs.~\eqref{eq:meanvaretag}, and that $\mu_u$ and $\sigma^2_u$ can then be obtained from Eqs.~\eqref{eq:ustats}.

It should be noted that the mutual information result does not guarantee that the analytical expression of the cavity rescaled field distribution in Eq.~\eqref{eq:Peta}---derived under the assumption of independence between $g$ and $\eta$---accurately reproduces the true marginal. This expression relies on the additional approximation introduced in Eq.~\eqref{eq:approxsumeta}.
However, the self-consistent distribution of Eq.~\eqref{eq:P_u}, which incorporates the Gaussian approximation of Eq.~\eqref{eq:Peta}, closely matches the marginal distribution of the cavity Localization Landscapes obtained from population dynamics.

As discussed in Sec.~\ref{subsubsec:divergence-cluster-size}, the percolation transition at high connectivity is governed by the statistics of the cavity Localization Landscapes. Since the self-consistent prediction of Eq.~\eqref{eq:P_u} accurately reproduces the numerical results, we computed the high-connectivity critical curve by sampling from the factorized distribution $P_g(g)P_\eta(\eta)$, determined self-consistently from Eqs.~\eqref{eq:Pgacc} and \eqref{eq:Peta}. The resulting critical energies (solid black line in Fig.~\ref{fig:critcurve}) show excellent agreement with the exact ones obtained from sampling the full joint distribution.

\begin{figure}[!ht]
\vspace{0.5cm}
    \begin{center}
          \includegraphics[width=\textwidth]{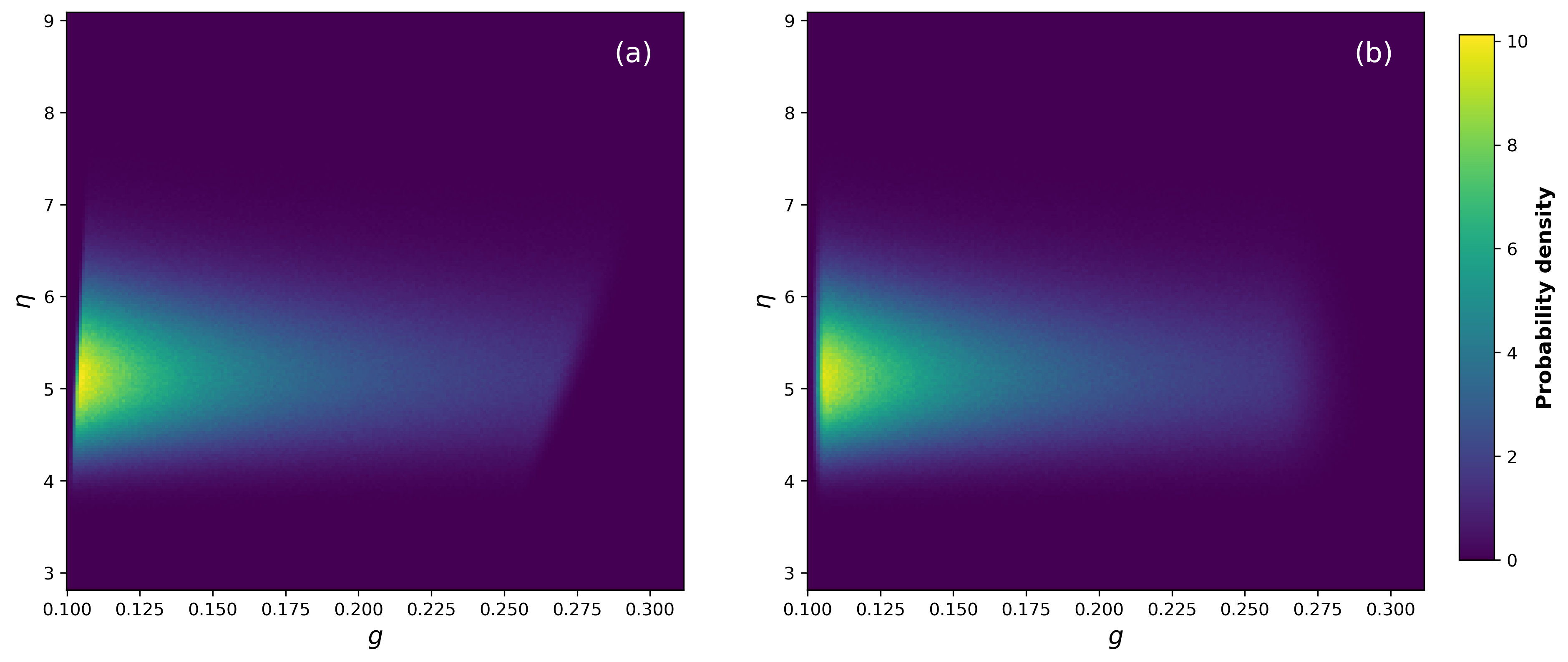}
   \end{center}
       \vspace{-0.5cm}
    \caption{\small{Plots of $P(g,\eta)$ and its mean-field approximation for $K=5,t=1$ and $W=6$. Left: plot of the histogram of the joint probability distribution obtained through population dynamics (see Sec. \ref{subsec:population}}) with $N=10^7$. Right: plot of the mean field approximation obtained by shuffling the cavity Green's functions of the pairs $(g,\eta)$ in the histogram on the left. As it can be seen from the plot, the main qualitative differences are in the tails of the distribution, and they're due to the fact that for extreme values of $\mathcal{G}_{i \to j}$ the set of values that $\eta_{i \to j}$ can take are restricted by Eqs. \eqref{eq:gsys} and \eqref{eq:etasys}.}
    \label{fig:Pgetas}
\end{figure}

In Fig.~\ref{fig:Pgetas} we represented the colormap plots of the histograms of the joint probability distribution $P(g,\eta)$, and its factorized approximation $P_g(g)P_\eta(\eta)$. The first one has been computed numerically through population dynamics, while the second one has been obtained by shuffling the $g$-variables between the pairs $(g,\eta)$ in the latter distribution, erasing the dependencies of the two variables. As we can see from the plots the two distributions are qualitatively similar. The main difference between the two is that the tails of the distributions have a slightly different shape in the $(g,\eta)$ plane. This is due to the fact that the values that $\eta_{i \to j}$ can take once $\mathcal{G}_{i \to j }$ is given are restricted by the coupled equations \eqref{eq:gsys} and \eqref{eq:etasys}. As revealed from the mutual information measurements, this feature does not result in significant effects on the global dependency of the two variables. 

\subsubsection{Lower bound for the critical disorder and isolated eigenvalue}
\label{subsubsec:lowW}

It is important to note that for physical consistency, the expectation value of the cavity rescaled fields must be non-negative, as the opposite would imply that the Localization Landscapes are negative, which is impossible as explained in the End Matter of the paper. 
Therefore, Eqs.~(\ref{eq:etastats}) must be only valid in the regime where
\begin{equation}
\label{eq:condboundW}
    0\leq \mu_\eta<+\infty \iff \mu_g <  \frac{1}{Kt}
    \; .
\end{equation}
A more formal derivation of this condition will be discussed in the following.

In the high-connectivity limit, Eq.~(\ref{eq:roughappr}) indicates that $\mu_g$ is a decreasing function of the disorder strength for $W>W_{\rm min}$. For $W<W_{\rm min}$, however, the value of $\mu_g$ depends explicitly on $E_{\rm iso}(W)$. This quantity was computed numerically as a function of $W$ in Ref.~\cite{biroli2010anderson} using the property that the Gaussian integral
\begin{equation}
    \label{eq:gaussy}
    \langle x_i\rangle = \frac{1}{Z_0}\int {\cal D}{\mathbf x} \;  x_i \; e^{-S_0[{\mathbf x}]}\,,  
\end{equation}
where
\begin{equation}
    S_0[\mathbf{x}]=\frac{1}{2}\mathbf{x^t}\left(\hat{\mathcal{H}}-E\hat{\mathcal{I}}\right)\mathbf{x}\,,
\end{equation}
and $Z_0$ is a normalization constant, must diverge when $E>E_{\rm iso}$. By introducing a source term $Jx_i$ in the action, one can follow a procedure analogous to that in the End Matter of the paper to compute this expectation value. Parametrizing a generic cavity marginal distribution as
\begin{equation}
    \mu_{i\to j}(x_i) \propto e^{-\frac{x_i^2}{2 \, \mathcal{G}_{i \to j}} + y_{i \to j} x_i} \, ,
\end{equation}
the cavity equations to be enforced are:
\begin{equation}
    \mathcal{G}_{i\to j}^{-1} = \varepsilon_i - t^2 \sum_{k \in \partial i \setminus j} \mathcal{G}_{k \to i}    
    \; , 
    \qquad\qquad
    y_{i \to j}  = t \sum_{k \in \partial i \setminus j} 
    \mathcal{G}_{k \to i} y_{k \to i}   \, . 
\end{equation}
Here, the $E$-dependence is contained in the random variables $\varepsilon_i$, which are uniformly distributed in the interval $[-W/2-E, W/2-E]$.
The recursion for the variables $y_{i \to j}$ is linear and has only two fixed points: $0$ and $+\infty$ (if initialized with a distribution with strictly positive domain). The fixed point at $0$ corresponds to the case where the Gaussian integral in Eq.~\eqref{eq:gaussy} converges. By identifying the value of $E$ at which this recursion first diverges, one can determine $E_{\rm iso}$.

Understanding when the recursion converges results pretty simple after taking a few assumptions that can be verified numerically. We present here an argument based on studying the behavior of the population dynamics algorithm for this recursion (see Sec.~\ref{subsec:population}). We consider the original formulation of the population dynamics algorithm, i.e. the one used in \cite{biroli2010anderson}, where at each timestep we update the full population.  For the sake of simplicity we initialize the population of $y$-variables at time $\tau=0$ according to an initial distribution with finite mean $\mu_y^{(0)}>0$ and finite variance $\sigma_y^{2\,(0)}$. Then, for each of the y-variables in the population we associate a cavity Green's function drawn from its stationary marginal distribution.
With this choice, using the same notation of Sec.~\ref{subsec:population}, our initial population is $\{(\mathcal{G}_i^{(0)},\,y_i^{(0)})\}_{i\leq N}$.
At each iteration step $\tau$ the algorithm draws randomly with replacement $N$ sets of $K$ nearest neighboring pairs from the population $\{(\mathcal{G}_i^{(\tau-1)},\,y_i^{(\tau-1)})\}_{i\leq N}$, and uses them to compute the new population $\{(\mathcal{G}_i^{(\tau)},\,y_i^{(\tau)})\}_{i\leq N}$. Pointing our attention on a generic variable $y_i$ in the population, after $T$ iterations of the algorithm its value can be written as
\begin{equation}
\label{eq:popdyny}
    y_i^{(T)} = t\sum_{k_{T-1}}\mathcal{G}^{(T-1)}_{k_{T-1}}y_{k_{T-1}}^{(T-1)} =\dots =t^T\sum_{k_{T-1}}\dots \sum_{k_0}\mathcal{G}^{(T-1)}_{k_{T-1}}\dots\mathcal{G}^{(0)}_{k_{T-1},\dots,k_{0}}\,y_{k_{T-1},\dots,k_{0}}^{(0)}
\end{equation}
In this equation one can see the summation as a sum over simple paths on a Cayley tree with $T$ generations, where the leaves are associated to initial conditions, and have time index $0$, and the root is associated to $y_i^{(T)}$. Defining as $\mathcal{P}(\tau)$ the set of all paths connecting a site at time $\tau$ to a site at time $T-1$ we can rewrite $y_i^{(T)}$ as
\begin{equation}
    y_i^{(T)} = t^T\sum_{\rho \in \mathcal{P}(0)} \mathcal{G}_{\rho_{T-1}}^{(T-1)}\dots\mathcal{G}_{\rho_0}^{(0)}y_{\rho_0}^{(0)}\,. 
\end{equation}
With this representation the statistical dependence of the variables in the sum is much more easily understood. The expectation value of $y_i^{(T)}$ can be computed as
\begin{equation}
\label{eq:muy1}
    \mu_y^{(T)}\equiv\mathbb{E}[y_{i}^{(T)}]=\mu_y^{(0)}t^T\sum_{\rho \in \mathcal{P}(0)} \mathbb{E}\left[\mathcal{G}_{\rho_{T-1}}^{(T-1)}\dots\mathcal{G}_{\rho_0}^{(0)}\right]= \mu_y^{(0)}t^T\sum_{\rho \in \mathcal{P}(1)} \sum_{\rho_0}\mathbb{E}\left[\mathcal{G}_{\rho_{T-1}}^{(T-1)}\dots\mathcal{G}_{\rho_1}^{(1)}\mathcal{G}_{\rho_0}^{(0)}\right]\,.
\end{equation} 
In principle, since the statistical dependence propagates from the leaves to the root, one cannot write the latter expectation value as a product of the expectation values of the cavity Green's function on the path $p$, however, in the high-connectivity limit, as $\mathcal{G}_{\rho_1}^{(1)}$ depends explicitly on the sum $\sum_{\rho_0} \mathcal{G}_{\rho_0}^{(0)}$, and the cavity Green's functions have a thin tailed stationary distribution (as we have argued in Sec.~\ref{subsubsec:marginals}), one can assume that the correlations between  $\mathcal{G}_{\rho_1}^{(1)}$ and each individual $\mathcal{G}_{\rho_0}^{(0)}$ are negligible. Thus, in Eq. \eqref{eq:muy1} one has:
\begin{equation}
    \sum_{\rho_0}\mathbb{E}\left[\mathcal{G}_{\rho_{T-1}}^{(T-1)}\dots\mathcal{G}_{\rho_1}^{(1)}\mathcal{G}_{\rho_0}^{(0)}\right] = \mathbb{E}\left[\mathcal{G}_{\rho_{T-1}}^{(T-1)}\dots\mathcal{G}_{\rho_1}^{(1)}\right]K\mu_g^{(0)}
\end{equation}
 Since the Green's function were initialized in their stationary distribution, $\mu_g$ remains constant and equal to its stationary value under iteration of the algorithm, and iterating the previous argument in Eq.~\eqref{eq:muy1} one finally obtains
\begin{equation}
    \mu_y^{(T)}= \mu_y^{(0)}(Kt\mu_g)^T \,.
\end{equation}

This implies that:
\begin{equation}
\label{eq:itery}
    \mu_y {=} \lim _{T\to \infty} \mu_y^{(T)} = \begin{cases} 
        0 \quad&\text{if}\quad \mu_g<1/Kt\,,\\
        \infty \quad&\text{if}\quad \mu_g> 1/Kt\,.
    \end{cases}
\end{equation}
Finally, $E_{\rm iso}(W)$ must be the one such that the fixed point of the recursion for the $y$-variables passes from $0$ to $+\infty$, i.e. the one for which the expectation value of the cavity Green's function for $W<W_{\rm min}$ is equal to $1/Kt$.  
We can now use the high-connectivity results of Sec.~\ref{subsubsec:marginals} to obtain the value of $E_{\rm iso}$ using the roughest approximation of $\mu_g$ from Eq.~\eqref{eq:roughappr}.
We have that
\begin{equation}
    \frac{1}{Kt}=\frac{1}{W}\ln \left(1-\frac{W}{W/2+E_{\rm iso}(W)+t}\right)\,,
\end{equation}
thus
\begin{equation}
\label{eq:Eiso}
    E_{\rm iso}(W)= -\frac{W}{2}-t+\frac{W}{1-e^{W/Kt}}\,.
\end{equation}
It is important to note that this equation correctly reproduces the zero-disorder value of the isolated eigenvalue, $E_{\rm iso}(0)=-t(K+1)$. Figure~\ref{fig:critcurve} shows the resulting prediction for $-E_{\rm iso}(W)$ for the case $K=5, \,t=1$.
Another important consequence is that when we return to the Localization Landscape percolation equations with the translated Hamiltonian from Sec.~\ref{subsec:cavity-derivation}, we find that $\mu_g(W)\equiv1/Kt$ for all $W\leq W_{\rm min}$. Consequently, according to the condition established earlier in Eq.~\eqref{eq:condboundW}, the percolation transition ceases to exist precisely at $W=W_{\rm min}$. This means that the minimal critical disorder for the Localization Landscape percolation transition, below which the system is always in the percolating phase, coincides with the disorder strength at which the isolated eigenvalue enters the bulk of the spectrum. This value $W_{\rm min}$ can be computed from Eq.~\eqref{eq:roughappr} as:
\begin{equation}
\label{eq:bound}
    W_{\rm\min}=Kt \ln\left| 1+\frac{W_{\rm min}}{t(2\sqrt K-1)} \right|\,,
\end{equation}
that can be solved self-consistently by iteration.

Note that if we chose to translate the Hamiltonian of a value $-E>-E_{\rm min}$ this would have decoupled $W_{\rm min}$ from the lower critical disorder for percolation.
A further confirmation of the equivalence of $W_{\rm min}$ and the lower critical disorder for percolation is given by observing that, following the same procedure of Eq.~\eqref{eq:popdyny}, we can compute the explicit expression for an $\eta_{i\to j}$ at time $T$ obtained through population dynamics from $\{(\mathcal{G}_i^{(0)},\,\eta_i^{(0)})\}_{i\leq N}$, according to the cavity equations~\eqref{eq:gsys},\eqref{eq:etasys}. What we find is
\begin{equation}
    \eta_i^{(T)}=1 +t\sum_{k_{T-1}}\mathcal{G}^{(T-1)}_{k_{T-1}}\eta_{k_{T-1}}^{(T-1)} =\dots 
    =1+\sum_{\tau=1}^{T-1}t^{T-\tau}  \sum_{\rho \in \mathcal{P}(\tau)} \mathcal{G}_{\rho_{T-1}}^{(T-1)}\dots\mathcal{G}_{\rho_\tau}^{(\tau)}+t^T \sum_{\rho \in \mathcal{P}(0)} \mathcal{G}_{\rho_{T-1}}^{(T-1)}\dots\mathcal{G}_{\rho_0}^{(0)}\eta_{\rho_0}^{(0)}\,.
\end{equation}
After taking the expectation value, using the same argument as for Eq.~\eqref{eq:muy1} we have
\begin{equation}
    \mu_\eta^{(T)}= \sum_{\tau=0}^{T-1} (Kt\mu_g)^\tau+\mu_\eta^{(0)}(Kt\mu_g)^T\,,
\end{equation}
thus,
\begin{equation}
    \mu_\eta = \lim_{T\to \infty} \mu_\eta^{(T)}=
    \begin{cases}
        1/(1-Kt\mu_g) \quad &\text{if} \quad \mu_g< 1/Kt\,,\\
        \infty \quad &\text{if} \quad \mu_g \geq 1/Kt\,.
    \end{cases}
\end{equation}
This result reproduces the result of Eq.~\eqref{eq:etastats} and contains, equivalently to Eq.~\eqref{eq:itery}, the information about the position of the isolated eigenvalue $E_{\rm iso}$, since in the large $T$ limit, neglecting the last term of the series, the sequence of the expectation values $\mu_\eta^{(T)}$ is proportional to the sequence of the partial sums of $\mu_y^{(\tau)}$:
\begin{equation}
    \mu_\eta^{(T)}\propto \sum_{\tau=0}^{T-1} \mu_y^{(\tau)}
\end{equation}

\subsubsection{Linear stability analysis}
\label{subsubsec:linstab}

Following the same idea of \cite{abou1973selfconsistent}, and already used in the analysis of the Anderson problem in Sec.~\ref{sec:ALeigenval}, 
we start from the self-consistent distributional equation (\ref{eq:stochSC}). Using the integral representation of the Dirac delta
\begin{equation}
    \delta\bigg( p - \theta(g \eta - 1/E) \sum_k p_k \bigg)
    = \int_{-\infty}^{\infty} 
    \frac{d\lambda'}{2\pi} \,e^{-{\rm i} \lambda' \left( p - \theta(g\eta - 1/E) \sum_k p_k \right)} \, ,
\end{equation}
we can rewrite
\begin{align}
    P(g,\eta,p)&=\int \frac{d\lambda'}{2\pi} \, e^{i\lambda' p}\int d\varepsilon \, \gamma_+(\varepsilon)
    \int \prod_{k=1}^K \left[dg_kd\eta_kdp_k \,\hat{P}\big ( g_k,\eta_k ,\theta(g\eta-1/E)\lambda'\big )\right]
    \nonumber\\ 
    & \quad\qquad\qquad \times\delta \bigg ( g-\frac{1}{\varepsilon-t^2\sum_k g_k}\bigg ) 
     \delta \bigg (  \eta- 1-t\sum_k g_k \eta_k \bigg ).
\end{align}
Then, taking the Fourier transform of the latter with respect to $p$ and recognizing the integral representation of $\delta(\lambda-\lambda')$, we derive
\begin{align}
    \hat{P}(g,\eta,\lambda)&=\int d\varepsilon \,\gamma_+(\varepsilon) 
    \int \prod_{k=1}^K \left[dg_kd\eta_k \, \hat{P}\big( g_k,\eta_k ,\; \theta(g\eta-1/E)\lambda\big)  \right]
    \nonumber\\ 
    & \quad\qquad\qquad
    \times \delta \bigg ( g-\frac{1}{\varepsilon-t^2\sum_k g_k}\bigg )\delta \bigg (  \eta- 1-t\sum_k g_k \eta_k \bigg ).
\end{align}
The theta function in the argument of the characteristic function of the distribution distinguishes between two 
cases. For $g \eta>1/E$ we have to solve the integral equation
\begin{equation}
    \label{eq:chfunc}
    \hat{P}(g,\eta,\lambda)=\int d\varepsilon \,\gamma_+(\varepsilon)\int \prod_{k=1}^K \left[dg_kd\eta_k \, \hat{P}\big(g_k,\eta_k,\lambda\big) \right] \;
    \delta \bigg ( g-\frac{1}{\varepsilon-t^2\sum_k g_k}\bigg )\delta \bigg (  \eta- 1-t\sum_k g_k \eta_k \bigg ),
\end{equation}
while for $g\eta<1/E$ we simply have
\begin{equation}
    \hat{P}\big( g_k,\eta_k,\theta(g\eta-1/E)\lambda)=
    \hat{P}( g_k,\eta_k,0)=P( g_k,\eta_k ),
\end{equation}
thus
\begin{equation}
    P(g,\eta,p)=P(g,\eta)\delta(p),
\end{equation} 

In order to find an equation for the critical curve, we are interested in finding a solution for the probability distribution that admits a finite expectation value of the cavity percolation probability. Since its distribution necessarily has finite mean and variance (because $p \in [0,1]$), we can assume that at criticality the characteristic function for $g\eta>1/E$ can be expanded in powers of $\lambda$, and that the term of the smallest order is exactly of order one:
\begin{equation}
    \hat{P}(g,\eta,\lambda)=P(g,\eta)+{\rm i} \lambda f(g,\eta) +O(\lambda^2).
\end{equation}
Substituting in the integral Eq.~(\ref{eq:chfunc}), and discarding terms of order higher than one we obtain a self-consistent equation for $f(g,\eta)$ that reads
\begin{align}
    f(g,\eta)&=K\int d\varepsilon\, \gamma_+(\varepsilon) \,dg'd\eta' \,f(g',\eta') 
    \int \prod_{k=1}^{K-1}\left[dg_kd\eta_k \, P( g_k,\eta_k )\right] 
    \nonumber \\ &
   \qquad\qquad\quad  \times\delta \bigg ( g-\frac{1}{\varepsilon-t^2 \big(g'+\sum_{k=1}^{K-1} g_k\big)}\bigg )
    \delta \bigg (  \eta- 1-t \Big (g'\eta'+ \sum_{k=1}^{K-1} g_k \eta_k \Big ) \bigg )\,,
\end{align}
and can be recast as
\begin{equation}
\label{eq:eigintperc}
    f(g,\eta)=\int dg'd\eta'\,\mathcal{K}_{\rm perc}(g,\eta,g',\eta')f(g',\eta'),
\end{equation}
with
\begin{equation}
    \mathcal{K}_{\rm perc}(g,\eta;g',\eta')=\int d\varepsilon d\tilde{g}d\tilde{u} \,\gamma_+(\varepsilon)R_{\rm perc}(\tilde{g},\tilde{u})\delta \bigg ( g-\frac{1}{\varepsilon-t^2 (g'+\tilde{g})}\bigg )\delta \bigg (  \eta- 1-t \Big (g'\eta'+ \tilde{u} \Big ) \bigg )\,, 
\end{equation}
where we have introduced the joint probability distribution of the sums $\sum_{l=1}^{K-1}g_l$ and $\sum_{l=1}^{K-1}g_l\eta_l$:
\begin{equation}
    R_{\rm perc}(\tilde{g},\tilde{u})
    \equiv
    \int \prod_{k=1}^{K-1}\left[dg_kd\eta_k \,P( g_k,\eta_k )\right]
    \;  \delta\bigg(\sum_{k=1}^{K-1}g_k-\tilde{g}\bigg)\delta\bigg(\sum_{k=1}^{K-1}g_k\eta_k-\tilde{u}\bigg).
\end{equation}

Performing the integrals over the tilde variables, the kernel simplifies to 
\begin{equation}
\label{eq:kernel_perc}
    \mathcal{K}_{\rm perc}(g, \eta;\, g', \eta') =
    \frac{K}{g^2\, t^3} \int d\varepsilon \, \gamma_+(\varepsilon) \;
    R_{\rm perc}\left(-g' - \frac{1 - \varepsilon g}{g t^2},\;
    -g'\eta'-\frac{1-\eta}{t}\right).
\end{equation}
Now, in the non-percolating phase, the solution with zero percolation probability, i.e. $P(g,\eta,p)=P(g,\eta)\delta(p)$,  is the stable one. This means that any infinitesimal perturbation of its characteristic function, under iteration of the recursive equation defining $\hat{P}(g,\eta,\lambda)$ will vanish. Instead, in the percolating phase, we expect that the first-order perturbative term will increase under iteration. Accordingly, the critical curve will be 
identified as the curve in the space of parameters where the first order correction remains stable under iteration, i.e. where the kernel $\mathcal{K}_{\rm perc}$ has top eigenvalue equal to $1$ (following exactly the same argument at the end of Sec. \ref{sec:ALeigenval}). 

The kernels of Eqs.~\eqref{eq:kernelAL} and \eqref{eq:kernel_perc} are significantly different, and this explains quantitatively the difference between the two critical behaviors. 

The integral operator can be computed numerically, using population dynamics to evaluate the distribution $R_{\rm perc}(\tilde{g}, \tilde{u})$, and it can be diagonalized numerically to obtain the critical curve with very high precision. 
However, this procedure is highly computationally expensive.
Therefore, in order to compute the critical curve in the high-connectivity limit we used a completely different technique that we present in the next section, where we also present an argument to show that the linear stability analysis produces the same result.

\subsubsection{Divergence of the average cluster size}
\label{subsubsec:divergence-cluster-size}

An alternative method to determine the critical curve involves deriving an expression for the average cluster size $S$ and identifying the values $(E, W)$ for which $S$ diverges.

In the general case, a cluster consists of a connected component of lattice sites $i$ where $u_i \geq 1/E$. The statistical dependence between $u_i$ and $u_k$ (for $k \in \partial i$) is complex, but in the high-connectivity limit the problem simplifies significantly. Here, we can treat $u_k \overset{d}{=} u_{k \to i} \overset{d}{=} \mathcal{G}_{k \to i} \eta_{k\to i}$, effectively breaking the mutual dependence between sites $i$ and $k$: $u_i$ depends on all its neighbors, while $u_{k\to i}$ depends on all its neighbors except $i$.

In this high-connectivity limit, we can replace the occupation variable $O_i$ from Eq.~\eqref{eq:Oiperc} with
\begin{equation}
    \label{eq:cavOi}
    O_{i\to j} = \begin{cases}
        1 & \text{if } u_{i \to j} \geq 1/E\, \\
        0 & \text{otherwise}
    \end{cases}\,.
\end{equation}
By substituting in the definition of Eq.~(\ref{eq:corrfuncnobond}) the set of occupation variables $\{O_1,\dots,O_{r}\}$ with $\{O_{1 \to 0},\dots,O_{r \to r-1}\}$, we can factorize Eq.~\eqref{eq:corrfuncnobond} as
\begin{align}
\label{eq:corrfunchicon}
    C_{\rm perc}(r) &= \Pr \{O_0=1, O_{1 \to 0}=1, \dots, O_{r \to r-1}=1\} \nonumber \\
           &= \left[ \prod_{s=2}^r \Pr \{O_{s\to s-1} =1 \mid O_{s-1 \to s-2}=1\} \right] \Pr \{O_{1\to 0}=1 \mid O_0=1\} \Pr \{O_0=1\}\,.
\end{align}

Due to translational invariance, the conditional probability that a site is occupied given that one of its nearest neighbors is occupied is independent of the specific pair of sites considered. Thus, we define
\begin{equation}
\label{eq:qbar}
    \bar{q} \equiv \Pr \{ O_{k\to i} =1 \mid O_{i\to j}=1 \} \quad \forall i, \forall j \in \partial i,\forall k \in \partial i \setminus j\,.
\end{equation}
Similarly, the last factor in Eq.~\eqref{eq:corrfunchicon} is given by the unconditional occupation probability $q$ (see Eq.~\eqref{eq:q}).
Using these definitions, $C_{\rm perc}(r)$ takes the form
\begin{equation}
    C_{\rm perc}(r) = q \bar{q}^r = \frac{q}{K^r} e^{-r/\xi_{p}},
\end{equation}
where the correlation length is given by
\begin{equation}
    \xi_{\rm perc} = - \frac{1}{\ln K\bar q}.
\end{equation}

As in the independent-site percolation problem~\cite{stauffer2018introduction}, the correlation function decays exponentially, so the average cluster size, defined in Eq.~\eqref{eq:avgclsize}, simplifies to
\begin{equation}
\label{eq:S}
    S =  \frac{1+\bar q}{1 - K \bar{q}}.
\end{equation}
We immediately see that $S$ diverges when
\begin{equation}
\label{eq:hiconcrit}
    \bar{q} = \bar{q}_c = 1/K.
\end{equation}
This condition is analogous to that of the independent percolation problem, except that the occupation probability is now conditioned on the occupation of a neighboring site.
By numerically evaluating the pairs $(E, W)$ that satisfy $\bar q = 1/K$, we obtain the critical curve. 
The numerical procedure to compute $\bar q$ is explained in detail in Sec.~\ref{app:numeval}. 

The solution of the integral integral eigenvalue equation~\eqref{eq:eigintperc} amounts to determining the critical energy at which the recursion for $p_{i\to j}$, defined in the linearized version in Eq.~\eqref{eq:psys} diverges. In the following we show, using an argument analogous to the one of Sec.~\ref{subsubsec:lowW} for the population dynamics of $y_{i\to j}$ (see Eq.~\ref{eq:itery}), that this divergence occurs when $\bar q =1/K$. Let us consider a population of cavity variables as $\{\mathcal{G}_i^{(0)},\,\eta_i^{(0)},\,p_i^{(0)}\}_{i\leq N}$, and for a generic element of the population at a generic time $\tau$ we denote $O^{(\tau)}_i=\theta(\mathcal{G}_i^{(\tau)}\eta_i^{(\tau)}-1/E)$. Here we assume that $\{\mathcal{G}_i^{(0)},\,\eta_i^{(0)}\}_{i\leq N}$ are drawn from their joint stationary distribution, and $\{p_i^{(0)}\}_{i\leq N}$ from an initial distribution bounded to $[0,1]$  with mean $\mu_p^{(0)}$ and variance $\sigma_p^{2\,(0)}$. As we described in Sec.~\ref{subsubsec:lowW} for the value $y_i^{(0)}$ of a generic $y$-variable at iteration $T$ of population dynamics, one can write $p_i^{(T)}$ as the result of the iteration of its cavity equation in the population dynamics algorithm in terms of paths over a Cayley tree with $T$ generations, where the leaves are associated to initial conditions and the root to the final value. Defining as $\mathcal{P}(\tau)$ the set of all paths connecting a site at time $\tau$ to a site at time $T-1$ we can write $p_i^{(T)}$ as
\begin{equation}
        p_i^{(T)} =O_i^{(T)}\sum_k p_k^{(T-1)}=\dots =O_i^{(T)}\sum_{k_{T-1}}\dots \sum_{k_0} O_{k_{T-1}}^{(T-1)}\dots O_{k_{T-1,\dots,1}}^{(1)}p_{k_0}^{(0)}=O_i^{(T)}\sum_{\rho \in \mathcal{P}(0)} O_{\rho_{T-1}}^{(T-1)}\dots O_{\rho_1}^{(1)}p_{\rho_0}^{(0)}\,.
\end{equation}
Now, taking the expectation value we have
\begin{equation}
    \mu_p^{(T)}= \sum_{\rho \in \mathcal{P}(1)}\mathbb{E}\Bigg[O_i^{(T)} O_{\rho_{T-1}}^{(T-1)}\dots O_{\rho_1}^{(1)}\Bigg]K\mu_p^{(0)}=\sum_{\rho \in \mathcal{P}(1)}\mathbb{E}\Bigg[O_i^{(T)} O_{\rho_{T-1}}^{(T-1)}\dots O_{\rho_2}^{(2)}\Bigg|O_{\rho_1}^{(1)}=1\Bigg]qK\mu_p^{(0)} =\bar q(K\bar q)^{T-2}(Kq)(K\mu_p^{(0)}).
\end{equation}
In the second equality we used the same high-connectivity property that we used to factorize the correlation function in Eq.~\eqref{eq:corrfunchicon} (that this time is encoded in the algorithm itself), for which the statistical dependence of the occupation variables propagates from the leaves to the root, thus
\begin{equation}
     \Pr \{O_i, O_{\rho_{T-1}}, \dots, O_{\rho_0}\} 
           = \Pr \{O_i|O_{\rho_{T-1}}\} \prod_{\tau=0}^{T-1}\left[ \Pr \{O_{\rho_{\tau}} \mid O_{\rho_{\tau-1}}\} \right]  \Pr \{O_{\rho_{0}}\}\,.
\end{equation}
Finally, in the last equality we computed the expectation value summing from the leaves to the root and using definitions~\eqref{eq:q},\eqref{eq:qbar}.

The expectation value of $p_i^{(T)}$ as $T\to \infty$ reads
\begin{equation}
     \lim _{T\to \infty} \mu_p^{(T)} = \begin{cases} 
        0 \quad&\text{if}\quad \bar q<1/K\,,\\
        \infty \quad&\text{if}\quad \bar q> 1/K\,,
    \end{cases}
\end{equation}
indicating that the integral recursion of Eq.~\eqref{eq:eigintperc} diverges when $\bar q = 1/K$, marking the transition to the percolating phase.
 
\subsubsection{The phase diagram}
\label{subsubsec:PD}

\begin{figure}[!ht]
\vspace{0.3cm}
\centering
    \begin{minipage}{0.48\textwidth}
        \centering
        \includegraphics[width=\textwidth]{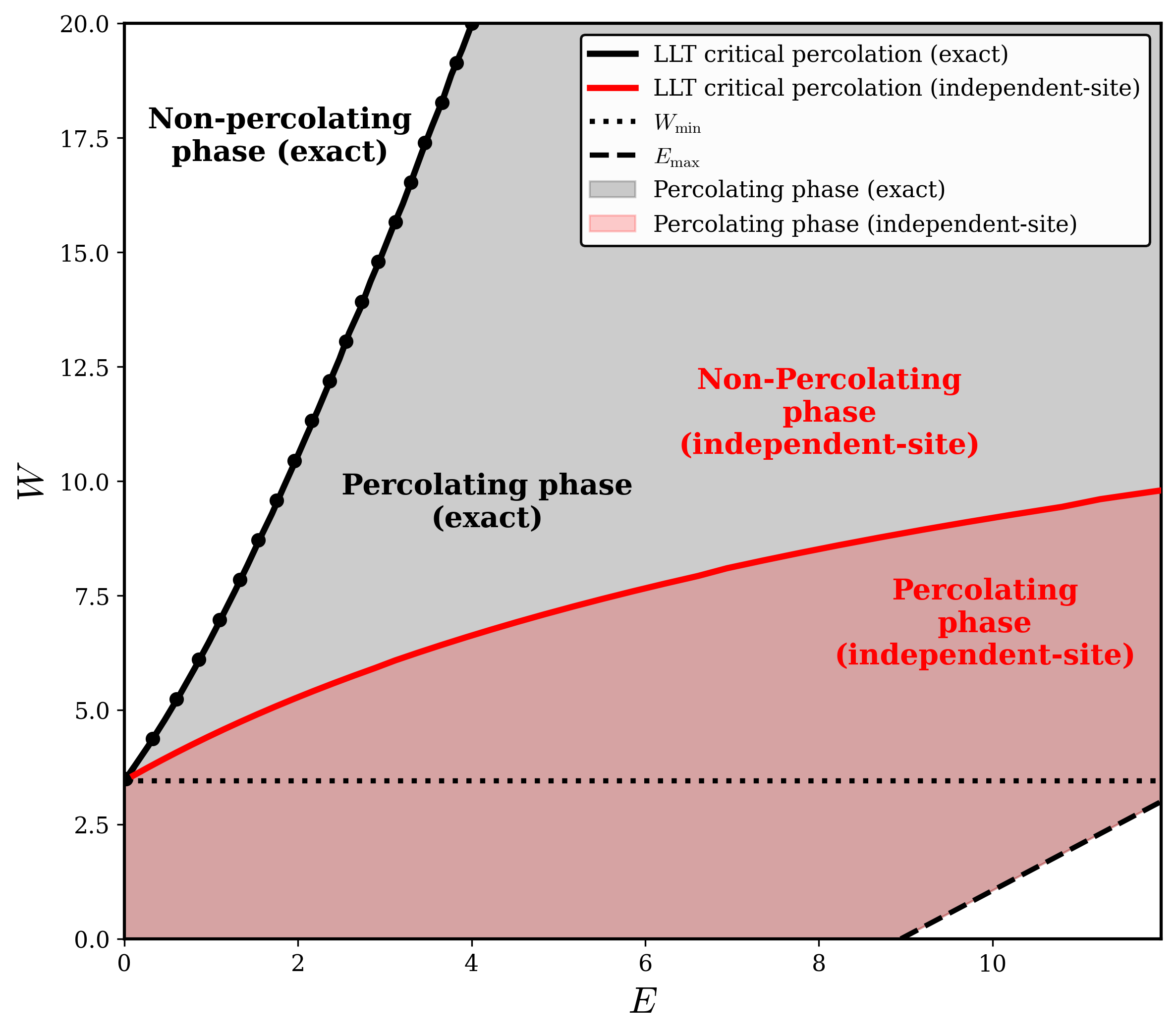}
    \end{minipage}
    \hfil
    \begin{minipage}{0.48\textwidth}
        \centering
        \includegraphics[width=\textwidth]{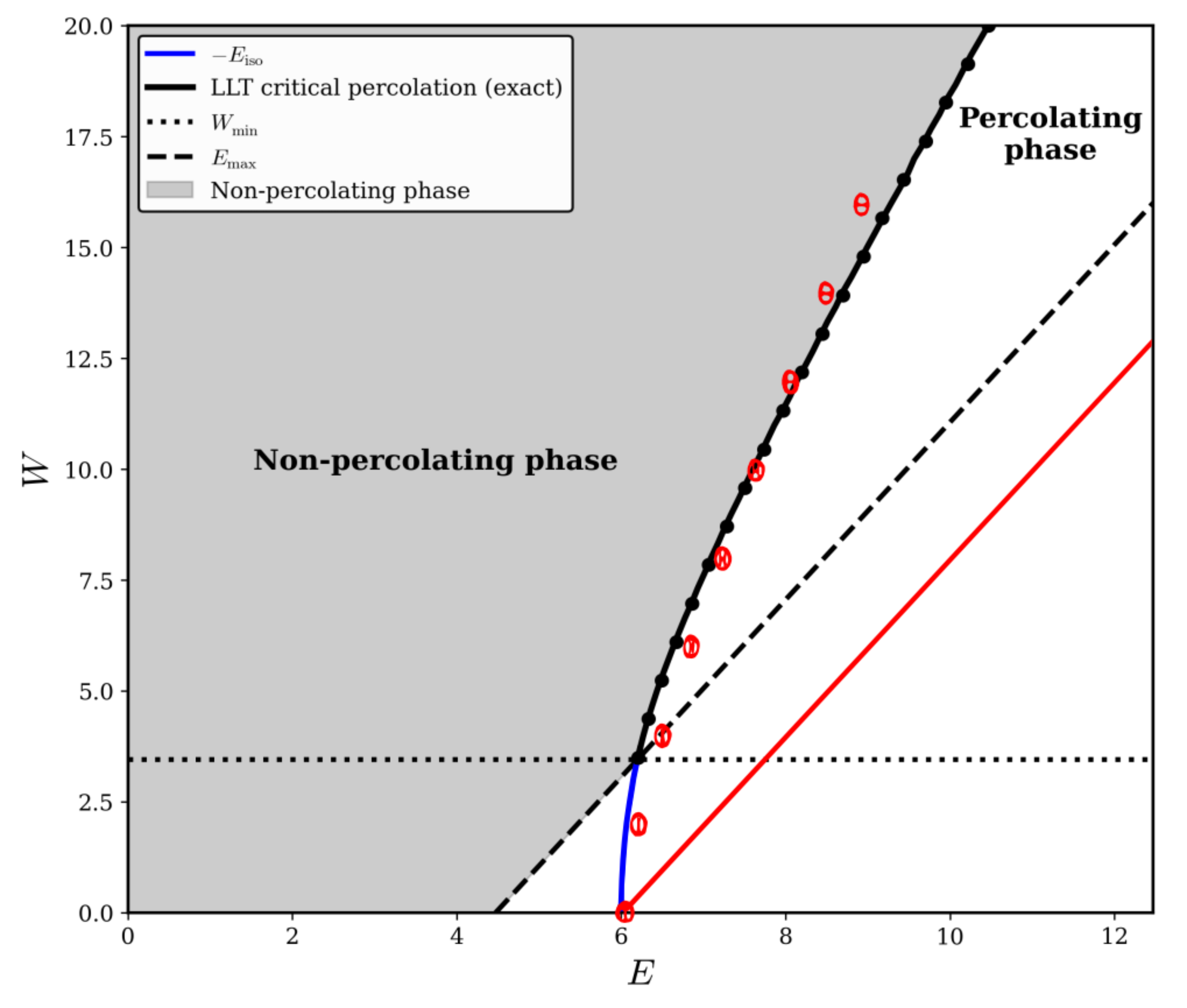}
    \end{minipage}
    \caption{\small{Percolation phase diagrams for $K=5$ and $t=1$. Left: positive definite Hamiltonian defined at the beginning of Sec.~\ref{subsec:cavity-derivation}. Red curve: critical curve separating percolating and non-percolating phases for high-connectivity in the independent-site approximation [Eq.~(\ref{eq:indsitescrit})].
    Black line: exact critical curve in the high-connectivity limit [Eq.~(\ref{eq:hiconcrit})]. Dashed line: upper boundary of the bulk of spectrum (see Sec.~\ref{subsec:cavity-derivation}). Dotted line: analytically predicted lower critical disorder [Eq.~(\ref{eq:bound})].
    Right: Hamiltonian with the statistically symmetric spectrum of Eq.~\eqref{eq:ham}. All the curves of the right plot have been obtained from the ones in the right plot by means of a translation and inversion of sign as explained in Sec.~\ref{subsubsec:PD}. The independent-site critical curve has not been plotted as it loses its physical interpretation in the symmetric case. Red dots: Localization Landscape percolation critical energies for the cubic lattice from \cite{filoche2024anderson}. Blue line: opposite of the isolated eigenvalue $-E_{\rm iso}$ for $W\leq W_{\rm min}$ from Eq.~\eqref{eq:Eiso}}}
    \label{fig:critcurve}
\end{figure}

Figure~\ref{fig:critcurve} presents the phase diagram obtained in the high-connectivity limit. The left panel displays both the curve derived from the independent-site approximation condition [Eq.~\eqref{eq:indsitescrit}] and the one obtained using the exact criterion [Eq.~\eqref{eq:hiconcrit}].

The left phase diagram is plotted in the $(E,W)$-plane for the positive definite Hamiltonian $\hat{\cal{H}}_+$ (see Sec.~\ref{subsec:cavity-derivation}). The right panel shows the positive-energy side of the phase diagram associated with the original Anderson Hamiltonian of Eq.~\eqref{eq:ham}. This was obtained from the left plot by translating the energies by $E_{\rm min}(W)$ to restore the statistically symmetric spectrum, and then switching $E \mapsto -E$ to display only the positive-energy region. Due to this transformation, the percolating and non-percolating phases are exchanged in the right image.

The exact curve for a given parameter pair $(K,t)$ requires sampling from the joint distribution $P(g,\eta)$, which properly accounts for correlations between cavity Green's functions and rescaled fields (see Sec.~\ref{sec:numerics} for numerical details). However, as anticipated from the results of Sec.~\ref{subsubsec:joint}, the phase diagram computed using the factorized approximation of Eq.~(\ref{eq:MF}) agrees excellently with that obtained from the joint distribution via population dynamics. In Fig.~\ref{fig:critcurve}, the solid black line corresponds to the curve derived from the self-consistent marginal distributions of Sec.~\ref{subsubsec:marginals}, while the black dots represent critical energies computed by sampling the joint distribution from population dynamics.

A key observation is that the independent-site curve deviates significantly from the exact one in the left panel, indicating that correlations between nearest-neighbor sites are essential to describe the physics of the system. Neglecting these correlations leads to incorrect results. This discrepancy is so pronounced that the transformed independent-site curve becomes physically meaningless in the symmetric case, and thus it is not plotted in the right panel.

The solid blue curve in the right panel shows the prediction of $-E_{\rm iso}(W)$ for $W \leq W_{\rm min}$ derived in Sec.~\ref{subsubsec:lowW}. As expected, it intersects the spectral boundary $E_{\rm max}(W) = 2t\sqrt{K} + W/2$ at the point where the percolation transition ceases to exist. The disorder value at which this occurs is indicated by a horizontal dotted line.

Although the high-connectivity equations are formally valid for $K \gg 1$, many probability distributions from Sec.~\ref{subsubsec:marginals} are accurate even at $K = 5$ (e.g., the marginals of the cavity Localization Landscapes and cavity Green's functions, as shown in Fig.~\ref{fig:PgPu}). Therefore, we plot both curves for $K = 5$, arguing that the exact high-connectivity condition captures many important features of the real curve at this connectivity, despite neglecting the influence of one neighbor out of six. Moreover, the solution on the Bethe lattice with $K = 5$ represents the Bethe approximation for the cubic lattice, allowing for direct comparison with the numerical results from Ref.~\cite{filoche2024anderson} (plotted as red circles in the right panel). The Bethe and cubic lattice transitions have a qualitative agreement. The main difference is that on the Bethe lattice, the spectral boundary prevents the curve from reaching zero disorder, as discussed in Sec.~\ref{subsubsec:lowW}.

\end{document}